\documentclass[paper]{JHEP3} 

\usepackage{amsfonts}
\usepackage{amsmath}
\usepackage{epsfig}
\usepackage{latexsym}
\usepackage{graphicx}

\usepackage{subfigure}
\usepackage{amssymb}
\numberwithin{equation}{section}
\allowdisplaybreaks

\def\be{\begin{equation}}
\def\ee{\end{equation}}
\def\bea{\begin{eqnarray}}
\def\eea{\end{eqnarray}}
\def\bequ{\begin{equation}}
\def\eequ{\end{equation}}

\def\del{\partial}

\renewcommand{\thefootnote}{\fnsymbol{footnote}}

\newcommand{\eq} {equation}
\newcommand{\eqa} {eqnarray}
\newcommand{\NN} {\mbox {$\nonumber$}}

\def\th{\theta}

\DeclareMathOperator{\Tr}{Tr}

\newcommand{\hf}{\frac{1}{2}}

\def\del{\partial}

\def\h#1{\widehat{#1}}

\def\rt#1{\sqrt{#1}}
\def\sitarel#1#2{\mathrel{\mathop{\kern0pt #1}\limits_{#2}}}

\newcommand{\nn}{\nonumber \\}

\title{Exact results on ABJ theory and the refined topological string}

\author{\large Masazumi Honda$^{a}$ and Kazumi Okuyama$^{b}$
\vspace*{0.5cm} \\
\llap{$^a$}Harish-Chandra Research Institute,\\
Chhatnag Road, Jhusi, Allahabad 211019, India\\
\llap{$^b$}Department of Physics, Shinshu University,\\
3-1-1 Asahi, Matsumoto 390-8621, Japan\\
\vspace*{0.5cm} \\
\email{masazumihonda@hri.res.in, kazumi@azusa.shinshu-u.ac.jp}}

\preprint{HRI/ST/1409}

\abstract{
We study the partition function of the ABJ theory,
which is the $\mathcal{N}=6$ superconformal Chern-Simons matter theory with gauge group $U(N)\times U(N+M)$
and Chern-Simons levels $(k,-k)$.
We exactly compute the ABJ partition function on a three sphere for various $k$, $M$ and $N$
via the Fermi gas approach.
By using these exact data,
we show that the ABJ partition function is completely determined by the refined topological string on local $\mathbb{P}^1 \times \mathbb{P}^1$,
including membrane instanton effects in the M-theory dual.
}


\begin{document}
\setcounter{footnote}{0}
\renewcommand{\thefootnote}{\arabic{footnote}}

\section{Introduction}
Progress in M-theory would hinge on deeper understanding of its non-perturbative effects.
A part of these effects appear as membrane instantons in M-theory \cite{Becker:1995kb}.
During the last couple of years, there has been remarkable progress in understanding the membrane instanton effects.
In \cite{Hatsuda:2013oxa}, the authors have completely determined
the non-perturbative structure of the free energy of M-theory on $AdS_4 \times S^7 /\mathbb{Z}_k$
based on many different previous results.
First of all, 
the M-theory on this background is expected to be dual to
low-energy effective theory of multiple M2-branes \cite{Maldacena:1997re}.
Meanwhile it turns out that
the M2-brane theory is described by an $\mathcal{N}=6$ supersymmetric Chern-Simons matter theory (CSM)
with gauge group $U(N)_k \times U(N)_{-k}$ commonly referred to as ABJM theory \cite{Aharony:2008ug}.
While M-theory region $(k\ll N^{1/5})$ cannot be accessed by the ordinary perturbative technique in the ABJM theory,
the supersymmetry localization \cite{Pestun:2007rz} reduces
the partition function of the ABJM theory on $S^3$ to a matrix integral \cite{Kapustin:2009kz} called ABJM matrix model.
This matrix model has been extensively studied 
by many approaches \cite{Marino:2009jd,Drukker:2010nc,Herzog:2010hf,Drukker:2011zy,Fuji:2011km,Okuyama:2011su,Marino:2011eh,Hanada:2012si,Hatsuda:2012hm,Putrov:2012zi,Hatsuda:2012dt,Calvo:2012du,Hatsuda:2013gj,Hatsuda:2013oxa,Kallen:2013qla}.
After the studies of the 't Hooft expansion \cite{Drukker:2010nc,Drukker:2011zy,Fuji:2011km,Marino:2009jd} and
the leading order in the M-theory limit \cite{Herzog:2010hf},
there  appeared a seminal work \cite{Marino:2011eh}, which rewrites the ABJM matrix model 
as an ideal Fermi gas system (see also \cite{Okuyama:2011su,Kapustin:2010xq}).
This Fermi gas approach enables us to reveal structures of 
the partition function \cite{Hatsuda:2012hm,Putrov:2012zi,Hatsuda:2012dt,Calvo:2012du,Hatsuda:2013gj,Hatsuda:2013oxa,Kallen:2013qla} and 
BPS Wilson loops \cite{Klemm:2012ii,Grassi:2013qva,Hatsuda:2013yua} in the ABJM theory,
including non-perturbative effects expected from the gravity side \cite{Cagnazzo:2009zh,Drukker:2011zy,Becker:1995kb}.
As a result, it turns out that
the ABJM free energy is completely determined by the refined topological string on 
the ``diagonal''
local $\mathbb{P}^1 \times \mathbb{P}^1$
whose K\"ahler parameters for two $\mathbb{P}^1$'s are equal \cite{Hatsuda:2013oxa}.

In this paper we extend the above analysis to the so-called ABJ theory \cite{Aharony:2008gk,Hosomichi:2008jb},
which is the $\mathcal{N}=6$ CSM with more general gauge group $U(N)_k \times U(N+M)_{-k}$.
The ABJ theory is  the low-energy effective theory of $N$ M2-branes on $\mathbb{C}^4 /\mathbb{Z}_k$
with $M$ fractional M2-branes at the singularity.
From the AdS/CFT perspective,
one expects that
this theory is dual to the M-theory on $AdS_4 \times S^7 /\mathbb{Z}_k$ with a discrete torsion, and
the type IIA superstring theory on $AdS_4 \times \mathbb{CP}^3$ with a nontrivial B-field holonomy.
Furthermore the recent studies \cite{Chang:2012kt,Giombi:2011kc} indicate that
this theory has also a relation to the $\mathcal{N}=6$ parity-violating Vasiliev theory on $AdS_4$ with a $U(N)$ gauge symmetry,
when $M,k\gg 1$ with $M/k$ and $N$ kept fixed.
Thus the ABJ theory is one of the most important supersymmetric gauge theories
having M/string/higher-spin theory dual.
The aim of this paper is to determine  
non-perturbative structure of free energy 
of the M-theory on $AdS_4 \times S^7 /\mathbb{Z}_k$ with the discrete torsion 
via the supersymmetry localization, Fermi gas formalism, and refined topological string.

Let us briefly summarize our result\footnote{
Recently an apparently similar conclusion was obtained in \cite{Matsumoto:2013nya} by a different approach.
While the values of the canonical partition function and finite part of grand potential presented in \cite{Matsumoto:2013nya}
are totally consistent with our values,
there is an important difference of physical interpretation between \cite{Matsumoto:2013nya} and ours.
We will clarify an important difference between conclusions of \cite{Matsumoto:2013nya} and our work in section \ref{sec:MM}.
}. Our starting point is a matrix integral representation for the ABJ partition on $S^3$ obtained by the localization method \cite{Kapustin:2009kz,Jafferis:2010un,Hama:2010av}:
\begin{\eqa}
Z^{(N,N+M)}(k) 
&=& \frac{i^{-\frac{1}{2}(N^2 -(N+M)^2 ){\rm sign}(k)}}{(N+M)!N!} \int_{-\infty}^\infty \frac{d^{N+M} \mu}{(2\pi )^{N+M}} \frac{d^N \nu}{(2\pi )^N}
e^{-\frac{ik}{4\pi}\left( \sum_{j=1}^{N+M} \mu_j^2 -\sum_{a=1}^N \nu_a^2 \right) } \NN\\
&&\times \Biggl[ 
  \frac{ \prod_{1\leq j<l \leq N+M}2\sinh{\frac{\mu_j -\mu_l }{2}} \prod_{1\leq a<b \leq N}2\sinh{\frac{\nu_a -\nu_b }{2}}}
        {\prod_{j=1}^{N+M} \prod_{b=1}^N 2\cosh{\frac{\mu_j -\nu_b }{2}}} \Biggr]^2 ,
\label{eq:ABJmat}
\end{\eqa}
which we call the ABJ matrix model.
After conjectured in \cite{Awata:2012jb}, 
the article \cite{Honda:2013pea} has proven that
this matrix model has another equivalent representation suitable for a Fermi gas approach as explained in the next section.
The Fermi gas expression of the ABJ matrix model is obtained by going to the 
grand canonical ensemble, {\it i.e.} by introducing the chemical potential
$\mu$ and summing over $N$, as usual in the statistical mechanics.
Moreover one can show that the computation of the grand canonical partition function boils down to 
construct a series of functions, which can be solved recursively.
Thus we can exactly calculate the canonical partition function for various $k$, $M$, up to 
a fairly high $N=N_{\rm max}$.
Indeed we have obtained exact values of the partition function
for $(k,M, N_{\rm max} )=(2,1,65 )$, $(3,1,31 )$, $(4,1,62 )$, $(4,2, 29 )$, $(6,1, 23 )$, 
$(6,2, 21 )$ and $(6,3,22 )$,
which are partly listed in appendix \ref{app:values}.
Then we compare these exact data with an expectation
from the refined topological string on local $\mathbb{P}^1 \times \mathbb{P}^1$ with general K\"ahler parameters,
which is a natural generalization of the ABJM case \cite{Hatsuda:2013oxa}.
We will argue that the grand potential $J_k^{(M)}(\mu )$ of the ABJ theory
is completely determined by the refined topological string free energy,
\begin{\eq}
J_k^{(M)}(\mu ) 
= F_{\rm top} (T_1^{\rm eff} ,T_2^{\rm eff} ; g_s) 
  +\frac{1}{2\pi i} \frac{\del}{\del g_s}\Biggl[ g_s F_{\rm NS} \left( \frac{T_1^{\rm eff} }{g_s}, \frac{T_2^{\rm eff} }{g_s} ;\frac{1}{g_s} \right) \Biggr] ,
\label{eq:conclusion}
\end{\eq}
where the parameter $g_s$ is related to the Chern-Simons level $k$ as
\begin{\eq}
g_s =\frac{2}{k} ,
\label{ourgs}
\end{\eq}
and $T_{1,2}^{\rm eff}$ are the effective K\"ahler parameters given by 
\begin{\eq}
T_1^{\rm eff} (\mu ) = \frac{4\mu_{\rm eff}}{k} +2\pi i \left( \frac{1}{2} -\frac{M}{k} \right) ,\quad
T_2^{\rm eff} (\mu ) = \frac{4\mu_{\rm eff}}{k} -2\pi i \left( \frac{1}{2} -\frac{M}{k} \right) ,
\label{eq:Teff0}
\end{\eq}
with the effective chemical potential $\mu_{\rm eff}$ defined in \eqref{eq:mueff}.
Here $F_{\rm top}$ and $F_{\rm NS}$ are the free energies of the un-refined topological string and
the refined topological string in the Nekrasov-Shatashvili limit, respectively \cite{Nekrasov:2009rc}.
Although each individual terms in \eqref{eq:conclusion} have apparent divergences for 
physical integer $k$,
a careful treatment shows that these divergences are actually canceled as in the ABJM case \cite{Hatsuda:2013gj}.
Then it remains a finite part, which will be presented in \eqref{eq:evenk} for even $k$ and \eqref{eq:oddk} for odd $k$, respectively.
Given the grand potential, we find that the ABJ partition function is written as
the following sum of the Airy functions,
\begin{\eqa}
 Z^{(N,N+M)}(k) 
&=& C^{-\frac{1}{3}} e^{A+i\theta (N,M,k) }  Z_{\rm CS}^{(M)}(k)   \NN\\
&& \sum_{\ell ,m=0}^\infty  g_{\ell ,m}\left( -\frac{\del}{\del N} \right) {\rm Ai}\Biggl[ C^{-\frac{1}{3}}\left( N-B+2\ell +\frac{4m}{k}\right) \Biggr] ,
\label{ZAisum}
\end{\eqa}
where $g_{\ell, m}$ is a polynomial explicitly determined by the grand potential $J_k^{(M)}(\mu )$,
and  $C$ and $B$ are 
the coefficients appearing in the perturbative part of the grand potential, 
\begin{\eq}
 C= \frac{2}{\pi^2 k} ,\quad B =\frac{1}{3k} -\frac{k}{12} +\frac{k}{2}\left( \frac{M}{k} -\frac{1}{2} \right)^2 .
\label{eq:pertC}
\end{\eq}
The overall constants $Z_{\rm CS}^{(M)}(k)$, $\theta (N,M,k)$ and $A$ will be given by
\eqref{eq:pureCS}, \eqref{eq:phase} and \eqref{eq:constmap}, respectively.
The indices $(\ell ,m)$ label the number of D2-brane instantons and worldsheet instantons from 
the type IIA string viewpoint,
while these are lifted up to membrane instantons wrapping two different three cycles in the M-theory.
We will see that 
the large $N$ asymptotic behavior of \eqref{ZAisum} reproduces the results of the classical supergravity (SUGRA) 
and the one-loop quantum SUGRA \cite{Bhattacharyya:2012ye},
and the correct weights of the instantons.

This paper is organized as follows.
In section \ref{sec:Fermi}, we present the Fermi gas formalism of the ABJ partition function on $S^3$.
In section \ref{sec:gravity}, we compare the exact values of the partition function with the results on the gravity side.
In section \ref{sec:topo}, we describe our conjecture \eqref{eq:conclusion} for the grand potential 
in terms of the refined topological string 
and perform a nontrivial test of \eqref{eq:conclusion} by using the exact data.
Section \ref{sec:final} is devoted to discussion.

\section{ABJ theory as a Fermi Gas}
\label{sec:Fermi}
In this section we will show that
the partition function of the ABJ theory on $S^3$ is described by an ideal Fermi gas system 
as in the ABJM case \cite{Marino:2011eh}.
This picture enables us to compute the partition function exactly.

\subsection{Rewriting canonical partition function}
Let us consider the $U(N)_k \times U(N+M)_{-k}$ ABJ partition function on $S^3$.
We begin with the ABJ matrix model \eqref{eq:ABJmat} obtained by the localization method \cite{Kapustin:2009kz,Jafferis:2010un,Hama:2010av}.
It is not so obvious
(except $M=0$ corresponding to the ABJM case)
whether this expression of the ABJ partition function has an ideal Fermi gas description or not.
However, it has been proven \cite{Honda:2013pea}, after conjectured in \cite{Awata:2012jb}, that
the partition function has another equivalent representation, which is suitable for 
a Fermi gas approach.
Although this representation takes seemingly different forms between $M\leq |k|/2$ and $M>|k|/2$,
we will show below that the ABJ partition function can be rewritten for\footnote{
For $M>|k|$, the ABJ partition function vanishes \cite{Awata:2012jb,Honda:2013pea} since $Z_{\rm CS}^{(M)}(k)=0$ for this case.
This is consistent with an expectation that
spontaneously breaking of the supersymmetries occurs in this regime \cite{Aharony:2008gk,Kitao:1998mf,Bergman:1999na}.
Hence we consider the case for $M\leq |k|$ below.
} all $(k,M)$ as 
\begin{\eqa}
Z^{(N,N+M)}(k) =   e^{i\theta (N,M,k)}  Z_{\rm CS}^{(M)}(k) \hat{Z}^{(N,N+M)}(k)  ,
\label{eq:Hermite}
\end{\eqa}
where $Z_{\rm CS}^{(M)}(k)$ is the $U(M)$ pure CS partition function on $S^3$ given by \cite{Marino:2002fk,Aganagic:2002wv,Marino:2004uf,Tierz:2002jj}
\begin{\eqa}
Z_{\rm CS}^{(M)}(k)
&=&  |k|^{-\frac{M}{2}}  \prod_{s=1}^{M-1} \left( 2\sin{\frac{\pi s}{|k|} } \right)^{M-s}  ,
\label{eq:pureCS}
\end{\eqa}
and $\theta (N,M,k)$ is the phase of the partition function,
\begin{\eqa}
 e^{i\theta (N,M,k)}  =i^{NM} e^{-\frac{i\pi}{6k}M(M^2 -1) }  .
\label{eq:phase}
\end{\eqa}
The quantity $\hat{Z}^{(N,N+M)}(k)$ mainly used below,
is the absolute value of the partition function divided by
the pure CS contribution,
\begin{align}
 \hat{Z}^{(N,N+M)}(k)=\left|\frac{Z^{(N,N+M)}(k) }{Z_{\rm CS}^{(M)}(k)}\right|.
\label{hatdef}
\end{align}
It turns out that $\hat{Z}^{(N,N+M)}(k)$ has a simple 
integral representation
\begin{\eq}
\hat{Z}^{(N,N+M)}(k)   
=  \frac{1}{N!} \int_{-\infty}^\infty \frac{d^N y}{(4\pi |k|)^N}   
    \prod_{a<b} \tanh^2{\frac{y_a -y_b }{2k}}  \prod_{a=1}^N V(y_a ) ,
\label{Zhatint}
\end{\eq}
with
\begin{\eq}
V(x) =  \frac{1}{e^{\frac{x}{2}} +(-1)^M e^{-\frac{x}{2}}  } \prod_{s=-\frac{M-1}{2}}^{\frac{M-1}{2}}  \tanh{\frac{x +2\pi i s   }{2|k|}}  .
\end{\eq}
Here the product over $s$ runs with the step $\Delta s =1$.
By using the Cauchy identity
\[
\prod_{1\leq a<b \leq N} \tanh^2{\frac{z_a -z_b }{2}} 
= \sum_{\sigma\in S_N} (-1)^\sigma \prod_{a=1}^N \frac{1}{\cosh{\frac{z_a -z_{\sigma (a)} }{2}}} ,
\]
we can rewrite the partition function as an ideal Fermi gas system,
\begin{\eq}
\hat{Z}^{(N,N+M)}(k)   
=  \frac{1}{N!} \sum_{\sigma\in S_N} (-1)^\sigma \int_{-\infty}^\infty \frac{d^N y}{(4\pi k)^N}   \prod_{a=1}^N \rho (y_a ,y_{\sigma (a)}) ,
\label{hatZperm}
\end{\eq}
where the density matrix $\rho (x,y)$ is given by
\begin{\eq}
\rho (x,y) = \frac{\sqrt{V(x)V(y)}}{\cosh{\frac{x-y}{2k}}} .
\label{eq:density}
\end{\eq}
In the rest of this section, we will prove that
the ABJ partition function \eqref{eq:ABJmat} can be written as \eqref{eq:Hermite} for all $k$ and $M$.

\subsubsection{For $M\leq |k|/2$}
\begin{figure}[t]
\begin{center}
\includegraphics[width=7.4cm]{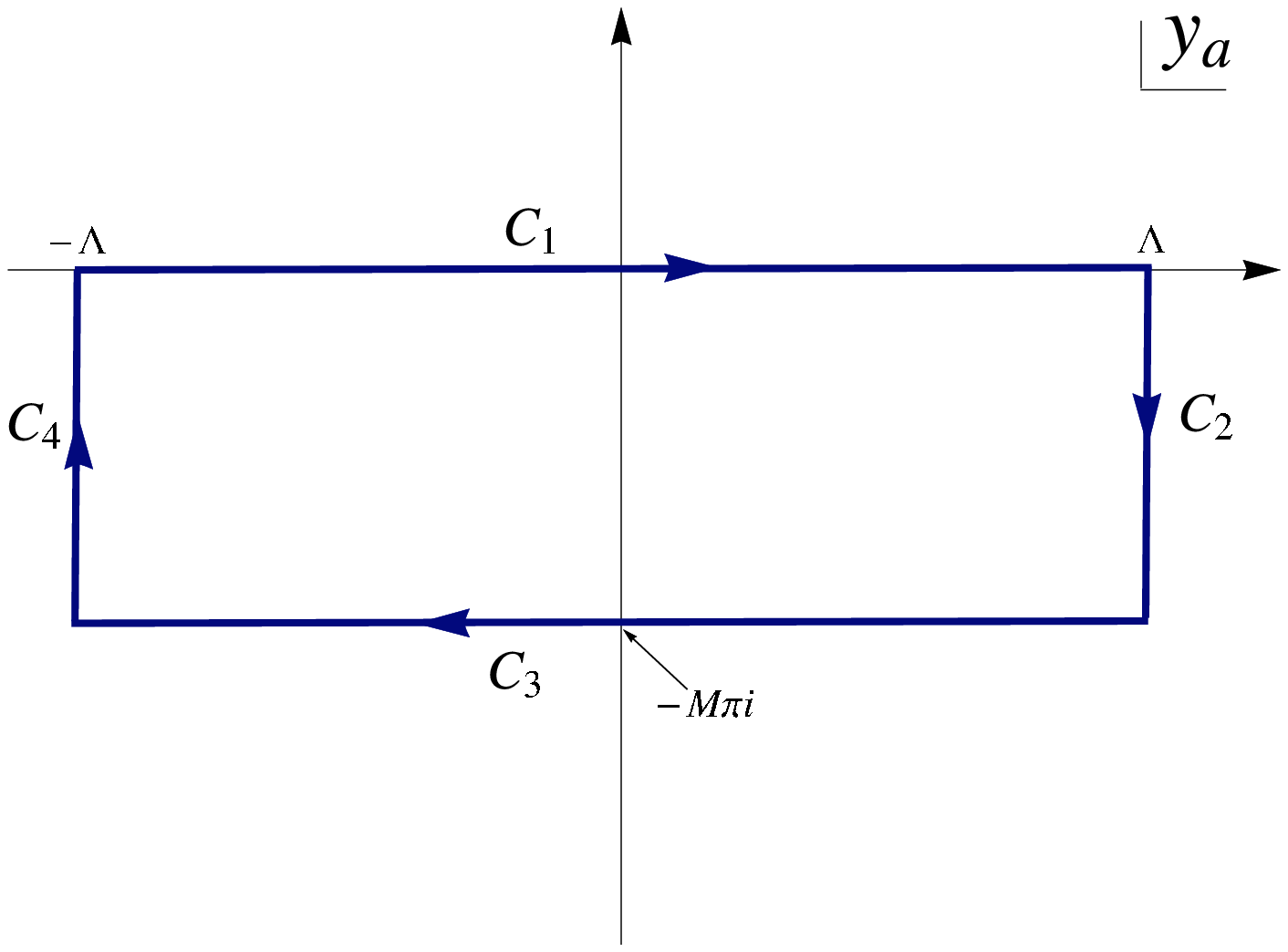}
\includegraphics[width=7.4cm]{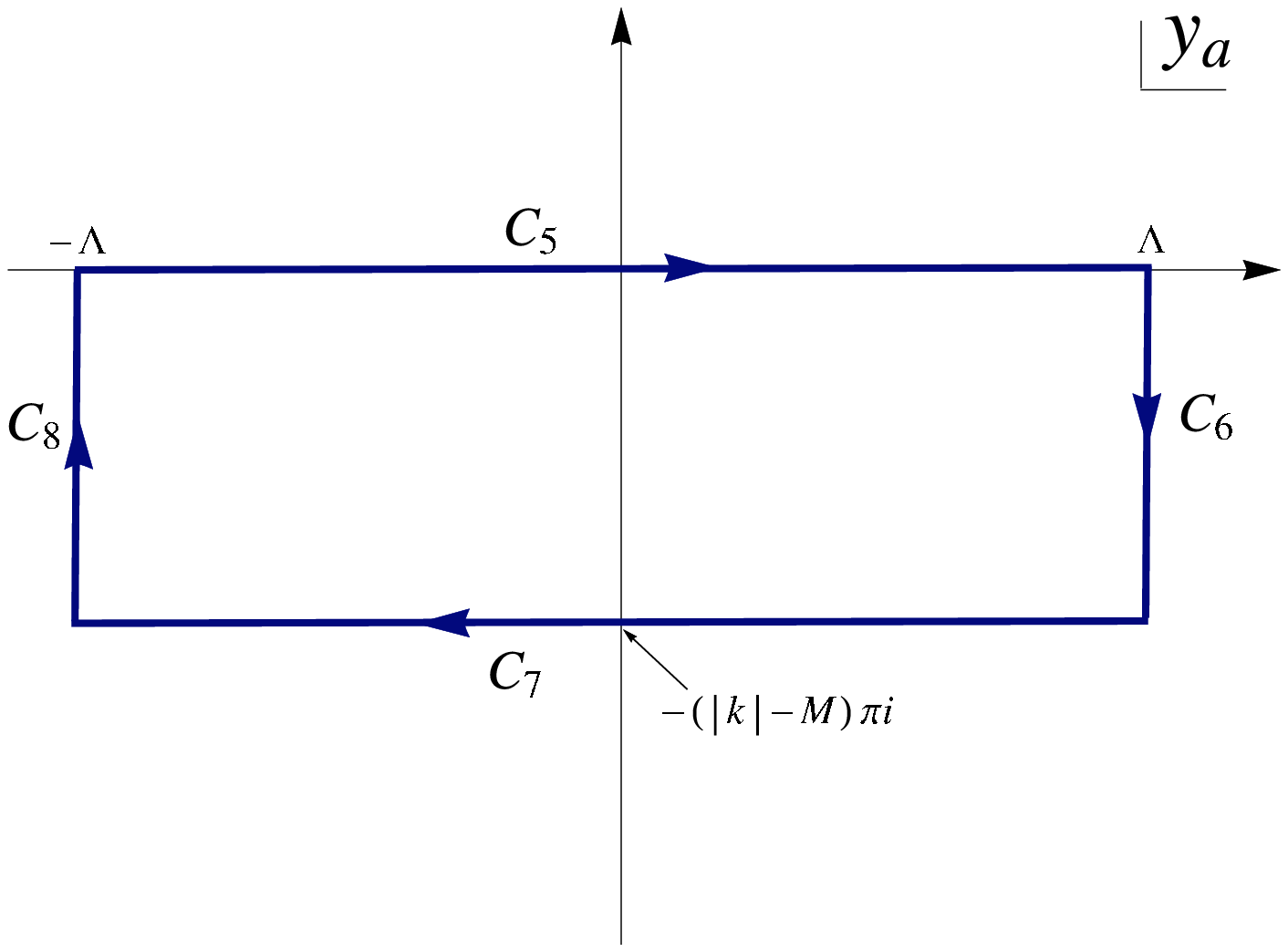}
\end{center}
\caption{
(Left) Explanation of the integral contours $C_1 ,C_2 ,C_3$ and $C_4$. 
The real positive parameter $\Lambda$ is taken to infinity.
(Right) Explanation of the integral contours $C_5 ,C_6 ,C_7$ and $C_8$.}
\label{fig:contour}
\end{figure}

For $M\leq |k|/2$, 
the previous study has shown \cite{Honda:2013pea} that 
the ABJ partition function \eqref{eq:ABJmat} is equivalent to
\begin{\eqa}
Z^{(N,N+M)}(k)  
&=&    \frac{1}{ N! } e^{i\theta (N,M,k)}  Z_{\rm CS}^{(M)}(k)   
 \int_{-\infty }^{\infty } \frac{d^N y}{(4\pi |k|)^N}     \prod_{1\leq a<b\leq N} \tanh^2{\frac{y_a -y_b }{2k}} \NN\\
&& \times \prod_{a=1}^N \Biggl[ \frac{1}{2\cosh{\frac{y_a }{2}}} 
    \prod_{s=-\frac{M-1}{2}}^{\frac{M-1}{2}}  \tanh{\frac{y_a +2\pi i (s +M/2 )  }{2|k|}}  \Biggr] .
\end{\eqa}
If we integrate this integrand along the contour $C_1 +C_2 +C_3 +C_4$ depicted in fig.~\ref{fig:contour} (Left),
then we can easily see that the integration vanishes by the Cauchy integration theorem\footnote{
Note that poles of $1/\cosh{\frac{y_a }{2}}$ at $y_a =-(2m+1)\pi i$ with $m=1,2,\cdots, [(M-1)/2]$ 
are canceled out due to zeros of $\prod \tanh{\frac{y_a +2\pi i (s +M/2 )  }{2k}}$.
Also, poles of $\tanh{\frac{y_a +2\pi i (s +M/2 )  }{2|k|}}$ are not located
inside of $C_1 +C_2 +C_3 +C_4$.
}.
Since the integrals along $C_2$ and $C_4$ becomes zero in the limit $\Lambda\rightarrow\infty$,
we find
\begin{\eq}
Z^{(N,N+M)}(k)  = \int_{C_1} d^N y_a (\cdots ) = \int_{C_3} d^N y_a (\cdots ).
\end{\eq}
The right hand side is nothing but \eqref{eq:Hermite}.

\subsubsection{For $M> |k|/2$}
For $M> |k|/2$, 
the ABJ matrix model \eqref{eq:ABJmat} is also written as \cite{Awata:2012jb,Honda:2013pea}
\begin{\eqa}
Z^{(N,N+M)}(k)  
&=&    \frac{1}{ N! } e^{i\theta (N,M,k)}  Z_{\rm CS}^{(M)}(k)   
 \int_{-\infty }^{\infty } \frac{d^N y}{(4\pi |k|)^N}     \prod_{1\leq a<b\leq N} \tanh^2{\frac{y_a -y_b }{2k}} \NN\\
&& \times \prod_{a=1}^N \Biggl[ \frac{1}{2\cosh{\frac{y_a }{2}}} 
    \prod_{s=-\frac{|k|-M-1}{2}}^{\frac{|k|-M-1}{2}}  \tanh{\frac{y_a +2\pi i (s +|k|/2-M/2 )  }{2|k|}}  \Biggr] .
\end{\eqa}
By considering an integral contour $C_5 +C_6 +C_7 +C_8$ in fig.~\ref{fig:contour} (Right) with the same integrand,
the partition function becomes
\begin{\eq}
Z^{(N,N+M)}(k)  = \int_{C_5} d^N y_a (\cdots ) = \int_{C_7} d^N y_a (\cdots ),
\end{\eq}
as in the case for $M\leq |k|/2$.
Thus we obtain
\begin{\eqa}
Z^{(N,N+M)}(k)  
&=&    \frac{1}{ N! } e^{i\theta (N,M,k)}  Z_{\rm CS}^{(M)}(k)   
 \int_{-\infty }^{\infty } \frac{d^N y}{(4\pi |k|)^N}     \prod_{1\leq a<b\leq N} \tanh^2{\frac{y_a -y_b }{2k}} \NN\\
&& \times \prod_{a=1}^N \Biggl[ \frac{1}{e^{\frac{y_a }{2}} +(-1)^{|k|-M} e^{-\frac{y_a }{2}} } 
    \prod_{s=-\frac{|k|-M-1}{2}}^{\frac{|k|-M-1}{2}}  \tanh{\frac{y_a +2\pi i s  }{2|k|}}  \Biggr] .
\label{eq:Zcase2}
\end{\eqa}
Noting that
\[
\prod_{s=-\frac{|k|-1}{2}}^{\frac{|k|-1}{2}}\tanh{\frac{y+2\pi i s}{2|k| }}
= \left\{ \begin{matrix}
1 & {\rm for}\ k:{\rm even} \cr  \tanh{\frac{y}{2}} & {\rm for}\ k:{\rm odd}
\end{matrix} \right. ,
\]
we find the following identity
\begin{\eq}
\frac{1}{e^{\frac{y }{2}} +(-1)^{|k|-M} e^{-\frac{y }{2}} }  \prod_{s=-\frac{|k|-M-1}{2}}^{\frac{|k|-M-1}{2}}  \tanh{\frac{y +2\pi i s  }{2|k|}}
=\frac{1}{e^{\frac{y }{2}} +(-1)^{M} e^{-\frac{y }{2}} }  \prod_{s=-\frac{M-1}{2}}^{\frac{M-1}{2}}  \tanh{\frac{y +2\pi i s  }{2|k|}} .
\label{eq:dualtanh}
\end{\eq}
Plugging this into \eqref{eq:Zcase2}, we obtain the same expression \eqref{eq:Hermite}.

\subsubsection{Remarks on Seiberg-like duality}
From the representation \eqref{eq:Hermite},
one can see that the ABJ partition function
transforms properly under the Seiberg-like duality \cite{Aharony:2008gk} between the ABJ theories 
with different gauge groups
\begin{\eq}
U(N)_k \times U(N+M)_{-k} \quad {\rm and} \quad U(N+|k|-M)_k \times U(N)_{-k} .
\label{eq:Seiberg}
\end{\eq}
This duality comes \cite{Kapustin:2010mh} from the Giveon-Kutasov duality \cite{Giveon:2008zn},
which has been proven for the $S^3$ partition function \cite{Willett:2011gp}.

Let us check this duality in terms of \eqref{eq:Hermite}.
First, it is well-known that 
the pure CS partition function $Z_{\rm CS}^{(M)}(k)$ enjoys the level-rank duality (See e.g. \cite{Kapustin:2010mh} for a proof)
\begin{\eq}
Z_{\rm CS}^{(M)}(k) =Z_{\rm CS}^{(|k|-M)}(-k) .
\label{trfCS}
\end{\eq}
Next, using \eqref{eq:dualtanh} one can easily see that
the integral representation of
$\hat{Z}^{(N,N+M)}(k)$ in \eqref{Zhatint}
is also duality invariant,
\begin{\eq}
\hat{Z}^{(N,N+M)}(k)  =\hat{Z}^{(N,N+|k|-M)}(-k) . 
\label{trfhat}
\end{\eq}
Finally the phase factor $e^{i\theta (N,M,k)}$ given in \eqref{eq:phase} is not duality invariant 
but appropriately transforms as discussed in \cite{Kapustin:2010mh,Willett:2011gp}.

In what follows, we will assume $k>0$ without loss of generality.
Since $Z_{\rm CS}^{(M)}(k)$ and
$\hat{Z}^{(N,N+M)}(k)$
depend only on the absolute value of $k$,
\eqref{trfCS} and \eqref{trfhat} imply that
they are invariant under the exchange
$M\leftrightarrow k-M$:
\begin{align}
 Z_{\rm CS}^{(M)}(k)=Z_{\rm CS}^{(k-M)}(k),\qquad
\hat{Z}^{(N,N+M)}(k)=\hat{Z}^{(N,N+k-M)}(k) .
\label{seiberg}
\end{align}

\subsection{Grand canonical formalism}
Let us switch to the grand canonical formalism.
We define\footnote{
One could define the grand potential in terms of the whole partition function $Z^{(N,N+M)}(k)$ rather than $\hat{Z}^{(N,N+M)}(k)$.
Then the $N$-dependent factor of the phase \eqref{eq:phase} can be absorbed 
by redefining the chemical potential as $\mu \rightarrow \mu +iM\pi /2$,
while the $N$-independent factor of the phase and pure CS partition function $Z_{\rm CS}^{(M)}(k)$
can also be absorbed by redefinition of the perturbative coefficient $A$ in \eqref{eq:constmap}.
Particularly, in our definition of the grand partition function \eqref{granddef},
we will see that corresponding topological string free energy is invariant under the Seiberg-like duality \eqref{eq:Seiberg}.
If we did not drop the phase in defining the grand partition function,
then corresponding topological string free energy do not become duality invariant.
} the grand partition function as
the generating function of 
$\hat{Z}^{(N,N+M)}(k)$ 
in \eqref{hatdef} ,
\begin{\eq}
\Xi_k^{(M)} (z)  = \sum_{N=0}^\infty  z^N \hat{Z}^{(N,N+M)}(k)  
=\sum_{N=0}^\infty  z^N \left|\frac{Z^{(N,N+M)}(k) }{Z_{\rm CS}^{(M)}(k)}\right|,
\label{granddef}
\end{\eq}
where $z$ is related to the chemical potential $\mu$ by
\begin{\eq}
z =e^\mu .
\end{\eq} 
From the representation 
\eqref{hatZperm} of $\hat{Z}^{(N,N+M)}(k)$ as the sum over permutations,
one can show that the grand partition function is written as a Fredholm determinant,
\begin{\eq}
\Xi_k^{(M)} (z)   ={\rm Det}(1+z\rho ) .
\end{\eq}
It is also convenient to introduce the grand potential as
\begin{\eq}
\bar{J}_k^{(M)} (\mu )  = \log{\Xi_k^{(M)} (z )} .
\end{\eq} 
Given the grand partition function $\Xi_k^{(M)} (z)$, we can easily come back to the canonical partition function $\hat{Z}^{(N,N+M)}(k)$ by
\begin{\eq}
\hat{Z}^{(N,N+M)}(k) = \oint \frac{dz}{2\pi i}  \frac{1}{z^{N+1}} \Xi_k^{(M)} (z) .
\label{comeback}
\end{\eq} 
As explained in \cite{Hatsuda:2012dt,Honda:2014ica},
the grand potential consists of a primary non-oscillatory part and an oscillatory part.
The non-oscillatory part\footnote{
The oscillatory part is defined by
\[
\log{\Biggl[ 1 +\sum_{n=-\infty , n\neq 0 }^\infty e^{J_k^{ (M) }(\mu +2\pi in ) -J_k^{ (M) }(\mu ) } \Biggr]} ,
\] 
although we will not use this expression explicitly below.
} $J_k^{ (M) }(\mu )$
satisfies the following relation
\begin{\eq}
e^{\bar{J}_k^{ (M) }(\mu )} = \sum_{n=-\infty}^\infty e^{J_k^{ (M) }(\mu +2\pi in )}.
\end{\eq}
Then we can deform the integral contour in \eqref{comeback} to the whole imaginary axis,
by just replacing the total grand potential $\bar{J}_k^{ (M) }(\mu )$ with its non-oscillatory part $J_k^{ (M) }(\mu )$, namely,
\begin{\eq}
\hat{Z}^{(N,N+M)}(k) = \int_{-i\infty}^{i\infty} \frac{d\mu}{2\pi i}  \exp{\Bigl[ J_k^{ (M) }(\mu ) -\mu N \Bigr]} .
\label{eq:comeback2}
\end{\eq} 

Now let us describe our method for the exact computation of
the partition functions.
As discussed in \cite{Hatsuda:2012dt} (see appendix \ref{app:algorithm} for details),
thanks to the Tracy-Widom's lemma \cite{Tracy:1995ax}, the grand canonical
partition function is determined by a series of functions $\phi_l^+(y)$,
\begin{\eq}
\Xi_k^{(M)} (z)  
= \exp{\Biggl[ -\sum_{n=1}^\infty \frac{z^{2n}}{n} {\rm Tr}\rho_+^{2n} \Biggr]}  \cdot \sum_{l=0}^\infty \phi_l^+ (0) z^l ,
\label{eq:grandpart}
\end{\eq}
where $\rho_+$ and  $\phi_l^+(y)$ are given by
\begin{\eqa}
&&\rho_+ (x,y) =\frac{\rho (x,y) +\rho (x,-y) }{2} =\frac{E_+ (x) E_+ (y)}{\cosh{\frac{x}{k}} +\cosh{\frac{y}{k}} } ,\quad
E_+ (x) = \cosh{\frac{x}{2k}} \sqrt{V(x)}, \NN\\
&&\rho_+^n (x,y) =\frac{E_+ (x)E_+ (y)}{\cosh{\frac{x}{k}} +(-1)^{n-1}\cosh{\frac{y}{k}} } \sum_{l=0}^{n-1} (-1)^l \phi_l^+ (x) \phi_{n-l-1}^+(y ) ,\NN\\
&&\phi_{l+1}^+ (y) 
= \frac{1}{\cosh{\frac{y}{2k}}} \int \frac{dy^\prime}{2\pi k} \frac{\cosh{\frac{y^\prime}{2k}}}{2\cosh{\frac{y-y^\prime}{2k}}} 
   V(y^\prime ) \phi_l^+ (y^\prime ) ,  \quad \phi_0^+ (y)=1 ,\NN\\
&& {\rm Tr}\rho_+^{2n} 
= \int \frac{dy}{2\pi} \frac{E_+ (y)^2}{\sinh{\frac{y}{k}}} \sum_{l=0}^{2n-1} (-1)^l \frac{d\phi_l^+ (y)}{dy} \phi_{2n-l-1}^+ (y) .
\label{eq:TBA}
\end{\eqa}
In particular, $\phi_l^+(y)$  can be constructed recursively starting from
$\phi_0^+(y)=1$. 
Using the above formula, we have computed the exact partition functions
of the ABJ theory 
for $(k,M, N_{\rm max} )=(2,1,65 )$, $(3,1,31 )$, $(4,1,62 )$, $(4,2, 29 )$, $(6,1, 23 )$, 
$(6,2, 21 )$ and $(6,3,22 )$.
The explicit values of the partition functions are listed in appendix D. 

\section{Test of AdS/CFT correspondence}
\label{sec:gravity}
In this section, using our exact data of the partition function, we test the AdS/CFT correspondence 
between the ABJ theory and M-theory on $AdS_4 \times S^7 /\mathbb{Z}_k$ with a discrete torsion.

\subsection{Expectation from the gravity side}
Here we describe some expectations from the gravity side.

\subsubsection{Classical supergravity}
The $U(N)_k \times U(N+M)_{-k}$ ABJ theory \cite{Aharony:2008gk,Hosomichi:2008jb}
describes the low-energy effective theory of $N$ M2-branes with $M$ fractional M2-branes.
In the near-horizon limit, the geometry associated with this brane configuration becomes $AdS_4 \times S^7 /\mathbb{Z}_k$ 
with $M$ units of discrete torsion realized by a discrete holonomy of the 3-form field \cite{Aharony:2008gk}.

When $k\gg N^{1/5}$, the curvature becomes very small and 
we expect that the eleven-dimensional classical SUGRA on $AdS_4 \times S^7 /\mathbb{Z}_k$ provides 
a good approximation of the gravity side.
The free energy of the classical SUGRA with the boundary $S^3$ 
obeys the famous $N^{3/2}$-law \cite{Klebanov:1996un}, given by (see e.g. \cite{Marino:2011nm} for a
derivation)
\begin{\eq}
F_{\rm SUGRA} = -\frac{\pi \sqrt{2k}}{3} N^{3/2} .
\label{eq:SUGRA}
\end{\eq}
In the next subsection we compare this with our exact result on the ABJ side in the large $N$ regime with fixed $k$.

\subsubsection{One-loop quantum supergravity}
The one-loop correction of the SUGRA can also be analyzed
by the technique 
successful in computing the logarithmic correction to 
the black hole entropy \cite{Banerjee:2010qc,Banerjee:2011jp,Sen:2011ba,Sen:2012cj,Sen:2012dw}.
The authors in \cite{Bhattacharyya:2012ye} have shown that
the free energy of the eleven dimensional SUGRA on $AdS_4 \times X_7$,
where $X_7$ is a seven-dimensional manifold including $S^7 /\mathbb{Z}_k$,
contains
the following universal logarithmic correction,
\begin{\eq}
-\frac{1}{4} \log{N} .
\label{eq:oneloop}
\end{\eq}
For $\mathcal{N}=3$ necklace quiver CSM with coincident rank of gauge groups,
it has been shown that this logarithmic behavior comes from the large $N$ asymptotics of 
the Airy function \cite{Marino:2011eh}
in the perturbative part of partition function (See also \cite{Marino:2012az} for $\mathcal{N}=2$ case).

Strictly speaking, in the presence of the discrete torsion,
there might be further massless degrees of freedom and 
the logarithmic correction could change\footnote{
We would like to thank Ashoke Sen for explaining this point.
}.
However, as we will see in section 4, the exact ABJ partition functions
show a nice agreement with the Airy function, and hence
the ABJ partition function also exhibit this 1-loop behavior \eqref{eq:oneloop}.

\subsubsection{Non-perturbative effects}
One expects two kinds of non-perturbative effects on the gravity side as in the ABJM case \cite{Cagnazzo:2009zh,Drukker:2011zy}.
If we identify a direction of M-theory circle with the orbifolding direction and consider 
a large $k$ regime,
then the eleventh dimension in the geometry $AdS_4 \times S^7 /\mathbb{Z}_k$ shrinks and
the bulk theory reduces to the type IIA string theory on $AdS_4 \times \mathbb{CP}^3$.
Since $\mathbb{CP}^3$ has a nontrivial 2-cycle $\mathbb{CP}^1$ and a Lagrangian submanifold $\mathbb{RP}^3$,
we expect that
the dual type IIA string theory has the worldsheet instanton \cite{Cagnazzo:2009zh}
and the D2-brane instanton \cite{Drukker:2011zy} corrections characterized by the following weights, respectively,
\begin{\eq}
\exp{\Bigl[ -T_{\rm F1} {\rm Vol}(\mathbb{CP}^1 ) \Bigr]} = \exp{\left( -2\pi \sqrt{\frac{2N}{k}} \right)} ,\quad
\exp{\Bigl[ -T_{\rm D2} {\rm Vol}(\mathbb{RP}^3 ) \Bigr]} = \exp{\left( -\pi \sqrt{2kN} \right)} .
\label{eq:weight}
\end{\eq} 
As discussed in \cite{Drukker:2010nc},
the worldsheet instanton also receives the following contribution from the
coupling to the background NSNS 2-form field
\begin{align}
 \int_{\mathbb{CP}^1} B_{\rm NS}=\frac{1}{2}-\frac{M}{k}.
\label{Bfield}
\end{align}
We also expect that the D2-instanton has a coupling to the
background RR 3-form field.
From the viewpoint of M-theory, these instantons correspond to M2-branes wrapping three cycles 
known as membrane instantons \cite{Becker:1995kb}.
The worldsheet instanton effects in \eqref{eq:weight} and 
\eqref{Bfield} have been successfully reproduced from the ABJ matrix model in the 't Hooft limit
\cite{Drukker:2010nc}.
In section.~\ref{sec:topo},
we will show that 
the ABJ partition function contains all the expected instanton effects and
determine the structure from the refined topological string.

\subsection{Comparison with the gravity side}
\begin{figure}[t]
\begin{center}
\includegraphics[width=7.5cm]{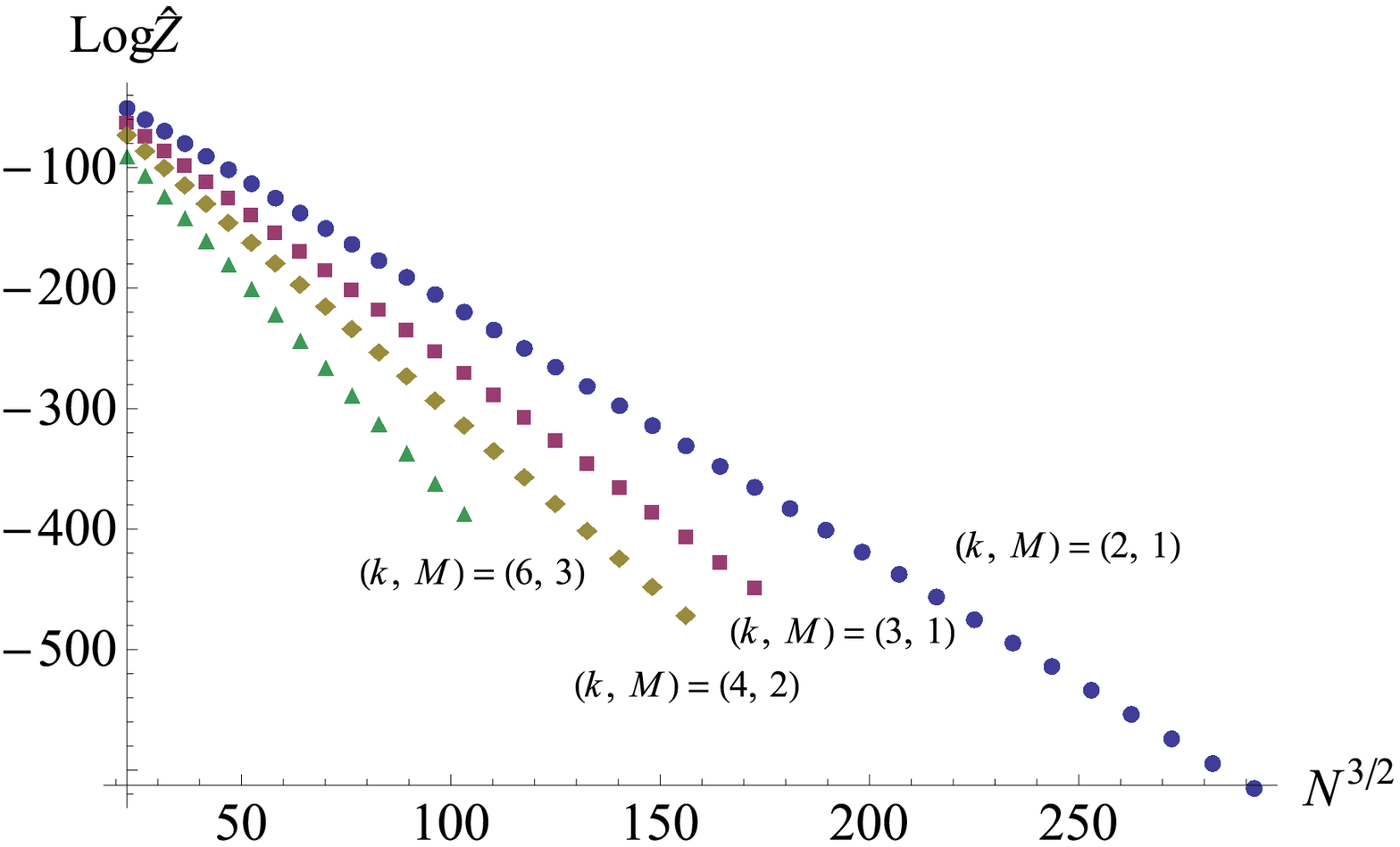}
\includegraphics[width=7.5cm]{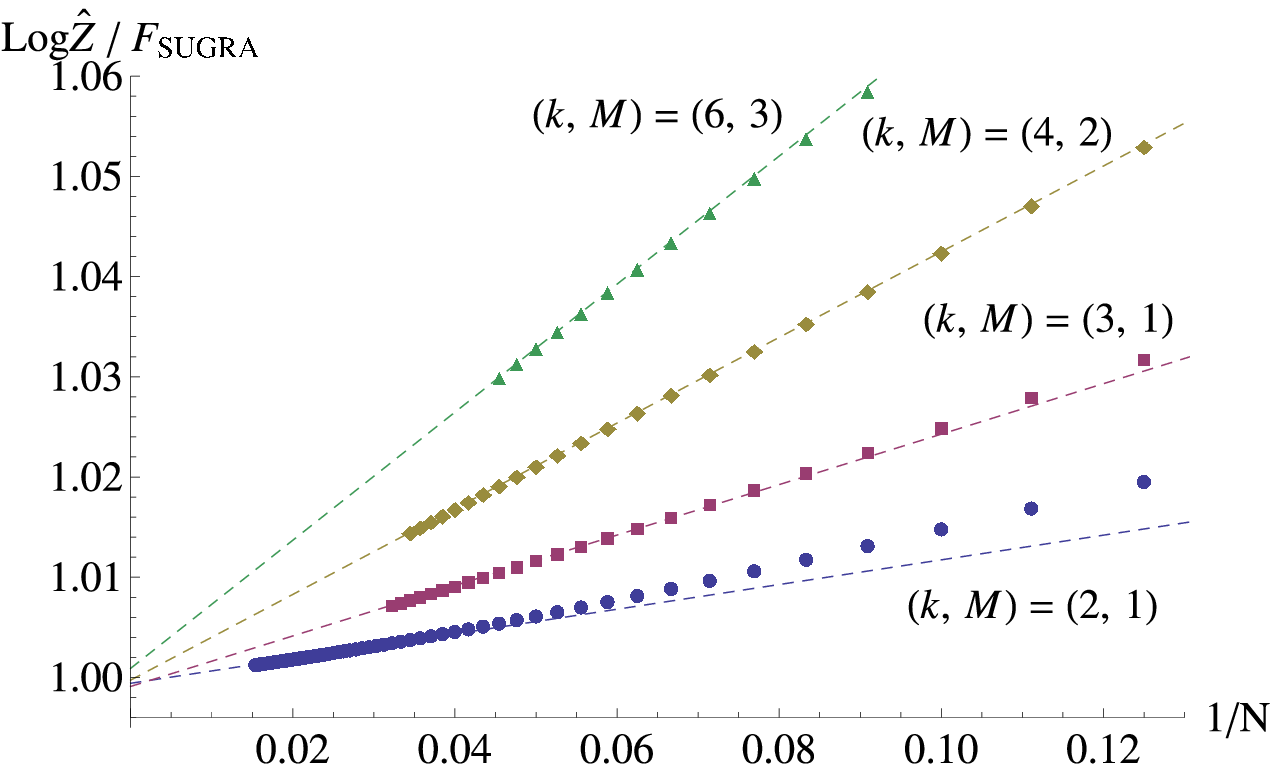}
\end{center}
\caption{
(Left) The free energy $\log{\hat{Z}}^{(N,N+M)}(k)$ is plotted to $N^{3/2}$.
(Right) The ratio between the exact free energy $\log{\hat{Z}}^{(N,N+M)}(k)$ and the classical SUGRA free energy $F_{\rm SUGRA}$
is plotted against $1/N$.
The symbols and dashed lines denote the exact data and fitting functions in the large $N$ region, respectively.
}
\label{fig:SUGRA}
\end{figure}

Let us compare our exact result of the partition function 
with the classical SUGRA result \eqref{eq:SUGRA}.
In fig.~\ref{fig:SUGRA} (Left),
we plot the exact values of the free energy $\log{\hat{Z}}^{(N,N+M)}(k)$ listed in appendix \ref{app:values} against $N^{3/2}$ for some cases.
As expected from the gravity side,
we observe that the free energy is approximately proportional to $N^{3/2}$ in the large $N$ regime.
Fig.~\ref{fig:SUGRA} (Right) compares our result with the classical SUGRA result.
In this plot, we fit the ratio between the exact free energy and classical SUGRA free energy
by linear functions of $1/N$, whose intercepts should be almost $1$ 
if the AdS/CFT correspondence is correct. 
We easily see that the data points  are well fitted by the fitting functions 
in the large $N$ region, 
and indeed the values of these intercepts are almost $1$.
Although we have plotted only for $(k,M)=(2,1),(3,1),(4,2)$ and $(6,3)$,
we have checked that similar result holds also for $(k,M)=(4,1),(6,1)$ and $(6,2)$.

We can reproduce the classical SUGRA result also from the thermodynamic limit of the Fermi gas system 
as in the ABJM case \cite{Marino:2011eh}.
In the large $N$ limit, the free energy is approximated by a saddle point of the $\mu$-integration in \eqref{eq:comeback2} as
\begin{\eq}
\log{\hat{Z}^{(N,N+M)}(k)} \simeq   J_k^{ (M) }(\mu_\ast ) -\mu_\ast N ,\quad \ \ \  {\rm with}\ \ 
 \frac{\partial J_k^{ (M) }(\mu )}{\partial \mu} \Biggr|_{\mu =\mu_\ast} =N .
\label{eq:saddle}
\end{\eq} 
Hence we can estimate the large $N$ free energy if we find the large $\mu$ asymptotic behavior of the grand potential,
which is captured by a semi-classical expansion of the Fermi gas system.
By introducing the quantum mechanical operators $\hat{q}$ and $\hat{p}$ satisfying $[\hat{q},\hat{p}] = 2\pi i k$,
and $\hat{q}|x\rangle  =x|x\rangle $,
we can rewrite the density matrix $\rho (x, y)$ as a quantum mechanical operator
\begin{\eq}
\rho (x, y) = \langle x |  e^{-\hat{H}(\hat{q},\hat{p})} |y\rangle ,\quad
e^{-\hat{H}(\hat{q},\hat{p})} = \sqrt{V(\hat{q})} \frac{1}{2\cosh{\frac{\hat{p}}{2}}} \sqrt{V(\hat{q})} ,
\label{Hamiltonian}
\end{\eq}
where $\hat{H}$ is the Hamiltonian of the Fermi gas system.
The asymptotic behavior of the classical Hamiltonian for $|q|,|p|\gg 1$ is given by
\begin{\eq}
H_{\rm cl}(q,p)\sim \frac{|q|+|p|}{2} ,
\end{\eq}
which is exactly the same as in the ABJM case \cite{Marino:2011eh}.
Thus the same analysis as \cite{Marino:2011eh}
gives the large $\mu$ grand potential and the saddle point $\mu_\ast$ as 
\begin{\eq}
J_k^{(M)}(\mu ) \simeq \frac{2}{3\pi^2 k} \mu^3 ,\quad
\mu_\ast = \pi \sqrt{\frac{kN}{2}} .
\label{eq:mustar}
\end{\eq}
We can easily see that plugging this into \eqref{eq:saddle} reproduces the result of the classical supergravity.

Using the expression of  the saddle point value $\mu_*$ in \eqref{eq:mustar},
we can translate the instanton effects \eqref{eq:weight} 
on the gravity side into the language of the grand canonical formalism.
In terms of $\mu_\ast$, we can express the instanton factors as
\begin{\eq}
 \exp{\left( -2\pi \sqrt{\frac{2N}{k}} \right)} =\exp{\left( -2\mu_\ast \right)} ,\quad 
\exp{\left( -\pi \sqrt{2kN} \right)} =\exp{\left( -\frac{4\mu_\ast}{k} \right)} .
\label{instmu}
\end{\eq} 
If we assume that the free energy on the gravity side receives a series of multi-instanton corrections,
we would expect that the ABJ grand potential have the following expansion, 
\begin{\eq}
J_k^{ (M) }(\mu )
=\sum_{l,m=0 }^\infty 
  f_{l,m}(\mu ) \exp{\Biggl[ -\left( 2l +\frac{4m}{k}\right) \mu  \Biggr] } ,
\label{eq:expectJ}
\end{\eq}
where $f_{l,m}(\mu )$ is a polynomial  of $\mu$, whose coefficients depend on $M$ and $k$.
Although there is no currently available technique to
determine the coefficients $f_{l,m}(\mu )$
of instantons from the computation of the gravity side,
the ABJ matrix model and its relation to the refined topological string
give a very concrete prediction for these 
coefficients,
as we will see in the next section.

\section{ABJ partition function from the refined topological string}
\label{sec:topo}
In this section we propose that
the ABJ free energy including the non-perturbative effects
is completely determined by the refined topological string on local $\mathbb{P}^1 \times \mathbb{P}^1$.
As explained in \cite{Marino:2009jd},
the partition function of the ABJ theory on $S^3$ can be analytically continued\footnote{
There is much strong evidence for this relation \cite{Marino:2009jd,Drukker:2011zy,Hanada:2012si,Hatsuda:2012dt,Awata:2012jb,Honda:2013pea,Yost:1991ht,Dijkgraaf:2003xk,Dijkgraaf:2002pp}.
} to the partition function of the 
$L(2,1)$ lens space matrix model \cite{Marino:2002fk,Aganagic:2002wv}
which comes from the pure Chern-Simons theory on $S^3 /\mathbb{Z}_2$.
The lens space matrix model corresponds to
the topological string on local $\mathbb{P}^1 \times \mathbb{P}^1$ via a topological version of
the large $N$ duality \cite{Gopakumar:1998ki}.
Thus we expect that structure of the ABJ free energy is captured by the topological string
on local $\mathbb{P}^1 \times \mathbb{P}^1$.

For the ABJM case, as the special case of the ABJ theory, this expectation has been confirmed 
quite successfully.
In \cite{Hatsuda:2013oxa} it is discussed that
the grand potential of the ABJM theory corresponds to the free energy of the topological string on 
the ``diagonal'' local $\mathbb{P}^1 \times \mathbb{P}^1$
in large radius frame, where the K\"ahler parameters of two $\mathbb{P}^1$'s are equal.
More precisely, it has been shown, 
based on the exact and numerical results \cite{Hatsuda:2012hm,Putrov:2012zi,Hatsuda:2012dt,Hatsuda:2013gj}, 
that the perturbative and worldsheet instanton part of the ABJM grand potential are given by
the free energy of the un-refined topological string on the 
local $\mathbb{P}^1 \times \mathbb{P}^1$ \cite{Marino:2011eh,Hatsuda:2012dt},
while the D2-brane instanton part and its mixed contribution with the worldsheet instanton are captured by the
Nekrasov-Shatashvili limit \cite{Nekrasov:2009rc} of the refined topological string
on the same local $\mathbb{P}^1 \times \mathbb{P}^1$.

In this paper we generalize this argument to the ABJ theory.
From the topological string viewpoint, this amounts to consider non-diagonal local $\mathbb{P}^1 \times \mathbb{P}^1$,
or equivalently local $\mathbb{P}^1 \times \mathbb{P}^1$ with general K\"ahler parameters.
For this purpose, we decompose
$J(\mu )$ into the
perturbative part $J_{\rm pert} (\mu )$ and 
the non-perturbative part $J_{\rm np} (\mu )$ as
\begin{\eqa}
&& J_k^{(M)}(\mu ) = J_{\rm pert} (\mu ) +J_{\rm np} (\mu ) .
\end{\eqa}
We propose that 
structure of the ABJ grand potential, or equivalently the ABJ partition function, is completely determined by
the topological string including the non-perturbative effect and test this proposal by using our
exact data of the partition functions. 

\subsection{Perturbative part}
In order to study the non-perturbative structure of the ABJ grand potential in terms of the exact data,
we shall first determine the perturbative part of the ABJ partition function
to subtract this from the exact data.
In this subsection, we determine the perturbative part of the ABJ grand potential by using the 
un-refined topological string.

As explained in \cite{Lockhart:2012vp},
the perturbative part of the free energy 
of un-refined topological string on a Calabi-Yau manifold (CY) $X$
 is given by
\begin{\eq}
F_{\rm pert} 
= \frac{1}{6g_s^2} \int_X J\wedge J\wedge J -\frac{1}{24} \left( 1 -\frac{1}{g_s^2} \right) \int_X J\wedge c_2 ,
\end{\eq}
where $J$ is the K\"ahler form and $c_2$ is the second Chern class.
From this general expression,
 we expect that the perturbative part of the grand potential of ABJ theory is given by 
\begin{\eq}
J_{\rm pert} (\mu ) 
= \frac{T_1^3 +T_2^3 -3T_1^2 T_2 -3T_1 T_2^2}{6g_s^2 (4\pi i)^2}
   +\frac{1}{24}\left( 1 -\frac{1}{g_s^2} \right) (T_1 +T_2 ) +A ,
\label{JpertT}
\end{\eq}
where $T_{1,2}$ are the K\"{a}hler parameters of local
$\mathbb{P}^1\times\mathbb{P}^1$, whose relation to $\mu$ will be explained shortly.
Here $g_s$ denotes the string coupling $g_s =2/k$ defined in \eqref{ourgs}
and $A$ is given by
\begin{\eqa}
&& A = -\log{|Z_{\rm CS}^{(M)} (k)  |}
           -\frac{\zeta (3) k^2}{8\pi^2} +\frac{1}{6}\log{\frac{4\pi}{k}} +2\zeta^\prime (-1) \NN\\
&&~~~~~~~    -\frac{1}{3} \int_0^\infty dx \left( \frac{3}{x^3} -\frac{1}{x} -\frac{3}{x\sinh^2{x}} \right) \frac{1}{e^{kx} -1} .
\label{eq:constmap}
\end{\eqa}
Apart from the first term in \eqref{eq:constmap}, 
$A$ is the so-called constant map contribution analyzed in \cite{Hanada:2012si} in detail.
The first term $-\log|Z_{\rm CS}^{(M)} (k)  |$ in \eqref{eq:constmap}
comes from our definition of the grand partition function \eqref{granddef}. 
In order to reproduce the
worldsheet instanton factor $e^{-4\mu/k}$ \eqref{instmu} 
together with the effect of the B-field \eqref{Bfield},
it is natural to make the following identification of the K\"ahler parameters
\begin{\eq}
T_1 (\mu ) = \frac{4\mu}{k} +2\pi i \left( \frac{1}{2} -\frac{M}{k} \right) ,\quad
T_2 (\mu ) = \frac{4\mu}{k} -2\pi i \left( \frac{1}{2} -\frac{M}{k} \right) .
\label{eq:kahler}
\end{\eq}
With this identification,  the Seiberg-like duality \eqref{seiberg} is naturally 
realized as the exchange of two $\mathbb{P}^1$'s of local $\mathbb{P}^1\times \mathbb{P}^1$
\begin{align}
 M\leftrightarrow k-M\quad\Leftrightarrow\quad T_1\leftrightarrow T_2 .
\end{align}
Then the perturbative grand potential in \eqref{JpertT} becomes
\begin{\eq} 
J_{\rm pert} (\mu ) =\frac{C}{3} \mu^3 +B\mu +A ,
\label{eq:pert}
\end{\eq}
where $C$ and $B$ are defined in \eqref{eq:pertC}.
This expression \eqref{eq:pert} 
of the perturbative part is consistent with the result for the ABJM case \cite{Marino:2011eh} and
the same as the proposal in \cite{Matsumoto:2013nya}.
Then, as in the ABJM case, 
the perturbative canonical partition function $\hat{Z}_{\rm pert}^{(N,N+M)} (k)$ is given by
the Airy function,
\begin{\eq}
\hat{Z}_{\rm pert}^{(N,N+M)} (k) 
= \int_{-i\infty}^{i\infty} \frac{d\mu}{2\pi i} e^{J_{\rm pert} (\mu ) -N\mu} = C^{-1/3} e^A {\rm Ai} [C^{-1/3}(N-B) ] .
\end{\eq}
Below, we will show that this is indeed the correct perturbative partition function by
comparing with our exact data.  
In the large $N$ limit with fixed $k$, 
the perturbative part of the free energy becomes
\begin{\eqa}
\log{\hat{Z}_{\rm pert}^{(N,N+M)} (k)}
= -\frac{2}{3} C^{-1/2} N^{3/2} +C^{-1/2}B N^{1/2} -\frac{1}{4}\log{N} +\mathcal{O}(1) .
\end{\eqa}
The first term reproduces the classical SUGRA result \eqref{eq:SUGRA}
and the third term agrees with the logarithmic behavior \eqref{eq:oneloop}
of the one-loop quantum supergravity \cite{Bhattacharyya:2012ye}.

\begin{figure}[t]
\begin{center}
\includegraphics[width=7.4cm]{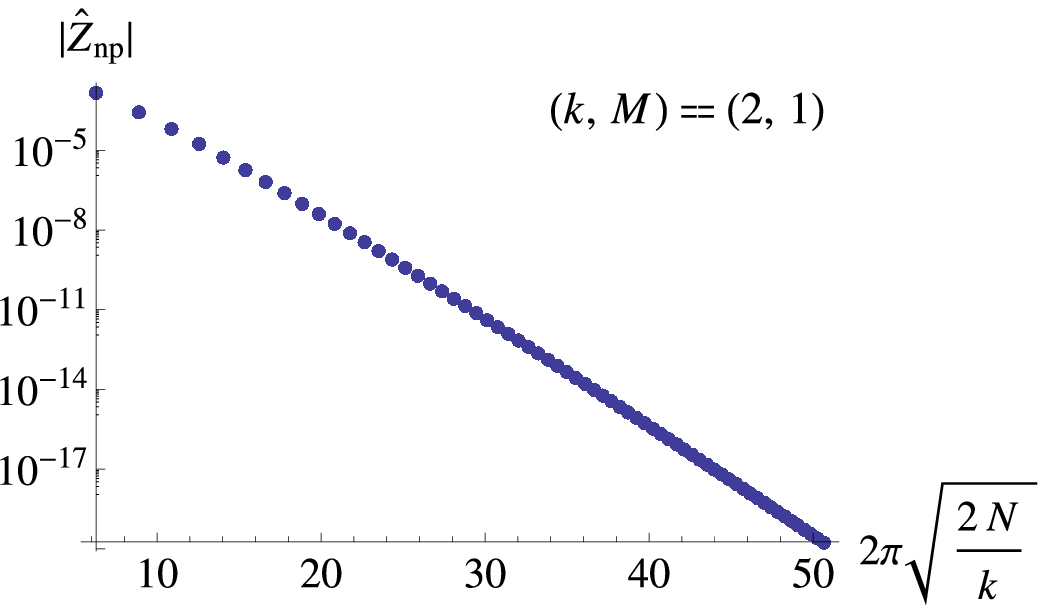}
\includegraphics[width=7.4cm]{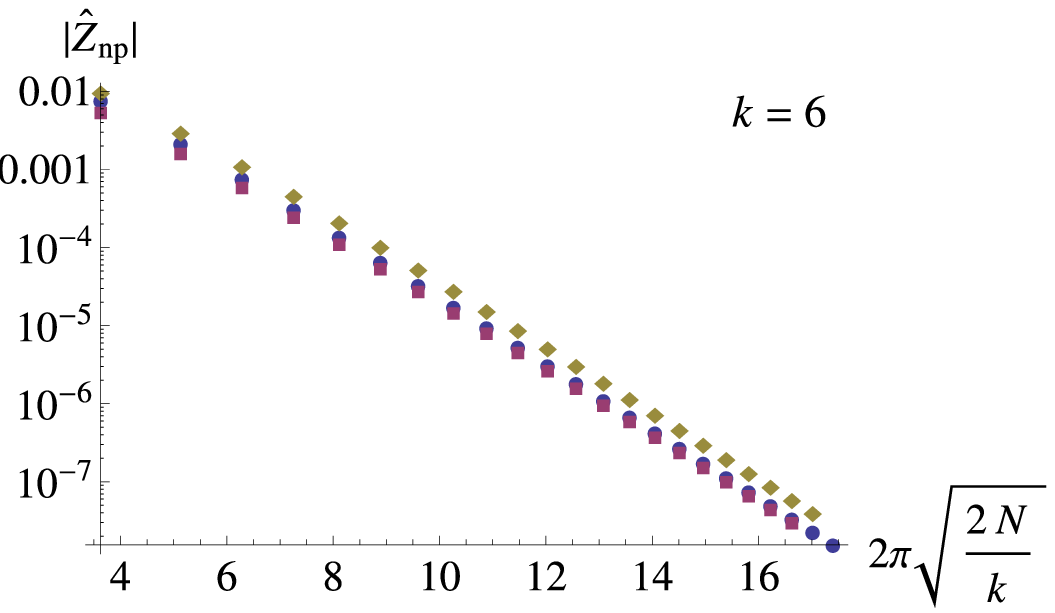}
\end{center}
\caption{
$\hat{Z}_{\rm np}^{(N,N+M)} (k)$ is plotted to $2\pi \sqrt{2N/k}$ in semi-log scale.
(Left) A plot for $(k,M)=(2,1)$.
(Right) Plots for $(k,M)=(6,1),(6,2)$ and $(6,3)$.
The blue circle, purple square and yellow diamond symbols show
the cases for $M=1$, $M=2$ and $M=3$, respectively.
}
\label{fig:pert}
\end{figure}

Let us compare $\hat{Z}_{\rm pert}^{(N,N+M)} (k)$ with our exact data of partition functions.
For this purpose we decompose the canonical partition function as
\begin{\eq}
\hat{Z}^{(N,N+M)} (k) = \hat{Z}_{\rm pert}^{(N,N+M)} (k) \left( 1 +\hat{Z}_{\rm np}^{(N,N+M)} (k) \right) ,
\end{\eq}
where
\begin{\eq}
\hat{Z}_{\rm np}^{(N,N+M)} (k) = \frac{\hat{Z}^{(N,N+M)} (k)}{\hat{Z}_{\rm pert}^{(N,N+M)} (k)} -1 .
\end{\eq}
If $\hat{Z}_{\rm pert}^{(N,N+M)} (k)$ is the correct perturbative part,
then $\hat{Z}_{\rm np}^{(N,N+M)} (k)$ 
should correspond to the non-perturbative part of the partition function
and it is expected to behave as
\begin{\eq}
\hat{Z}_{\rm np}^{(N,N+M)} (k) = \mathcal{O} \Bigl( e^{-2\pi \sqrt{\frac{2N}{k} }} \Bigr) ,
\end{\eq}
in the large $N$ regime.
Fig.~\ref{fig:pert} shows the plots of $\hat{Z}_{\rm np}^{(N,N+M)} (k)$ against $2\pi \sqrt{2N/k}$ in semi-log scale
for $(k,M)=(2,1), (6,1), (6,2)$ and $(6,3)$.
We can easily see that the data points are on the straight lines in the large $N$ regime,
which imply the exponentially suppressed behavior of $\hat{Z}_{\rm np}^{(N,N+M)} (k)$ by $e^{-2\pi \sqrt{\frac{2N}{k} }}$.
Thus we conclude that 
$J_{\rm pert} (\mu )$ given in \eqref{eq:pert} is indeed the correct
perturbative part of the grand potential.
Since this consistency check shows the presence of the ``$-\frac{1}{4}\log{N}$'' term \eqref{eq:oneloop}
in the canonical partition function, 
we have also tested the AdS/CFT correspondence at 1-loop level of the dual quantum supergravity \cite{Bhattacharyya:2012ye}.
Although we have not explicitly shown,
similar result holds also for $(k,M)=(3,1),(4,1)$ and $(4,2)$.

\subsection{Non-perturbative effects and the refined topological string}
In this section we briefly review \cite{Hatsuda:2013oxa}
and write down an expected form of the ABJ grand potential from the refined topological string
on local $\mathbb{P}^1 \times \mathbb{P}^1$ with  non-diagonal K\"ahler parameters.

\subsubsection{The refined topological string}
Let us first recall
the structure of the un-refined topological string on CY $X$ with K\"ahler parameters $T_I$.
The free energy of the un-refined topological string in the large radius frame is given by \cite{Gopakumar:1998jq}
\begin{\eq}
F_{\rm top} (T_I ; g_s)
= \sum_{g\geq 0} \sum_{w\geq 1} \sum_{\mathbf{d}}
\frac{(-1)^{g-1}}{w} n_g^{\mathbf{d}} \left( q_s^{w/2} -q_s^{-w/2} \right)^{2g-2} \mathbf{Q}^{w\mathbf{d}} ,
\end{\eq}
where 
\begin{\eq}
q_s =e^{\hat{g}_s} ,\quad \mathbf{d}=(d_1 ,d_2 , \cdots ), \quad \mathbf{Q}=\prod_{I} Q_I ,\quad Q_I =e^{-T_I} ,\quad 
\mathbf{Q}^{\mathbf{d}} = \prod_I Q_I^{d_I} .
\end{\eq}
Here $\hat{g}_s$ is the topological string coupling defined by $\hat{g}_s=2\pi ig_s$.
The integer numbers $n_g^{\mathbf{d}}$ are the so-called Gopakumar-Vafa invariants of $X$.

The free energy of the refined topological string is obtained by a
one parameter deformation of the standard topological string free energy
originated in the Nekrasov instanton partition function \cite{Nekrasov:2002qd}.
The refined free energy is determined by the 
supersymmetric index $N_{j_L ,j_R}^{\mathbf{d}}$
which counts the BPS states in the M-theory compactified on $X$ down to 5 dimensions
\cite{Iqbal:2007ii}.
These BPS states come from M2-branes wrapping a two-cycle of $X$ 
with degree $\mathbf{d}$. They  behave as massive particles
in the 5 dimensions, carrying spins $(j_L ,j_R )$ of 
the little group $SO(4)\simeq SU(2)\times SU(2)$.
The refined free energy takes the form of
\begin{\eq}
F_{\rm ref}(T_I ;\epsilon_1 ,\epsilon_2 )
=\sum_{j_L ,j_R \geq 0} \sum_{w\geq 1} \sum_{\mathbf{d}}
  \frac{(-1)^{s_L +s_R}}{w} N_{j_L ,j_R}^{\mathbf{d}} \frac{\chi_{j_L}(q_L^w) \chi_{j_R}(q_R^w) }{(q_1^{w/2} -q_1^{-w/2}) (q_2^{w/2} -q_2^{-w/2}) }
   \mathbf{Q}^{w\mathbf{d}} ,
\end{\eq}
where
\begin{\eq}
 s_{L,R} = 2j_{L,R} +1 ,\quad q_L = e^{\frac{\epsilon_1 -\epsilon_2}{2}} ,\quad q_R = e^{\frac{\epsilon_1 +\epsilon_2}{2}} ,\quad
    q_{1,2} =e^{\epsilon_{1,2}}, \quad   \chi_j (q) = \frac{q^{2j+1} -q^{-2j-1}}{q-q^{-1}} .
\end{\eq}
It is often useful to introduce another invariant $n_{g_L .g_R}^{\mathbf{d}}$ by
\begin{\eq}
\sum_{j_L ,j_R \geq 0} (-1)^{s_L +s_R} N_{j_L ,j_R }^{\mathbf{d}} \chi_{j_L}(q_L ) \chi_{j_R}(q_R ) 
 =\sum_{g_L ,g_R \geq 0} n_{g_L ,g_R }^{\mathbf{d}}  (q_L^{\frac{1}{2}} -q_L^{-\frac{1}{2}})^{2g_L} (q_R^{\frac{1}{2}} -q_R^{-\frac{1}{2}})^{2g_R} .
\end{\eq} 
The un-refined topological string corresponds to the following particular limit:
\begin{\eq}
\epsilon_1 = -\epsilon_2 = \hat{g}_s ,
\end{\eq}
with an identification
\begin{\eq}
n_g^{\mathbf{d}} = n_{g,0}^{\mathbf{d}} .
\end{\eq}

As discussed in \cite{Hatsuda:2013oxa}, the Nekrasov-Shatashvili (NS) limit
\begin{\eq}
\epsilon_1 =\hbar ,\quad \epsilon_2 \rightarrow 0.
\end{\eq}
of the refined topological string plays an important role in the D2-instanton corrections.
Since the refined free energy has a simple pole ($\sim 1/\epsilon_2$) in this limit,
we introduce the NS free energy as
\begin{\eqa}
F_{\rm NS}(T_I ;\hbar ) 
&=& \lim_{\epsilon_2 \rightarrow 0} \epsilon_2 F(T_I ;\hbar ,\epsilon_2 ) \NN\\
&=& \sum_{j_L ,j_R \geq 1} \sum_{w\geq 1} \sum_{\mathbf{d}}
  \frac{(-1)^{s_L +s_R}}{w^2} N_{j_L ,j_R}^{\mathbf{d}} \frac{\chi_{j_L}(q^{w/2}) \chi_{j_R}(q^{w/2}) }{q^{w/2} -q^{-w/2} }   \mathbf{Q}^{w\mathbf{d}} ,
\end{\eqa}
where
\begin{\eq}
q=e^\hbar=e^{2\pi i/g_s} .
\end{\eq}

\subsubsection{ABJ partition function}
As a natural generalization of the ABJM case,
we conjecture that
the ABJ grand potential
can be written 
in terms of the free energy of the refined topological string on local $\mathbb{P}^1 \times \mathbb{P}^1$.
For  later convenience, we decompose the non-perturbative grand potential $J_{\rm np} (\mu )$ 
into the worldsheet instanton part $J_{\rm WS} (\mu )$, the 
D2-brane instanton part $J_{\rm D2} (\mu )$ 
and the mixed contributions $J_{\rm WS+D2} (\mu )$,
\begin{\eq}
J_{\rm np} (\mu ) =J_{\rm WS} (\mu ) +J_{\rm D2} (\mu ) +J_{\rm WS+D2} (\mu ) .
\end{\eq}
As in the ABJM case \cite{Hatsuda:2013oxa,Drukker:2010nc,Hatsuda:2012dt}, 
we expect that the perturbative part plus the worldsheet instanton part is described by
the un-refined topological string,
\begin{\eq}
J_{\rm pert} (\mu ) + J_{\rm WS} (\mu ) =F_{\rm top} (T_1 ,T_2 ; g_s) ,
\end{\eq}
where $g_s=2/k$ in our definition \eqref{ourgs}.
More explicitly, the worldsheet instanton part $J_{\rm WS} (\mu )$ is given by\footnote{
Note that this expression is ill-defined for physical integer values of $k$.
Hence this formula should be understood as
the simple analytic continuation to unphysical values of $k$.
As we will see in section \ref{sec:cancel},
the divergences at integer values of $k$ are actually apparent and
canceled out if we take into account contributions from D2-brane instanton.
}
\begin{\eq}
J_{\rm WS} (\mu )
= \sum_{m=1}^\infty \sum_{d_{1,2},j_{L,R}} (-1)^{s_L +s_R +1} N_{j_L ,j_R}^{d_1 ,d_2} 
  \Biggl[ -\frac{  s_R \sin{\frac{4\pi ms_L }{k} } }{m (2\sin{ \frac{2\pi m}{k} })^2 \sin{ \frac{4\pi m}{k} }}
             e^{-m (d_1 T_1 +d_2 T_2) } \Biggr] ,
\label{eq:WSinst}
\end{\eq}
where $N_{j_L ,j_R}^{d_1 ,d_2}$ denotes the BPS index of local $\mathbb{P}^1 \times \mathbb{P}^1$ partly listed\footnote{
A part of the data is extracted from the table in section 5.5.2 of \cite{Iqbal:2007ii}.
We are grateful to Can Koz\c{c}az to tell us the other data of the BPS index, which is not listed in \cite{Iqbal:2007ii}.
} in table \ref{tab:index}.
\begin{table}[t]
\begin{center}
  \begin{tabular}{|c|c|}
\hline           $(d_1 ,d_2 )$  & $\sum_{j_L ,j_R} N_{j_L ,j_R}^{d_1 ,d_2}(j_L,j_R)$  \\
\hline\hline   $(1 ,n ), n\geq 0$ & $(0,n+\hf)$ \\
\hline           $(2,2)$              & $(\hf,4)\oplus(0,\frac{7}{2})\oplus(0,\frac{5}{2})$ \\
\hline           $(2,3)$             &  $(1,\frac{11}{2})\oplus(\hf,5)\oplus(\hf,4)\oplus2(0,\frac{9}{2})\oplus (0,\frac{7}{2})\oplus(0,\frac{5}{2})$ \\
\hline           $(2,4)$          & $(\frac{3}{2},7)\oplus(1,\frac{13}{2})\oplus(1,\frac{11}{2})\oplus 2(\hf,6)\oplus(\hf,5)\oplus2(0,\frac{11}{2})\oplus(\hf,4)$\\
                                       & $\oplus2 (0,\frac{9}{2})\oplus (0,\frac{7}{2})\oplus (0,\frac{5}{2})$ \\
\hline           $(3,3)$          & $(2,\frac{15}{2})\oplus (\frac{3}{2},7)\oplus (\frac{3}{2},6)\oplus 3(1,\frac{13}{2})\oplus (\hf,7)\oplus 2(1,\frac{11}{2})
                                          \oplus 3(\hf,6)\oplus (1,\frac{9}{2})$ \\
                                       & $\oplus 3(\hf,5)\oplus 4(0,\frac{11}{2})\oplus 2(\hf,4)\oplus 3(0,\frac{9}{2}) 
                                            \oplus (\hf,3)\oplus 3(0,\frac{7}{2})\oplus (0,\frac{5}{2})\oplus (0,\frac{3}{2})$  \\
\hline          $(2,5)$        & $(2,\frac{17}{2})\oplus (\frac{3}{2},8)\oplus (\frac{3}{2},7)\oplus 2(1,\frac{15}{2})\oplus (1,\frac{13}{2})\oplus 2(\frac{1}{2},7)
                                     \oplus (1,\frac{11}{2})\oplus 2(\frac{1}{2},6)$\\
                                   &$\oplus 3(0,\frac{13}{2}) \oplus (\frac{1}{2},5)\oplus 2(0,\frac{11}{2})\oplus (\frac{1}{2},4)\oplus 2(0,\frac{9}{2})\oplus
                                       (0,\frac{7}{2})\oplus (0,\frac{5}{2}) $ \\
\hline          $(3,4)$       & $(3,\frac{19}{2})\oplus (\frac{5}{2},9)\oplus (\frac{5}{2},8)\oplus 3(2,\frac{17}{2})\oplus (\frac{3}{2},9)\oplus
                                        2(2,\frac{15}{2})\oplus 4(\frac{3}{2},8)$\\
                                 & $\oplus (1,\frac{17}{2})\oplus (2,\frac{13}{2})\oplus 4(\frac{3}{2},7) \oplus 7(1,\frac{15}{2})\oplus 2(\frac{1}{2},8)
                                   \oplus (0,\frac{17}{2})\oplus 2(\frac{3}{2},6)$\\
                               &$\oplus 6(1,\frac{13}{2})\oplus 7(\frac{1}{2},7)\oplus (0,\frac{15}{2})\oplus (\frac{3}{2},5)\oplus 5(1,\frac{11}{2})
                                 \oplus 8(\frac{1}{2},6) \oplus 7(0,\frac{13}{2})$\\
                               &$\oplus 2(1,\frac{9}{2})\oplus 6(\frac{1}{2},5)\oplus 6(0,\frac{11}{2})\oplus (1,\frac{7}{2})\oplus 4(\frac{1}{2},4)
                                  \oplus 7(0,\frac{9}{2})\oplus 2(\frac{1}{2},3)$ \\
                             &$\oplus 4(0,\frac{7}{2})\oplus (\frac{1}{2},2) \oplus 3(0,\frac{5}{2})\oplus (0,\frac{3}{2})\oplus (0,\frac{1}{2}) $\\
 \hline
\end{tabular}
\caption{BPS index $N_{j_L ,j_R}^{d_1 ,d_2}$ of local $\mathbb{P}^1 \times \mathbb{P}^1$ up to $d_1 +d_2 =7$. }
\label{tab:index}
\end{center}
\end{table}

It is also expected that the D2-brane instanton part is captured by
the Nekrasov-Shatashvili limit with the effective shift of the chemical potential,
\begin{\eq}
J_{\rm pert} (\mu ) +J_{\rm D2} (\mu )
=J_{\rm pert} (\mu_{\rm eff} ) 
  +\frac{1}{2\pi i} \frac{\del}{\del g_s}\Biggl[ g_s F_{\rm NS} \left( \frac{T_1^{\rm eff} }{g_s}, \frac{T_2^{\rm eff} }{g_s} ;\frac{1}{g_s} \right) \Biggr] .
\label{eq:D2}
\end{\eq}
In this expression, we treat $T_{1,2}^{\rm eff}$
and $g_s$ as independent variables, namely
\begin{\eq}
\frac{\del T_{1,2}^{\rm eff} }{\del g_s} =0 .
\end{\eq}
As discussed in \cite{Hatsuda:2013oxa}, 
the effective K\"ahler parameters are determined by the
so-called quantum A-period, whose closed form is not known for
the general value of $g_s$. However,
for
the physical case of integer $k$,
we can write down the effective K\"ahler parameter
$T_{1,2}^{\rm eff}$ in terms of a hypergeometric function
(see appendix C for an explanation)
\begin{align}
\frac{T_I^{\rm eff}}{g_s} &= 
\frac{T_I}{g_s} -4 e^{-\frac{T_I}{g_s}}\ _4 F_3 \left( 1,1,\frac{3}{2},\frac{3}{2},2,2,2;16e^{-\frac{T_I}{g_s}} \right)   & {\rm for}& \quad g_s^{-1} \in \mathbb{Z} \nonumber\\
\frac{2T_I^{\rm eff}}{g_s}&=\frac{2T_I}{g_s} -4 e^{-\frac{2T_I}{g_s}}\ _4 F_3 \left( 1,1,\frac{3}{2},\frac{3}{2},2,2,2;16e^{-\frac{2T_I}{g_s}} \right)   
& {\rm for} &\quad g_s^{-1} \in \mathbb{Z}+\frac{1}{2} ,
\label{eq:Teff}
\end{align}
which gives the effective chemical potential $\mu_{\rm eff}$ as
\begin{\eq}
\mu_{\rm eff} = \left\{ \begin{matrix}
\mu  -2(-1)^{\frac{k}{2}-M}e^{-2\mu} \ _4 F_3 \left( 1,1,\frac{3}{2},\frac{3}{2},2,2,2; (-1)^{\frac{k}{2}-M}16e^{-2\mu} \right)   & {\rm for} & k:{\rm even} \cr
\mu +e^{-4\mu} \ _4 F_3 \left( 1,1,\frac{3}{2},\frac{3}{2},2,2,2; -16e^{-4\mu} \right)   
& {\rm for} & k:{\rm odd} \cr
\end{matrix} \right. .
\label{eq:mueff}
\end{\eq}
The second term of the right hand side in \eqref{eq:D2} takes the following explicit form
\begin{\eqa}
&&\frac{1}{2\pi i} \frac{\del}{\del g_s}\Biggl[ g_s F_{\rm NS} \left( \frac{T_1^{\rm eff} }{g_s}, \frac{T_2^{\rm eff} }{g_s} ;\frac{1}{g_s} \right) \Biggr] \NN\\
&&=   \sum_{m=1}^\infty \sum_{d_{1,2},j_{L,R}} \frac{(-1)^{s_L +s_R +1}}{4m} N_{j_L ,j_R}^{d_1 ,d_2}\NN\\
&&\ \  \Biggl[
   \frac{\sin{\frac{k\pi ms_L}{2}} \sin{\frac{k\pi ms_R}{2}}  }{\pi m\sin^3{\frac{k\pi m}{2}} }  
  -\frac{k}{2} \frac{ s_L \cos{\frac{k\pi ms_L}{2}} \sin{\frac{k\pi ms_R}{2}} +s_R \sin{\frac{k\pi ms_L}{2}} \cos{\frac{k\pi ms_R}{2}}  }
                     { \sin^3{\frac{k\pi m}{2}} } \NN\\ 
&& +\frac{k}{2\pi} \left( 3\pi \cot{\frac{k\pi m}{2}}  +d_1 T_1^{\rm eff} +d_2 T_2^{\rm eff}  \right) 
    \frac{\sin{\frac{k\pi ms_L}{2}} \sin{\frac{k\pi ms_R}{2}}  }{ \sin^3{\frac{k\pi m}{2}} } 
         e^{-\frac{km}{2}(d_1 T_1^{\rm eff} +d_2 T_2^{\rm eff}) }  \Biggr] .
\label{eq:NSlimit}
\end{\eqa}
This shift of the chemical potential also plays an important role
to describe the mixed contribution of the worldsheet instantons and D2-instantons by
\begin{\eq}
J_{\rm WS} (\mu ) +J_{\rm WS+D2} (\mu ) = J_{\rm WS} (\mu_{\rm eff} ) .
\end{\eq}
To summarize, we conjecture that the total ABJ grand potential is given by
\begin{\eqa}
J_k^{(M)}(\mu ) 
&=& F_{\rm top} (T_1^{\rm eff} ,T_2^{\rm eff} ; g_s) 
  +\frac{1}{2\pi i} \frac{\del}{\del g_s}\Biggl[ g_s F_{\rm NS} \left( \frac{T_1^{\rm eff} }{g_s}, \frac{T_2^{\rm eff} }{g_s} ;\frac{1}{g_s} \right) \Biggr] \NN\\
&=& \frac{C}{3}\mu_{\rm eff}^3 +B\mu_{\rm eff} +A \NN\\
&& +\sum_{m=1}^\infty \sum_{d_{1,2},j_{L,R}} (-1)^{s_L +s_R +1} N_{j_L ,j_R}^{d_1 ,d_2} \NN\\
&&\times \Biggl[ - \frac{  s_R \sin{\frac{4\pi ms_L }{k} } }{m (2\sin{ \frac{2\pi m}{k} })^2 \sin{ \frac{4\pi m}{k} }}
             e^{-m (d_1 T_1^{\rm eff} +d_2 T_2^{\rm eff}) } \NN\\
&&    +\frac{1}{4m}  \Biggl\{   \frac{\sin{\frac{k\pi ms_L}{2}} \sin{\frac{k\pi ms_R}{2}}  }{\pi m\sin^3{\frac{k\pi m}{2}} }  
  -\frac{k}{2} \frac{ s_L \cos{\frac{k\pi ms_L}{2}} \sin{\frac{k\pi ms_R}{2}} +s_R \sin{\frac{k\pi ms_L}{2}} \cos{\frac{k\pi ms_R}{2}}  }
                     { \sin^3{\frac{k\pi m}{2}} } \NN\\ 
&& +\frac{k}{2\pi} \left( 3\pi \cot{\frac{k\pi m}{2}}  +d_1 T_1^{\rm eff} +d_2 T_2^{\rm eff}  \right) 
    \frac{\sin{\frac{k\pi ms_L}{2}} \sin{\frac{k\pi ms_R}{2}}  }{ \sin^3{\frac{k\pi m}{2}} } \Biggr\}
         e^{-\frac{km}{2}(d_1 T_1^{\rm eff} +d_2 T_2^{\rm eff}) }  \Biggr] .\NN\\
\label{eq:apparent}
\end{\eqa}
We can rewrite this expression in the form of \eqref{eq:expectJ} and read off
the coefficient $f_{l,m}(\mu )$ of the term
$e^{-(2\ell+4m/k)\mu}$, as expected from the gravity side.

By exponentiating $J_k^{(M )}(\mu )$ in \eqref{eq:expectJ}, we find the following expansion
\begin{\eq}
e^{J_k^{(M )}(\mu )}
= e^A \sum_{l,m=0}^\infty g_{l,m}(\mu ) \exp{\Biggl[ \frac{C}{3}\mu^3 +\left( B -2l -\frac{4m}{k} \right)\mu \Biggr]} ,
\label{Xiexp}
\end{\eq}
where $g_{l,m}(\mu )$ is a polynomial of $\mu$ explicitly determined by 
the coefficient $f_{l,m}(\mu )$.
Then we come back to the canonical formalism by applying the integral transform
\eqref{eq:comeback2} to \eqref{Xiexp},
\begin{\eq}
\hat{Z}^{(N,N+M)}(k)
= C^{-\frac{1}{3}}e^A  \sum_{l,m=0}^\infty  g_{l,m}\left( -\frac{\del}{\del N} \right) 
{\rm Ai}\Biggl[ C^{-\frac{1}{3}}\left( N-B+2l +\frac{4m}{k}\right) \Biggr] .
\label{Zexp}
\end{\eq}
Since the Airy function satisfies the differential equation
\[
\frac{d^2}{dz^2} {\rm Ai}[z] = z {\rm Ai}[z] ,
\]
the canonical partition function given in
\eqref{Zexp}
can be rewritten as a combination of the Airy function and its first derivative alone.

The expression \eqref{eq:apparent} has apparent divergences at some values of the
Chern-Simons level $k$,
in particular at physical integer $k$. 
For example, for odd $k$
such divergences occur 
when $m$ in \eqref{eq:apparent} is a multiple of $k$ for the worldsheet instanton and even integer 
for D2-instanton, respectively.
In the next subsubsection, we discuss that these divergences are actually canceled out 
and we compute the finite part of the grand potential.

\subsubsection{Cancellation of divergence between instantons}
\label{sec:cancel}
Here we show the cancellation of apparent divergences in the grand potential \eqref{eq:apparent} and 
compute its finite part.
\subsubsection*{Even Chern-Simons level}
Let us first consider the even $k$ case;
we set $k=2n_0$ with some integer $n_0$.
By expanding\footnote{
This expansion should be handled with care. See the comments below and 
section \ref{sec:MM} for detail.
} the $m$-th worldsheet instanton term with $m=n_0 \ell$ ($\ell\in\mathbb{Z}$)
around $k=2n_0$, 
we find that the worldsheet instanton part \eqref{eq:WSinst} has poles at $k=2n_0$:
\begin{\eqa}
&& \frac{ (-1)^{s_L +s_R +(n+M)\ell d_-}  }{4n_0 \ell}  s_R s_L N_{j_L ,j_R}^{d_1 ,d_2}   e^{-2l d_+ \mu_{\rm eff} } \NN\\
&&\Biggl[ \frac{4n^2}{\pi^2 \ell^2 (k-2n_0 )^2} +\frac{4n }{\pi^2 \ell^2 (k-2n_0 ) }  
     +1 +\frac{1}{\pi^2 \ell^2} -\frac{2}{3}s_L^2  +\mathcal{O} (k-2n_0 ) \Biggr]  ,
\end{\eqa} 
where we have introduced the notation
\begin{\eq}
d_\pm = d_1 \pm d_2 .
\end{\eq}
The contribution \eqref{eq:NSlimit} from $F_{\rm NS}$, which gives the part of the D2-instanton correction, 
has also poles for arbitrary $m$:
\begin{\eqa}
&&\frac{(-1)^{(s_L +s_R +1 )(mn_0 +1)+mMd_- + d_- mn_0}   e^{-2m d_+ \mu_{\rm eff} } s_L s_R N_{j_L ,j_R}^{d_1 ,d_2} }{4m} \NN\\
&& \Biggl[ \frac{4n_0}{\pi^2 m^2 } \frac{1}{(k-2n_0 )^2}   +\frac{4}{\pi^2 m^2}  \frac{1}{k-2n_0 }   -\frac{n_0 }{2} +\frac{n_0}{6}(s_L^2 +s_R^2 )  \NN\\
&& -\frac{2d_+^2}{\pi^2 n_0} \mu_{\rm eff}^2  -\frac{2d_+}{\pi^2 mn_0} \mu_{\rm eff} +\frac{d_-^2}{2n_0}(n_0 -M)^2  
     +\mathcal{O} (k-2n_0 ) \Biggr]  .
\end{\eqa} 
It is easy to see that these poles are canceled with each other if
\begin{\eq}
(-1)^{s_L +s_R +1}=1 .
\label{spincond}
\end{\eq}
It turns out that the
non-zero values of the BPS index 
$N_{j_L ,j_R}^{d_1 ,d_2}$ for local $\mathbb{P}^1 \times \mathbb{P}^1$
appear only for the spins obeying the condition
\eqref{spincond}, and hence the poles are indeed canceled.
After the pole cancellation, we find the finite part of the grand potential as\footnote{
We have used the definition of polylogarithm 
${\rm Li}_s (z) = \sum_{p=1}^\infty \frac{z^p}{p^s}$.
}
\begin{\eqa}
J(\mu ) 
&=& \frac{C}{3}\mu_{\rm eff}^3 +B\mu_{\rm eff} +A \NN\\
&& +\sum_{m, 2m/k \neq \mathbb{Z} } \sum_{d_{1,2},j_{L,R}} (-1)^{s_L +s_R +1} N_{j_L ,j_R}^{d_1 ,d_2} 
  \Biggl[ - \frac{e^{ -2m\pi i d_- \left( \frac{1}{2} -\frac{M}{k} \right) } 
                                     s_R \sin{\frac{4\pi ms_L }{k} } }{m (2\sin{ \frac{2\pi m}{k} })^2 \sin{ \frac{4\pi m}{k} }}
             e^{-\frac{4m}{k} d_+ \mu_{\rm eff} } \Biggr] \NN\\
&& +\sum_{m=1 }^\infty \sum_{d_{1,2},j_{L,R}} \frac{  s_L s_R  }{2k}  N_{j_L ,j_R}^{d_1 ,d_2} \NN\\
&&  \Biggl[
          - \frac{2 d_+^2 }{\pi^2} \mu_{\rm eff}^2 {\rm Li}_1 ( (-1)^{(\frac{k}{2} +M)d_+} e^{-2d_+ \mu_{\rm eff} } )       
          - \frac{2 d_+ }{\pi^2} \mu_{\rm eff} {\rm Li}_2 ( (-1)^{(\frac{k}{2}+M)d_+} e^{-2d_+ \mu_{\rm eff} } )       \NN\\
&&      -\frac{1}{\pi^2} {\rm Li}_3 ( (-1)^{(\frac{k}{2}+M)d_+} e^{-2d_+ \mu_{\rm eff} } )  \NN\\
&&      +\left\{ \frac{ (16+k^2 ) s_L^2 +k^2 s_R^2 }{24} -\frac{8+k^2}{8} +\frac{d_-^2}{2} \left( \frac{k}{2}-M\right)^2 \right\}
          {\rm Li}_1 ( (-1)^{(\frac{k}{2}+M)d_+} e^{-2d_+ \mu_{\rm eff} } ) \Biggr] . \NN\\
\label{eq:evenk}
\end{\eqa}
 
Some comments are in order here. 
First, when taking the limit $k\to 2n_0$ in our general expression
\eqref{eq:apparent}, we should treat 
$k$ and $T^{\rm eff}_{1,2}$ as independent variables
since we have imposed the condition
$\del T_I^{\rm eff} /\del g_s =0$. This means that
$k$ in the expression of  $T^{\rm eff}_{1,2}$ in \eqref{eq:Teff0}
is set to $2n_0$ {\it before} taking the limit $k\to 2n_0$.
Second, the resulting finite part \eqref{eq:evenk} is
invariant under the Seiberg-like duality $M\leftrightarrow k-M$
for even integer $k$.
The same comments can be applied also to the odd $k$ case.

\subsubsection*{Odd Chern-Simons level}
When $k$ is odd,
there are also apparent divergences 
in the worldsheet instanton with $m=k\ell$ and the D2-instanton with $m=2\ell$ ($\ell\in\mathbb{Z}$).
As in the even $k$ case, similar calculation shows that 
these divergences are canceled again and the finite part of the grand potential is\footnote{
We have used $\sum_{p=1}^\infty \frac{x^{2p-1} }{2p-1} = {\rm Arctanh}x$.
}
\begin{\eqa}
J(\mu ) 
&=& \frac{C}{3}\mu_{\rm eff}^3 +B\mu_{\rm eff} +A \NN\\
&& +\sum_{m, 2m/k \neq \mathbb{Z} } \sum_{d_{1,2},j_{L,R}} (-1)^{s_L +s_R +1} N_{j_L ,j_R}^{d_1 ,d_2} 
  \Biggl[ - \frac{e^{ -2m\pi i d_- \left( \frac{1}{2} -\frac{M}{k} \right) } 
                                     s_R \sin{\frac{4\pi ms_L }{k} } }{m (2\sin{ \frac{2\pi m}{k} })^2 \sin{ \frac{4\pi m}{k} }}
             e^{-\frac{4m}{k} d_+ \mu_{\rm eff} } \Biggr] \NN\\
&& + \sum_{d_{1,2},j_{L,R}} \frac{  s_L s_R  }{4k}  N_{j_L ,j_R}^{d_1 ,d_2} 
  \Biggl[
          - \frac{2 d_+^2 }{\pi^2} \mu_{\rm eff}^2 {\rm Li}_1 ( (-1)^{d_+} e^{-4d_+ \mu_{\rm eff} } )       
          - \frac{ d_+ }{\pi^2} \mu_{\rm eff} {\rm Li}_2 ( (-1)^{d_+} e^{-4d_+ \mu_{\rm eff} } )       \NN\\
&&      -\frac{1}{4\pi^2} {\rm Li}_3 ( (-1)^{d_+} e^{-4d_+ \mu_{\rm eff} } )  \NN\\
&&      +\left\{ \frac{ (16+k^2 ) s_L^2 +k^2 s_R^2 }{24} -\frac{8+k^2}{8} +\frac{d_-^2}{2} \left( \frac{k}{2}-M \right)^2 \right\}
          {\rm Li}_1 ( (-1)^{d_+} e^{-4d_+ \mu_{\rm eff} } ) \Biggr]  \NN\\
&& + \frac{k}{8} \sum_{d_{1,2},j_{L,R}}  N_{j_L ,j_R}^{d_1 ,d_2} P_{d_+}  \left(  P_{s_R} s_R +  P_{s_L} s_L \right)   
 (-1)^{\frac{d_-}{2} +\frac{k}{2}(s_L +s_R +1) }    {\rm Arctanh}( e^{-2d_+ \mu_{\rm eff}} ) ,\NN\\
\label{eq:oddk}
\end{\eqa}
where $P_l$ denotes the projection to even $l$,
\begin{\eq}
P_l = \frac{1}{2} \Bigl( 1+(-1)^l \Bigr) .
\end{\eq}

\subsection{Test of our proposal}
\begin{figure}[!t]
\begin{center}
\includegraphics[width=7.5cm,height=4.6cm]{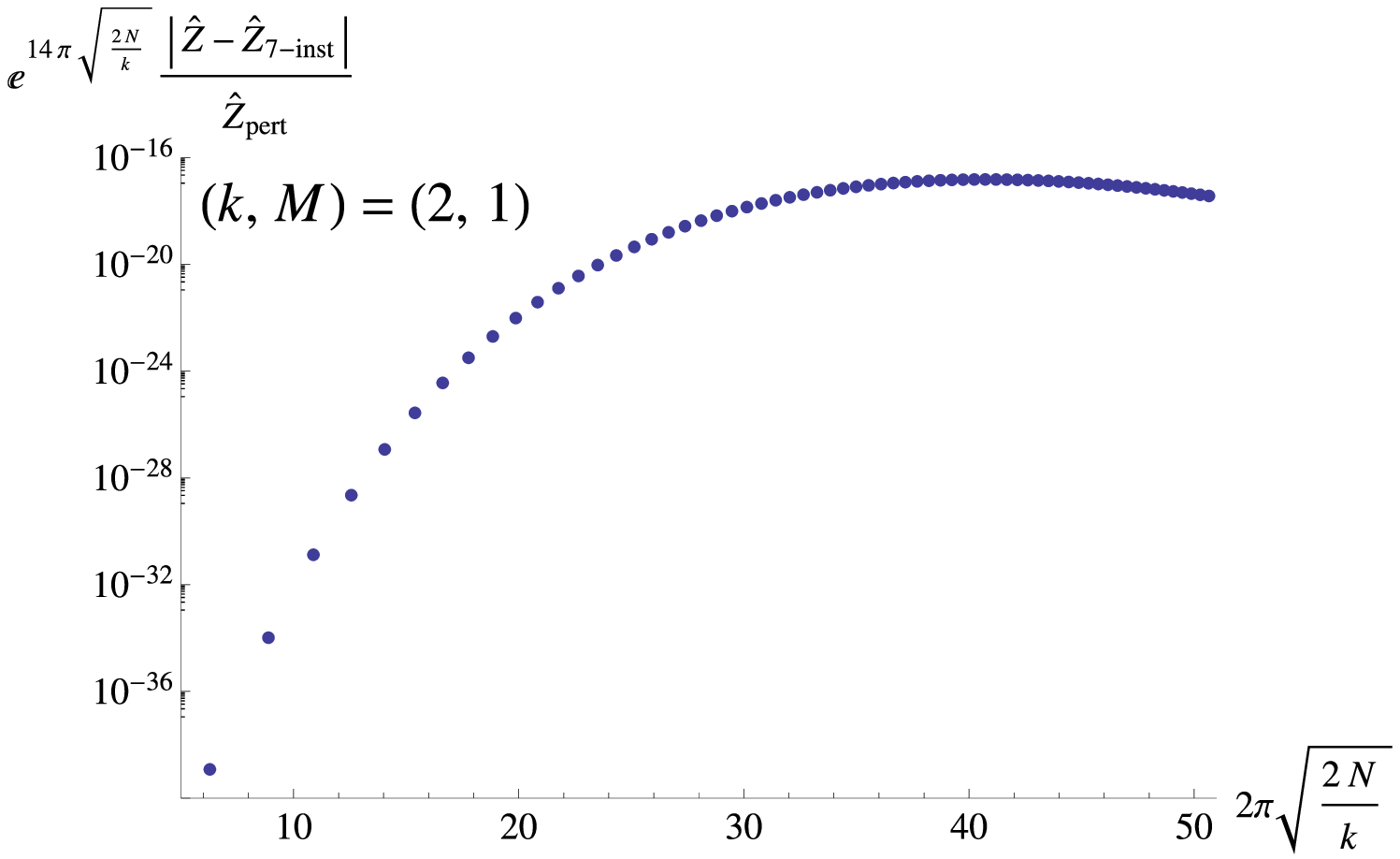}
\includegraphics[width=7.5cm,height=4.6cm]{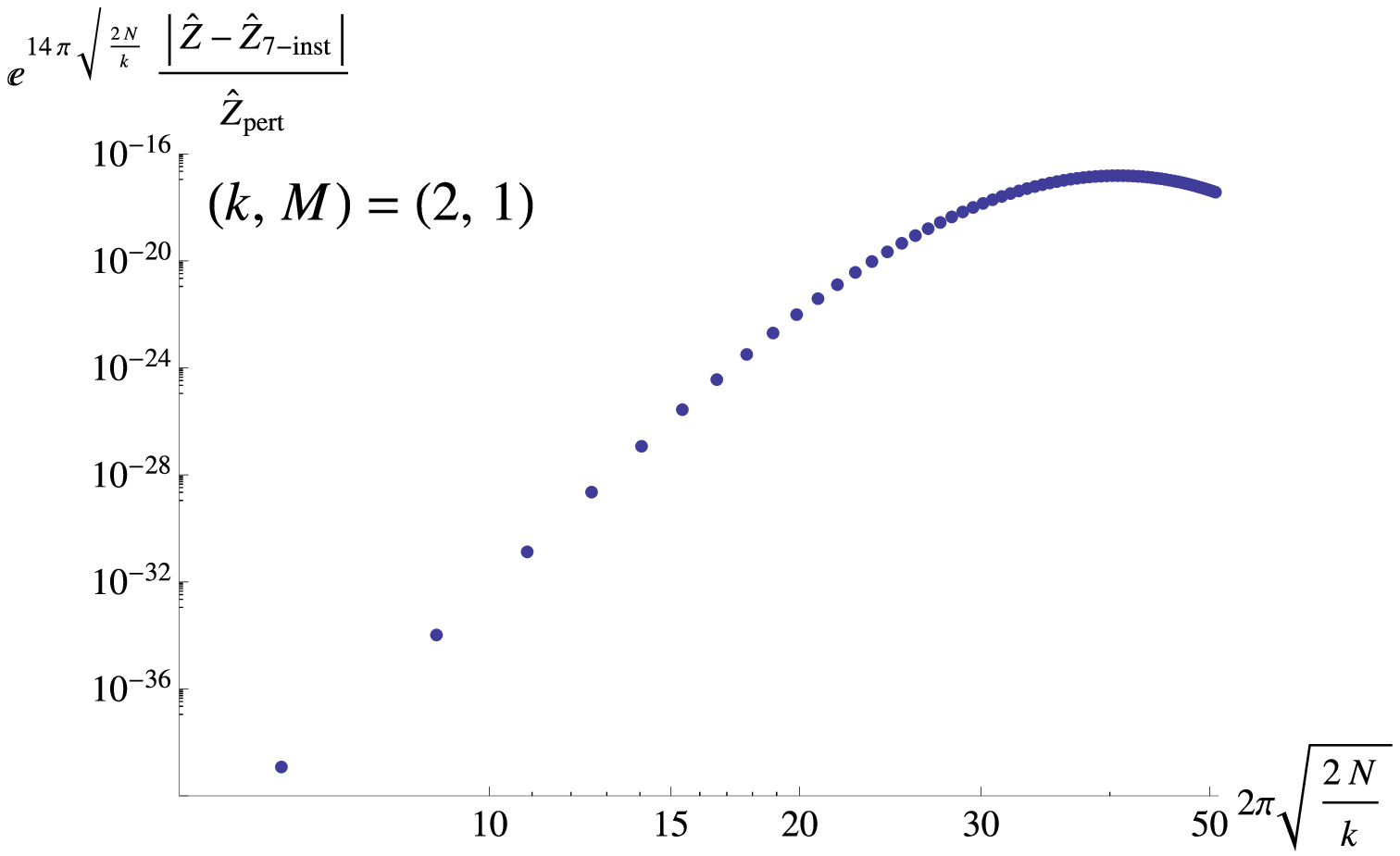}\\
\includegraphics[width=7.5cm,height=4.6cm]{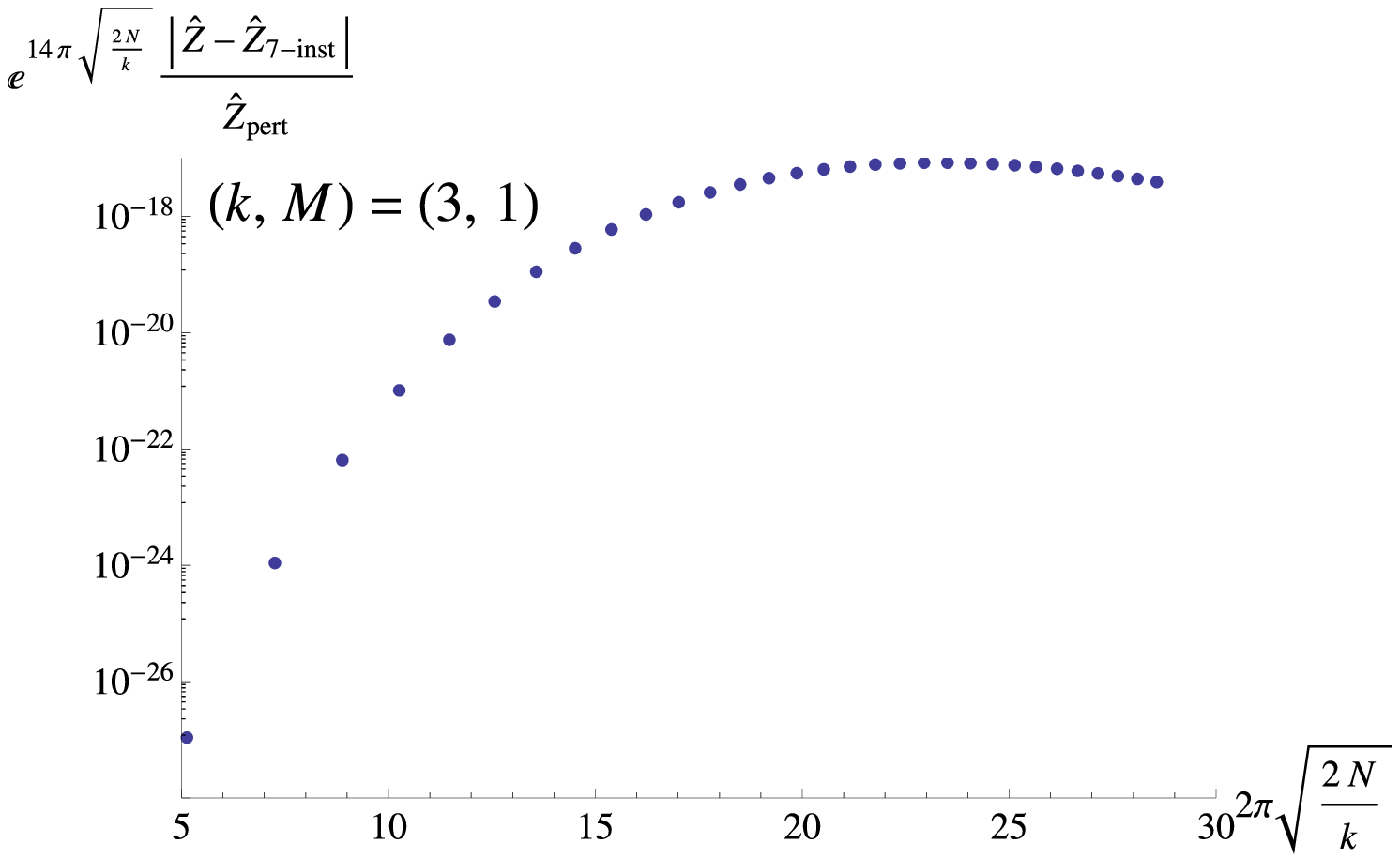}
\includegraphics[width=7.5cm,height=4.6cm]{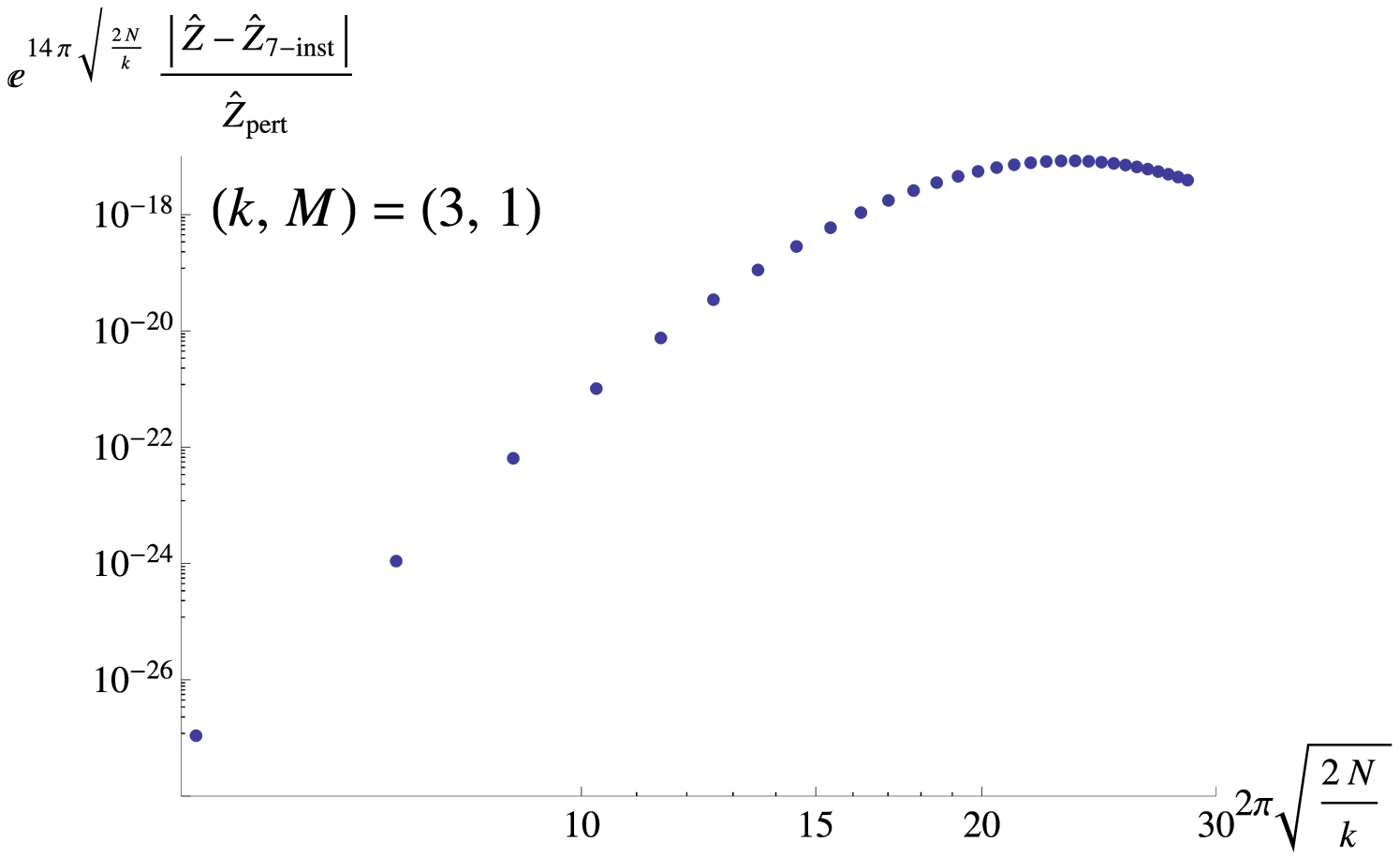}\\
\includegraphics[width=7.5cm,height=4.6cm]{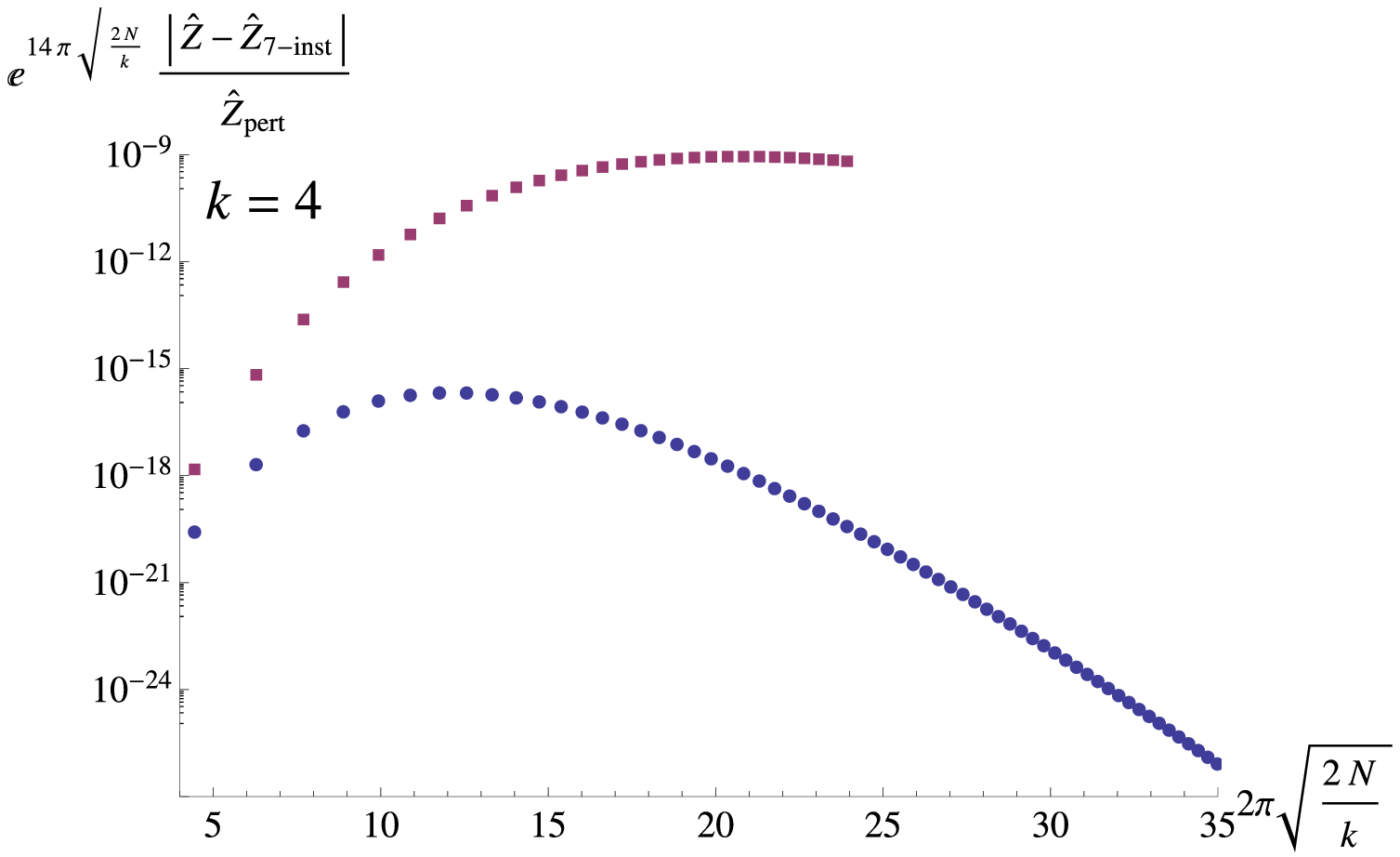}
\includegraphics[width=7.5cm,height=4.6cm]{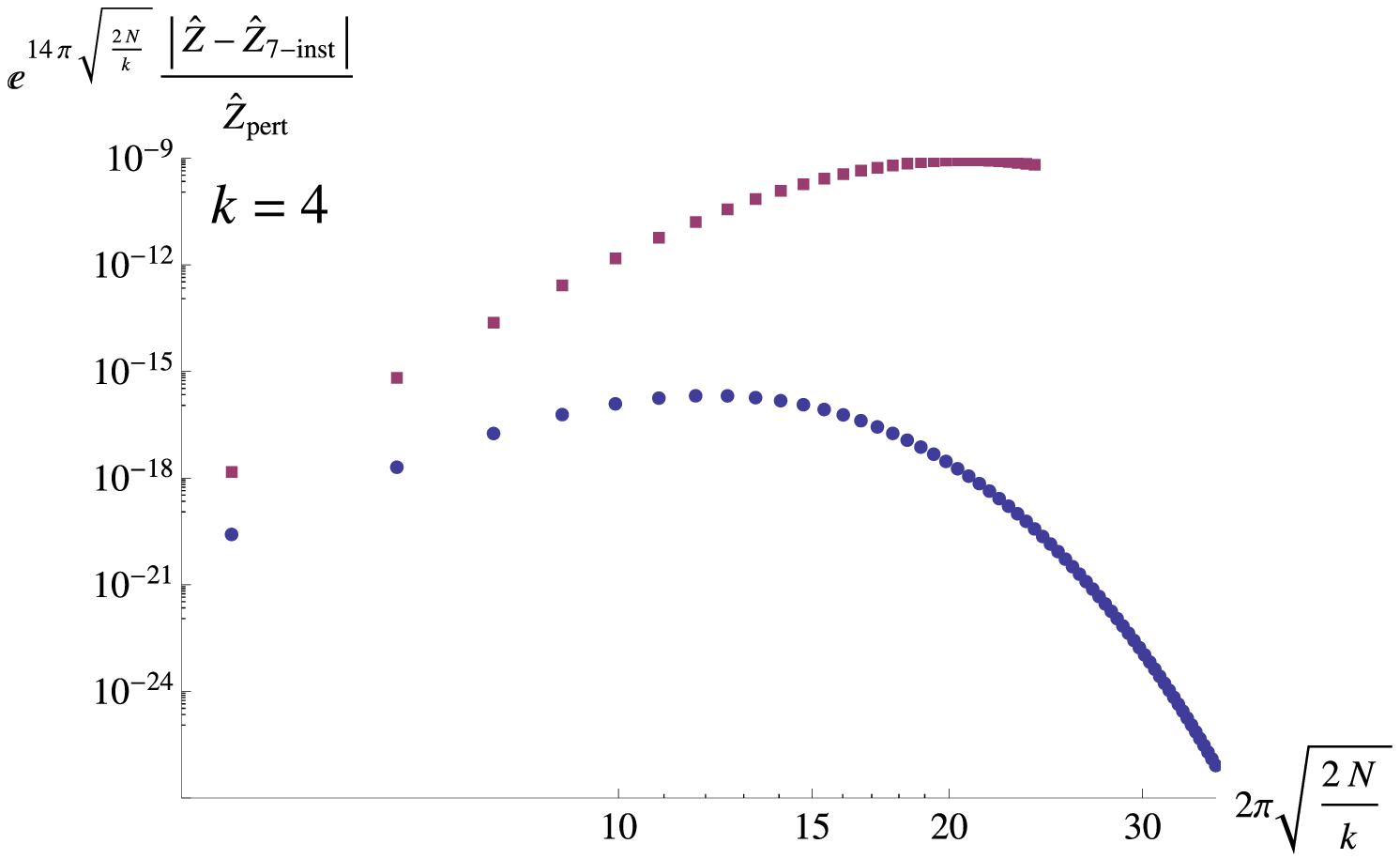}\\
\includegraphics[width=7.5cm,height=4.6cm]{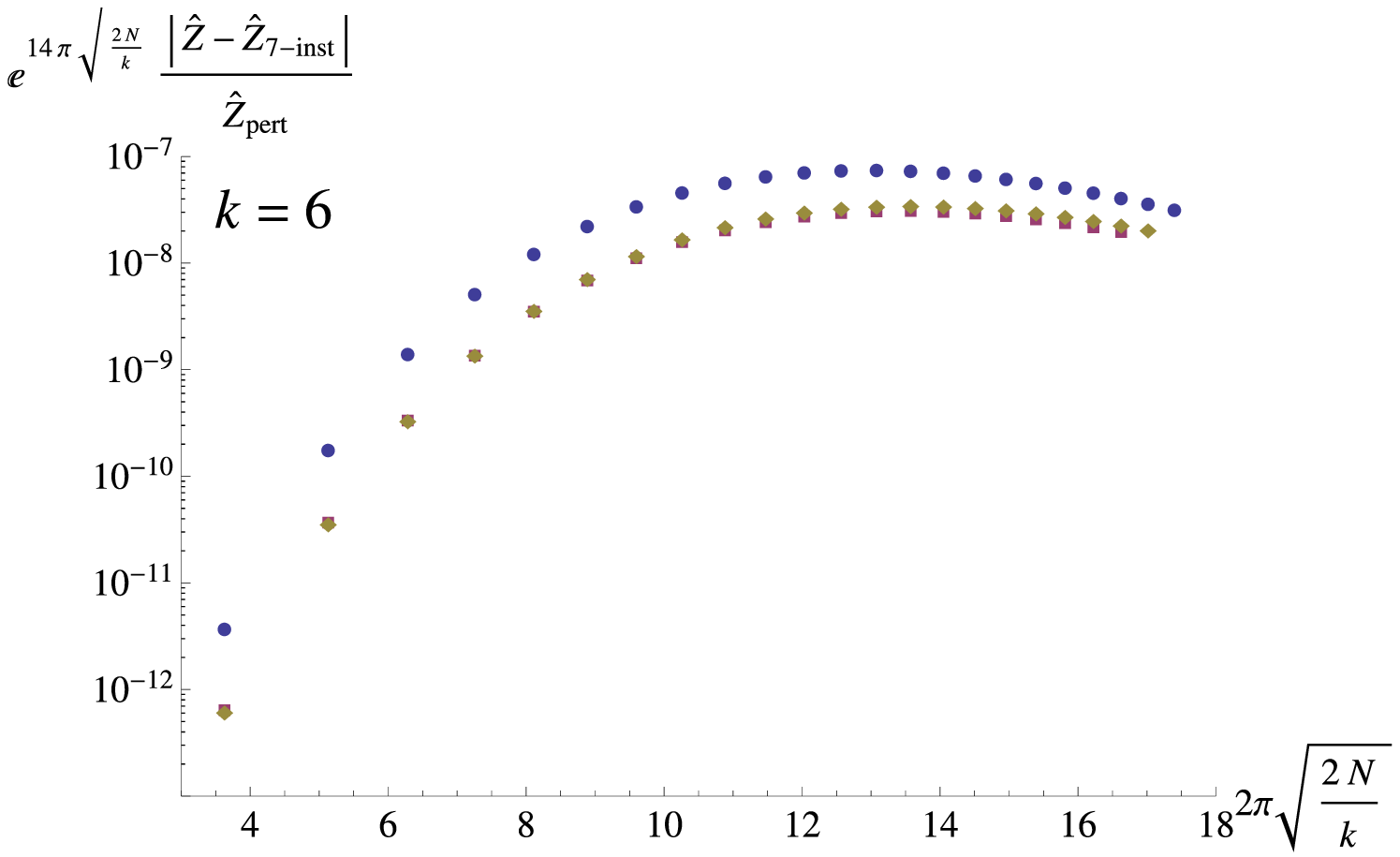}
\includegraphics[width=7.5cm,height=4.6cm]{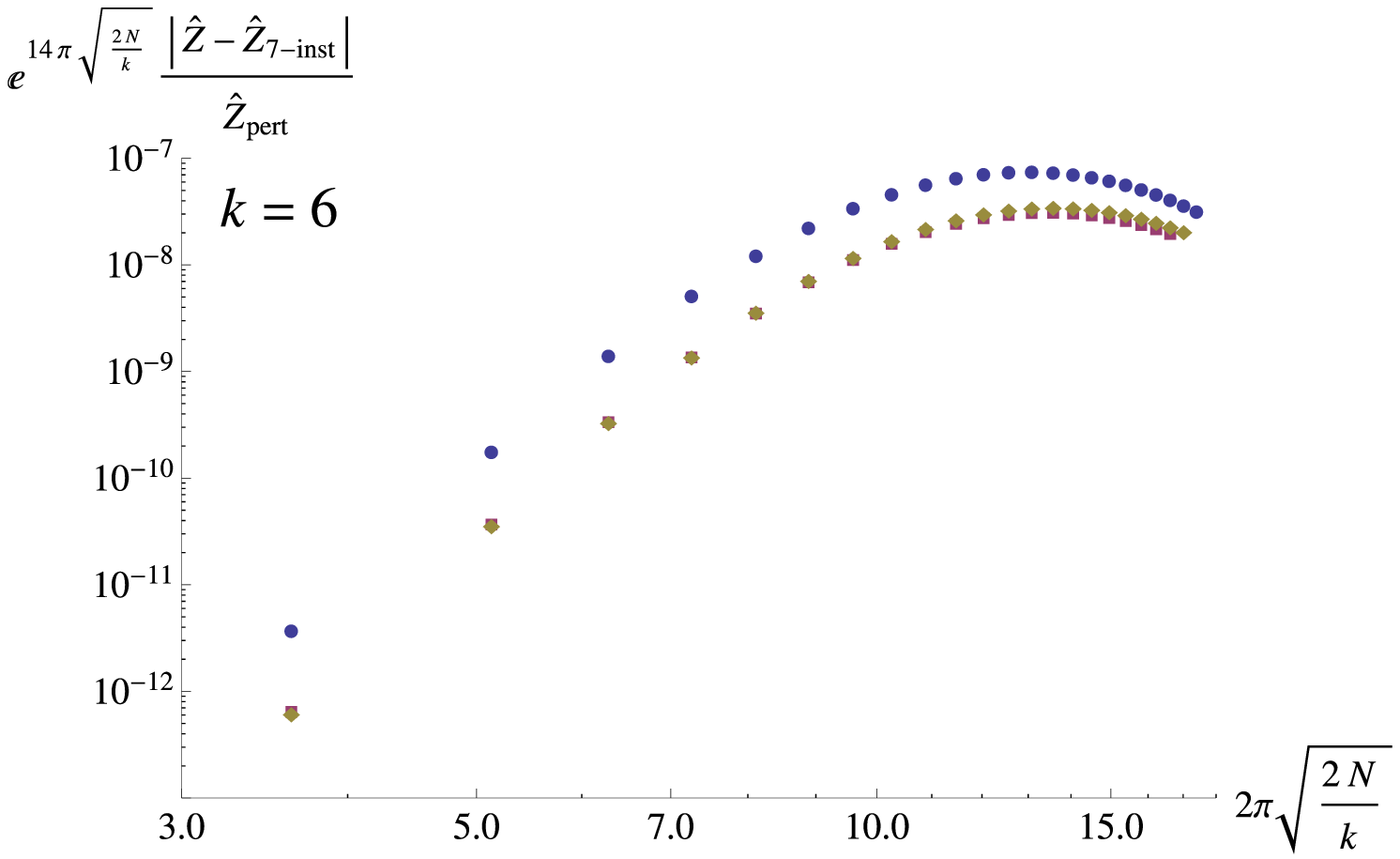}
\end{center}
\caption{
The quantity $e^{14\pi \sqrt{2N}{k}} |\hat{Z}_k^{(M)} -\hat{Z}_{\rm 7-inst}|/ \hat{Z}_{\rm pert}$ is plotted to $2\pi\sqrt{2N/k}$
both in semi-log scale (Left)  and log-log scale (Right) for $(k,M)=(2,1),(3,1),(4,1),(4,2),(6,1),(6,2)$ and $(6,3)$.
The blue circle, purple square and yellow diamond symbols show
the cases for $M=1$, $M=2$ and $M=3$, respectively.
}
\label{fig:inst}
\end{figure}
Let us test our conjecture of the finite part \eqref{eq:evenk} for even $k$ and
\eqref{eq:oddk} for odd $k$,
expected from the refined topological string.
Since the odd number of D2-instanton does not appear in the finite part of the expected grand potential for odd $k$ in \eqref{eq:oddk},
we can rewrite the expansion of the grand potential \eqref{Zexp} as\footnote{
Note that this expression of course includes also D2-instanton effects
since $k\ell /2$-th worldsheet instanton actually contributes by the same weight as $\ell$-th D2-instanton.
}
\begin{\eq}
e^{J_k^{(M )}(\mu )}
= e^A \sum_{m=0}^\infty \hat{g}_m (\mu ) \exp{\Biggl[ \frac{C}{3}\mu^3 +\left( B  -\frac{4m}{k} \right)\mu \Biggr]} ,
\end{\eq}
where $\hat{g}_m (\mu )$ is a polynomial of $\mu$.
Since we know the values of the BPS index $N_{j_L ,j_R}^{d_1 ,d_2}$ up to $d_1 +d_2 =7$ from table \ref{tab:index}, 
we can explicitly write down $\hat{g}_m (\mu )$ up to $m=7$.
Using these coefficients $\hat{g}_m (\mu )~(m=0,\cdots,7)$, we define the putative
partition function
\begin{\eq}
\hat{Z}_{\rm 7-inst}
= e^A  \sum_{m=0}^7  \hat{g}_m \left( -\frac{\del}{\del N} \right) {\rm Ai}\Biggl[ C^{-1/3}\left( N-B +\frac{4m}{k}\right) \Biggr].
\end{\eq}
If our conjecture is correct,
then this quantity should be the canonical partition function including the
instanton effects up to the order of 7-th worldsheet instanton correction.
Hence we expect
that the difference between the exact partition function
$\hat{Z}_k^{(M)}$ and $\hat{Z}_{\rm 7-inst}$ behaves as
\begin{\eq}
\frac{\hat{Z}_k^{(M)} -\hat{Z}_{\rm 7-inst}}{\hat{Z}_{\rm pert}}
=  
    \frac{\sum_{m=8}^\infty  \hat{g}_m \left( -\frac{\del}{\del N} \right) {\rm Ai}\Bigl[ C^{-1/3}\left( N-B +\frac{4m}{k}\right) \Bigr]}{{\rm Ai}\Bigl[ C^{-1/3}\left( N-B \right) \Bigr]} 
= \mathcal{O}(N^{\frac{\alpha}{2}} e^{-16\pi \sqrt{\frac{2N}{k}} }) ,
\label{eq:exp}
\end{\eq}
in the large $N$ regime.
Here $\alpha$ is some non-negative integer depending\footnote{
For example, if $(k,M)$ is $(2,1)$, the formula \eqref{eq:evenk} implies $\alpha =16$. 
This is because the grand partition function contains a term with $\mathcal{O}(\mu^{16}e^{-16\mu})$
coming from the $\mu^2$-terms in the first D2-instanton correction of the grand potential.
} on $k$ and $M$.
In fig.~\ref{fig:inst}, we plot the quantity 
\begin{align}
e^{14\pi \sqrt{2N}{k}} \frac{\left| \hat{Z}_k^{(M)} -\hat{Z}_{\rm 7-inst} \right|}{\hat{Z}_{\rm pert} },
\label{diffZ} 
\end{align}
against $2\pi\sqrt{2N/k}$
both in semi-log scale (Left)  and log-log scale (Right) for various values of $(k,M)$.
From fig.~\ref{fig:inst} one can see that this quantity \eqref{diffZ}
is extremely small,
and this already gives strong evidence for our conjecture.
We can argue more precisely as follows.
If $\hat{Z}_{\rm 7-inst}$ properly contains the instanton effects, then this quantity
 \eqref{diffZ} should be exponentially suppressed 
by $e^{-2\pi \sqrt{\frac{2N}{k}} }$ in the large $N$ regime.
On the other hand, if $\hat{Z}_{\rm 7-inst}$ had wrong information at 6-th or lower order of worldsheet instanton effects,
we could observe an exponential blow-up of this quantity.
The semi-log plots in fig.~\ref{fig:inst} (Left) show that
such exponential blow-up does not occur.
We also see from the log-log plots in fig.~\ref{fig:inst} (Right) that
the data points give a power law behavior in not so large $N$ region.
This behavior is consistent with our expectation \eqref{eq:exp}
since the data should show such a power law behavior before going to exponentially suppressed region.
Thus we conclude that 
the grand potential \eqref{eq:apparent} coming from the refined topological string is correct at least up to the
order of 6-th worldsheet instanton corrections.
In particular the case of $(k,M)=(4,1)$, plotted in the second bottom of fig.~\ref{fig:inst} (Left), clearly exhibits the exponentially suppressed behavior. 
This strongly supports our expectation \eqref{eq:exp}.

To summarize, all cases we studied strongly support our conjecture
that the ABJ grand potential is given by the
general expression \eqref{eq:conclusion}
in terms of the refined topological string on local $\mathbb{P}^1\times\mathbb{P}^1$,
with an appropriate identification of the K\"{a}hler parameters \eqref{eq:Teff0}.

\subsection{Comments on Matsumoto-Moriyama proposal}
\label{sec:MM}
Recently an apparently similar form of the grand potential was proposed in \cite{Matsumoto:2013nya}
by a different approach, which uses a relation between the ABJ partition function and 
the half-BPS Wilson loops of the ABJM theory.
The work \cite{Matsumoto:2013nya} have calculated the ABJ partition function exactly or numerically with high precision
for $(k,M, N_{\rm max} )=(2,1,4)$, $(3,1,3)$, $(4,1,2)$, $(4,2, 2)$, $(6,1, 3 )$,  $(6,2, 2 )$ and $(6,3,2 )$.
These values are totally consistent with our exact values listed in appendix \ref{app:values}.
However, there is an important difference of physical interpretation between \cite{Matsumoto:2013nya} and ours.

This difference comes from an ambiguity for taking the limit to a physical Chern-Simons level 
in the apparently divergent grand potential \eqref{eq:apparent}.
Suppose that we take the limit $k\rightarrow k_0\ (k_0 \in \mathbb{Z})$ in \eqref{eq:apparent}.
In section \ref{sec:cancel}, we have expanded \eqref{eq:apparent} around $k=k_0$ 
after fixing $k=k_0$ in the effective K\"ahler parameters $T_I^{\rm eff}$
since we have imposed $\del T_I^{\rm eff} /\del g_s =0$,
while the authors in \cite{Matsumoto:2013nya} expanded the total expression \eqref{eq:apparent} around $k=k_0$.
This difference of taking the limit affects the finite part\footnote{
Fortunately or unfortunately, this difference vanishes for $M=0$.
Therefore this ambiguity of how to take the limit is irrelevant for the ABJM case \cite{Hatsuda:2013oxa}.
} of the grand potential after the cancellation of divergence.
Since the latter prescription does not agree with the exact data,
the authors in \cite{Matsumoto:2013nya} added an extra new term, whose origin is unclear from the 
topological string perspective,
in such a way that the finite part agrees with their exact and numerical data.
It turns out that the finite part of our grand potential, corresponding to \eqref{eq:evenk} and \eqref{eq:oddk}, 
agrees with the one given in Figure 4 of \cite{Matsumoto:2013nya}.
However, our grand potential before cancellation \eqref{eq:apparent} is different 
from the proposal in \cite{Matsumoto:2013nya} with an extra term added.
We believe that 
our prescription is more natural since
everything can be completely explained by the refined topological string.

\section{Discussions}
\label{sec:final}
In this paper we have studied the partition function of the ABJ theory on $S^3$.
By using the Fermi gas formalism and Tracy-Widom's lemma \cite{Tracy:1995ax},
we have exactly computed the ABJ partition function 
for $(k,M, N_{\rm max} )=(2,1,65 )$, $(3,1,31 )$, $(4,1,62 )$, $(4,2, 29 )$, $(6,1, 23 )$, $(6,2, 21 )$ and $(6,3,22 )$.
These exact data enable us to test
the relation between the ABJ partition function and the free energy of the refined topological string on local $\mathbb{P}^1 \times \mathbb{P}^1$
with non-diagonal K\"ahler parameters.
We conclude that 
the ABJ partition function is completely determined by the refined topological string
including full series of the membrane instanton effects in the M-theory dual.

Although here we have focused on the usual AdS/CFT correspondence between
the ABJ theory and the M-theory on $AdS_4 \times S^7 /\mathbb{Z}_k$,
it is also interesting to study the higher spin limit \cite{Chang:2012kt,Giombi:2011kc},
corresponding to the regime where $M,k\gg 1$ with $M/k$ and $N$ kept fixed.
It would also be interesting to study the eigenvalue problem 
of the Hamiltonian \eqref{Hamiltonian} of the ABJ Fermi gas,
as in the case of the ABJM theory \cite{Kallen:2013qla}.
We leave this for future work.

Throughout many previous works about the ABJM theory and this work,
we have seen that
the (refined) topological string is a very powerful tool to determine the structures of the partition function and  the BPS Wilson loops in the ABJ(M) theory.
However, if we consider more general M2-brane theory beyond\footnote{
For previous studies, see e.g. \cite{Martelli:2011qj,Cheon:2011vi,Jafferis:2011zi,Gulotta:2011si,Gulotta:2011vp,Gulotta:2012yd} in the M-theory limit, 
\cite{Santamaria:2010dm,Suyama:2011yz,Suyama:2012uu,Suyama:2013fua} in the 't Hooft limit 
and \cite{Marino:2011eh,Marino:2012az,Mezei:2013gqa,Grassi:2014vwa,Honda:2014ica} in the Fermi gas approach.
} the ABJ(M) theory,
the
correspondence to the topological string is rather obscure, and it might be the case that there is no such relation in general.
Therefore even if we can exactly compute the partition function for the case
other than 
the ABJ(M) theory for some discrete values of parameters,
we do not have a good guiding principle 
to extend the non-perturbative effects to more general parameters.
Although this problem can be overcome for the ``orbifold ABJ(M) theory'' \cite{Honda:2014ica}
due to a relation to the ABJ(M) theory,
we need to develop a new approach to study more general M2-brane theories.
It is very interesting to study the instanton effects
in other M2-brane theories having $AdS_4$ dual.

\subsection*{Acknowledgment}
The authors would like to thank Yasuyuki Hatsuda, Johan K\"all\'en, Hiroaki Kanno, Can Koz\c{c}az, Sanefumi Moriyama, 
Kazuhiro Sakai, Ashoke Sen and Masato Taki for valuable discussions.
We would also like to thank Satoru Odake for
allowing us to use computers in the theory group, Shinshu University,
for the computation of the exact partition functions.
M.H. is grateful to KEK Theory Center and Kavli IPMU for warm hospitality.
The work of K.O. is supported in part by JSPS Grant-in-Aid for Young Scientists
(B) 23740178.

\appendix
\section{Algorithm for exact computation of ABJ grand partition function}
\label{app:algorithm}
In this appendix,
we show that 
the grand canonical partition function of the ABJ theory is 
determined by a series of functions $\phi_l^+(y)$ 
which can be computed recursively by the formula \eqref{eq:TBA}.
Our algorithm is essentially a simple generalization of the ABJM case \cite{Putrov:2012zi,Hatsuda:2012dt}.

As discussed in section \ref{sec:Fermi},
the ABJ grand partition function is given by
\begin{\eq}
\Xi_k^{(M)}(z) = \exp{\Biggl[ -\sum_{n=1}^\infty \frac{(-1)^n}{n} {\rm Tr}\rho^n \Biggr]} ,
\end{\eq}
with the density matrix \eqref{eq:density} .
Here we define multiplication and trace for certain matrices $\rho_1 ,\rho_2$ as
\begin{\eq}
\rho_1 \rho_2 (x ,y) = \int_{-\infty}^\infty \frac{dq}{2\pi k} \rho_1 (x,q) \rho_2 (q,y) ,\quad
{\rm Tr}\rho_1 = \int_{-\infty}^\infty \frac{dq}{2\pi k} \rho_1 (q,q),
\end{\eq}
respectively.
Then we decompose the density matrix into the even and odd parts:
\begin{\eqa}
\rho (x,y) = \rho_+ (x,y) +\rho_- (x,y),\quad  \rho_\pm (x,y) =\frac{\rho (x,y) \pm\rho (x,-y)}{2} .
\end{\eqa}
In terms of $\rho_\pm$, we rewrite the grand partition function as
\begin{\eq}
\Xi_k^{(M)}(z) = \Xi_+ (z) \Xi_- (z)  ,\quad \Xi_\pm (z) = {\rm Det}(1+z\rho_\pm ) .
\label{eq:pm}
\end{\eq}
Note that $\rho_\pm$ takes the form of
\begin{\eq}
\rho_\pm (x,y)
=  \frac{E_\pm (x)  E_\pm (y)  }{\cosh{\frac{x}{k}} +\cosh{\frac{y}{k}} } ,\quad
E_\pm (x) = \frac{\sqrt{V(x)}\left( e^{\frac{x}{2k}} \pm e^{\frac{x}{2k}}\right) }{2}.
\end{\eq}
Therefore, both $\rho_\pm$ have the form of 
\[
\frac{E(x)E(y)}{M(x)+M(y)} ,
\]
where the Tracy-Widom's lemma \cite{Tracy:1995ax} is applicable.
Then the power of $\rho_\pm$ is determined by 
a series of functions $\phi^\pm_l(y)$
\begin{\eqa}
&&\rho_\pm^{2n+1} (x,y) 
=\frac{E_\pm (x)E_\pm (y)}{\cosh{\frac{x}{k}} +\cosh{\frac{y}{k}} } 
  \sum_{l=0}^{2n} (-1)^l \phi_l^\pm (x) \phi_{2n-l}^\pm (y) ,\NN\\
&&\rho_\pm^{2n} (x,y) 
=\frac{E_\pm (x)E_\pm (y)}{\cosh{\frac{x}{k}} -\cosh{\frac{y}{k}} } 
  \sum_{l=0}^{2n-1} (-1)^l \phi_l^\pm (x) \phi_{2n-1-l}^\pm (y) ,\NN\\
&&\phi_l^\pm (x) = \frac{1}{E_\pm (x)} \int_{-\infty}^\infty \frac{dq}{2\pi k} \rho_\pm^l (x,q) E_\pm (q) ,\quad \phi_0^\pm (x)=1.
\end{\eqa}
At this stage, it seems that 
we should find the two series of the functions $\phi_l^+(y)$ and $\phi_l^-(y)$ to obtain the grand potential.
However, similar to the ABJM case \cite{Hatsuda:2012dt}, one can show the following convenient identity,
\begin{\eq}
\frac{\Xi_- (z) }{\Xi_+ (-z) } = \sum_{l=0}^\infty \phi_l^+ (0) z^l ,
\end{\eq}
which implies that we need information of only $\phi_l^+(y)$.
Plugging this into \eqref{eq:pm}, we arrive at \eqref{eq:grandpart}.
In the rest of this section, 
we present a concrete algorithm to compute $\phi_l^+(y)$ and ${\rm Tr}\rho_+^{2n}$ exactly.
\subsection{Algorithm for even Chern-Simons level}
By changing variables as $u=e^{\frac{y}{k}},v=e^{\frac{y^\prime}{k}}$, 
the recursion equation \eqref{eq:TBA} becomes 
\begin{\eqa}
\phi_{l+1}^+ (u) 
&=& \frac{1}{2\pi} \frac{u}{u+1} \int_0^\infty dv \frac{v^{\frac{k}{2}-1}}{u+v} \frac{v+1}{v^k +(-1)^M} 
  \Biggl[ \prod_{s=-\frac{M-1}{2}}^{\frac{M-1}{2}}  \frac{ v -e^{-\frac{2\pi s}{k}} } {v +e^{-\frac{2\pi s}{k}}  } \Biggr] \phi_l^+ (v ) .
\end{\eqa}
If we assume the form of $\phi_l^+ (u)$ as
\begin{\eq}
\phi_l^+ (u) =\sum_{j=0}^l A_l^{(j)} (u) (\log{u})^j ,
\label{eq:rationalexp}
\end{\eq}
with rational functions $A_l^{(j)}$, we find\footnote{
Here we have used an integral formula noted in \cite{Putrov:2012zi},
\[
\int_0^\infty dv C(v) \log^j v
= -\frac{(2\pi i)^j}{j+1} \oint_\gamma dv C(v) B_{j+1}\left( \frac{\log{v}}{2\pi i}\right) ,
\]
where $C(v)$ is a rational function and $B_{j+1}$ is the Bernoulli polynomial.
Choosing the branch cut of $\log{v}$ as the positive real axis,
the integral contour $\gamma$ goes from $+\infty$ to $0$ infinitesimally below the branch cut
and then to $+\infty$ infinitesimally above the branch cut.
}
\begin{\eqa}
\phi_{l+1}^+(u) 
&=& -\frac{1}{2\pi} \frac{u}{u+1} \sum_{j=0}^l \frac{(2\pi i)^{j+1}}{j+1} \NN\\
&&\sum_{\rm poles} {\rm Res}_v \Biggl[ A_l^{(j)}(v) B_{j+1}\left( \frac{\log{v}}{2\pi i} \right)
\frac{v^{\frac{k}{2}-1}}{u+v} \frac{v+1}{v^k +(-1)^M} 
   \prod_{s=-\frac{M-1}{2}}^{\frac{M-1}{2}}  \frac{ v -e^{-\frac{2\pi is}{k}} } {v +e^{-\frac{2\pi is}{k}}  } \Biggr] .\NN\\
\label{eq:recursion}
\end{\eqa}
The symbol $\sum_{\rm poles}$ indicates the summation over all the poles of the integrand.
From $\phi_l^+ (u)$, we can find $\rho_+^{2n} (u,u)$ as
\begin{\eq}
\rho_+^{2n} (u,u ) 
= \frac{1}{2} \frac{u^{\frac{k}{2}+1}(u+1)}{(u-1)(u^k +(-1)^M )}
\prod_{s=-\frac{M-1}{2}}^{\frac{M-1}{2}} \Biggl[ \frac{ u -e^{-\frac{2\pi is}{k}} } {u +e^{-\frac{2\pi is}{k}}  } \Biggr]     
   \sum_{l=0}^{2n-1} (-1)^l \frac{d\phi_l^+ (u)}{du} \phi_{2n-l-1}^+ (u)  .
\end{\eq}
By expanding $\rho_+^{2n} (u,u )/u $ in terms of the power of $\log u$
\begin{align}
 \frac{1}{u}\rho_+^{2n} (u,u )=\sum_{j=0}^{2n-1}R_n^{(j)}(u)(\log u)^j ,
\end{align}
where $R_n^{(j)}(u)$
is some rational function of $u$,
then the trace of $\rho_+^{2n}$ can be obtained as
a sum of residues
\begin{align}
 \Tr\rho_+^{2n}=-\frac{1}{2\pi} \sum_{j=0}^{2n-1} \frac{(2\pi i)^{j+1}}{j+1}
\sum_{\rm poles} {\rm Res}_u \Biggl[R_n^{(j)}(u) B_{j+1}\left( \frac{\log{u}}{2\pi i} \right)\Biggr].
\end{align}

\subsection{Algorithm for odd Chern-Simons level}
When $k$ is odd, by introducing the variables $u=e^{y/2k}, v=e^{y'/2k}$,
we can rewrite the recursion relation \eqref{eq:TBA} as 
\begin{\eqa}
\phi_{l+1}^+ (u) 
&=& \frac{1}{2\pi} \frac{u^2}{u^2+1} \int_0^\infty dv \frac{v^{k-1}}{u^2 +v^2} \frac{v^2 +1}{v^{2k} +(-1)^M} 
  \Biggl[ \prod_{s=-\frac{M-1}{2}}^{\frac{M-1}{2}}  \frac{ v^2 -e^{-\frac{2\pi is}{k}} } {v^2 +e^{-\frac{2\pi is}{k}}  } \Biggr] \phi_l^+ (v ) .
\end{\eqa}
By expanding $\phi_{l+1}^+$ as \eqref{eq:rationalexp},
we find
\begin{\eqa}
\phi_{l+1}^+(u) 
&=& -\frac{1}{\pi} \frac{u^2}{u^2+1} \sum_{j=0}^l \frac{(2\pi i)^{j+1}}{j+1} \NN\\
&&\sum_{\rm poles} {\rm Res}_v \Biggl[ A_l^{(j)}(v) B_{j+1}\left( \frac{\log{v}}{2\pi i} \right)
\frac{v^{k-1}}{u^2 +v^2} \frac{v^2 +1}{v^{2k} +(-1)^M} 
   \prod_{s=-\frac{M-1}{2}}^{\frac{M-1}{2}}  \frac{ v^2 -e^{-\frac{2\pi is}{k}} } {v^2 +e^{-\frac{2\pi is}{k}}  } \Biggr] .\NN\\
\end{\eqa}
Then $\rho_+^{2n} (u,u)$ is rewritten as
\begin{\eq}
\rho_+^{2n} (u,u ) 
= \frac{1}{4} \frac{u^{k+1}(u^2 +1)}{(u^2 -1)(u^{2k} +(-1)^M )}
\prod_{s=-\frac{M-1}{2}}^{\frac{M-1}{2}} \Biggl[ \frac{ u^2 -e^{-\frac{2\pi is}{k}} } {u^2 +e^{-\frac{2\pi is}{k}}  } \Biggr]     
   \sum_{l=0}^{2n-1} (-1)^l \frac{d\phi_l^+ (u)}{du} \phi_{2n-l-1}^+ (u)  .
\end{\eq}
In a similar way as the even $k$ case,
by expanding $\rho_+^{2n} (u,u )/u $ in terms of the power of $\log u$
\begin{align}
 \frac{1}{u}\rho_+^{2n} (u,u )=\sum_{j=0}^{2n-1}R_n^{(j)}(u)(\log u)^j ,
\end{align}
the trace of $\rho_+^{2n}$ can be obtained as
a sum of residues
\begin{align}
 \Tr\rho_+^{2n}=-\frac{1}{\pi} \sum_{j=0}^{2n-1} \frac{(2\pi i)^{j+1}}{j+1}
\sum_{\rm poles} {\rm Res}_u \Biggl[R_n^{(j)}(u) B_{j+1}\left( \frac{\log{u}}{2\pi i} \right)\Biggr] .
\end{align}
\section{$\pi^{-N}$ term of $\h{Z}^{(N,N+M)}(k)$}
In this appendix, we will briefly discuss the structure
of the highest transcendental term
of the partition functions of the ABJ theory listed in appendix D, and the partition functions
of the ABJM theory in \cite{Hatsuda:2012dt}.
As mentioned in \cite{Hatsuda:2012dt} for the $k=1$ ABJM theory,
the coefficient of the highest transcendental term $\pi^{-N}$ 
in the canonical partition function can be expressed
in terms of the Hermite polynomial using the relation
\begin{align}
 e^{2xz-z^2}=\sum_{n=0}^\infty \frac{H_n(x)}{n!}z^n .
\end{align}
We will see that the partition functions of ABJ theory  
also have a similar structure.

\subsection{$(k,M)=(2,1)$}
The partition function of $(k,M)=(2,1)$ case has the structure
\begin{align}
 \h{Z}^{(N,N+1)}(2)=\sum_{m=0}^N\frac{c_m^{(N)}}{\pi^m} .
\end{align}
Let us consider the highest transcendental term $\pi^{-N}$ in $\h{Z}^{(N,N+1)}(2)$.
We observe that such highest term of $\phi_l(u=1)$ comes from the $(\log u)^l$ term in the expansion
\eqref{eq:rationalexp}
\begin{align}
 \phi_l^+(u)&=\varepsilon_l\frac{\phi_1^+(u)^l}{l!}+\cdots,\qquad
\varepsilon_l=\rt{2}\sin\left(\frac{\pi(2l+1)}{4}\right)=\pm 1,\nn
\phi_l^+(1)_{\textrm{highest}}&=\varepsilon_l\frac{\phi_1^+(1)^l}{l!}
=\frac{\varepsilon_l}{(4\pi)^ll!},
\end{align}
where
\begin{align}
 \phi_1^+(u)=\frac{u}{2\pi(u^2-1)}\log u .
\end{align}
$\Tr\rho_+^{2m}$ with $m\geq2$ is always less transcendental compared to $\pi^{-2m}$,
and the only highest transcendental term comes from $\Tr\rho_+^2=\frac{1}{32\pi^2}$.
Thus we find that the generating function of the highest term $\pi^{-N}$ is given by
\begin{align}
 \sum_{N=0}^\infty z^N\frac{c_N^{(N)}}{\pi^N}&=e^{-\Tr\rho_+^2z^2}\left(\sum_{l=0}^\infty \phi_l^+(1)_{\textrm{highest}}z^l\right)\nn
&=e^{-\frac{z^2}{32\pi^2}}\left(\cos\frac{z}{4\pi}
+\sin\frac{z}{4\pi}\right)\nn
&=1+\frac{1}{4\pi}z-\frac{1}{16\pi^2}z^2-\frac{1}{96\pi^3}z^3+\frac{5}{3072\pi^4}z^4+\cdots.
\end{align} 
We further find
\begin{align}
\sum_{m=1}^\infty\frac{\Tr\rho_+^{2m}}{m}z^{2m}=\frac{z^2}{2^5\pi^2}+\frac{z^4}{2^93\pi^2}
+\frac{z^6}{2^{10}3^35\pi^2}+\frac{z^6}{2^{16}\pi^2}
-\frac{z^4}{2^{15}}\left(1+\cos\frac{z}{\pi}\right)+\cdots.
\end{align}
\subsection{$(k,M)=(4,1)$}
In a similar manner, the highest transcendental
term for the $(k,M)=(4,1)$ case is found to be
\begin{align}
 \sum_{N=0}^\infty z^N\frac{c_N^{(N)}}{\pi^N}&=e^{-\Tr(\rho_+^2)_{\textrm{highest}}z^2}\left(\sum_{l=0}^\infty \phi_l^+(1)_{\textrm{highest}}z^l\right)\nn
&=\exp\left(\frac{z^2}{64\pi^2}-\frac{z}{8\pi}\right)\nn
&=1-\frac{1}{8\pi}z+\frac{3}{128\pi^2}z^2-\frac{7}{3072\pi^3}z^3+\frac{25}{98304\pi^4}z^4+\cdots.
\end{align} 
The next-to-highest term is
\begin{align}
 \sum_{N=1}^\infty z^N\frac{c_{N-1}^{(N)}}{\pi^{N-1}}=\frac{z}{16}\exp\left(\frac{z^2}{64\pi^2}+\frac{3z}{8\pi}\right) .
\end{align}
Therefore, we find
\begin{align}
 \Xi(z)=e^{\frac{z^2}{64\pi^2}-\frac{z}{8\pi}}\left[1+\frac{z}{16}e^{\frac{z}{2\pi}}+\cdots\right].
\end{align}

\subsection{Summary of the Higher Transcendental Part of Grand Partition Functions}
By inspecting the exact partition functions, we find a similar structure for other cases.
Here we summarize the higher transcendental part of grand partition functions $\Xi^{(M)}_k(z)$
for various $k$ and $M$,
\begin{align}
 \Xi^{(0)}_1(z)&=\exp\left[\frac{z^4}{2^7\pi^2}+\frac{z^2}{2^4\pi}\cdots\right]\cdot\left[1+\frac{z}{4}e^{-\frac{z^2}{4\pi}}+
\frac{z^3}{64}e^{\frac{z^2}{4\pi}}\cdots\right]\nn
\Xi^{(0)}_3(z)&=\exp\left[\frac{z^4}{3\cdot2^7\pi^2}-\frac{z^2}{2^4\pi}+\cdots\right]\cdot\left[1+\frac{z}{12}e^{\frac{z^2}{12\pi}}+\cdots\right]\nn
 \Xi^{(0)}_2(z)&=\exp\left[\frac{z^2}{2\cdot2^4\pi^2}-\frac{z^4}{2^9\cdot3\pi^2}
+\frac{z^6}{2^{10}3^35\pi^2}+\frac{z^6}{2^{16}\pi^2}+\frac{z^4}{2^{14}}\cosh^2\frac{z}{2\pi}+\cdots\right] \nn
&~~~~~~~ \cdot\left[1+\frac{z}{8}\cosh\frac{z}{2\pi}+\cdots\right]\nn
\Xi^{(1)}_2(z)&=\exp\left[-\frac{z^2}{2\cdot2^4\pi^2}-\frac{z^4}{2^9\cdot3\pi^2}
-\frac{z^6}{2^{10}3^35\pi^2}-\frac{z^6}{2^{16}\pi^2}+\frac{z^4}{2^{14}}\cos^2\frac{z}{2\pi}+\cdots\right]\nn
&~~~~~~~ \cdot\left[\cos\frac{z}{4\pi}
+\sin\frac{z}{4\pi}\cdots\right]\nn
\Xi^{(0)}_4(z)&=\exp\left[-\frac{z^2}{4\cdot2^4\pi^2}+\cdots\right]\cdot\left[1+\frac{z}{16}-\frac{z^2}{128}\sin\frac{z}{2\pi}+\cdots\right]\nn
\Xi^{(1)}_4(z)&=\exp\left[\frac{z^2}{4\cdot2^4\pi^2}-\frac{z}{8\pi}+\cdots\right]\cdot\left[1+\frac{z}{16}e^{\frac{z}{2\pi}}+\cdots\right]\nn
\Xi^{(0)}_6(z)&=\exp\left[\frac{z^2}{6\cdot2^4\pi^2}+\cdots\right]\cdot\left[1+\frac{z}{24}\cosh\frac{z}{6\pi}+\frac{\rt{3}z^2}{2\cdot3^4}\sinh\frac{z}{6\pi}+\cdots\right]\nn
\Xi^{(1)}_6(z)&=\exp\left[-\frac{z^2}{6\cdot2^4\pi^2}+\cdots\right]\cdot\left[\cos\frac{z}{12\pi}+\sin\frac{z}{12\pi}+\cdots\right] .
\end{align}
We do not have a clear understanding of the origin of this 
structure. We leave this for a future work.
\section{Picard-Fuchs equation and effective K\"ahler parameters}
In this appendix, we will explain that the effective K\"{a}hler parameters
for integer $k$ are given by \eqref{eq:Teff}, which are essentially equal to
the classical periods
of diagonal local $\mathbb{P}^1\times\mathbb{P}^1$.

The Picard-Fuchs equation of local $\mathbb{P}^1\times\mathbb{P}^1$ for the classical
periods $\Pi$ is 
\begin{align}
 \th_I^2\Pi=z_I(2\th_1+2\th_2)(2\th_1+2\th_2+1)\Pi,\qquad(I=1,2)
\label{PFeq}
\end{align}
with
\begin{align}
 \th_I=z_I\frac{\del}{\del z_I}.
\end{align}
Note that $\th_I$ measures the exponent of $z^n_I$
\begin{align}
 \th_Iz_I^n=nz_I^n.
\end{align}
The classical A-period is a solution of the Picard-Fuchs equation
\eqref{PFeq}
\begin{align}
\Pi_{A_I}=\log z_I+ 2\sum_{k,l\geq0}\frac{(2k+2l-1)!}{(k!l!)^2}z_1^kz_2^l.
\end{align}
For the diagonal case $z_1=z_2=z$, this reduces to
\begin{align}
 \Pi_A=\log z+4z\sum_{n=0}^\infty \frac{(2n+1)!^2}{(n+1)!^3}\frac{z^n}{n!}=
\log z+4z{}_4F_3\left(1,1,\frac{3}{2},\frac{3}{2},2,2,2;16z
\right),
\label{clasicalP}
\end{align}
where we have used the identity
\begin{align}
 \sum_{m=0}^n\begin{pmatrix}n\\m \end{pmatrix}^2=\begin{pmatrix}2n\\n \end{pmatrix}.
\end{align}

As discussed in \cite{Hatsuda:2012dt},
the effective K\"{a}hler parameter is given by the quantum A-period.
For even $k$, 
the quantum parameter $q$
is equal to the classical value $q=e^{i\pi k}=1$,
and the instanton factors for the D2-brane
become ``diagonal''
\begin{align}
 z=e^{-T_{1,2}/g_s}=(-1)^{\frac{k}{2}-M}e^{-2\mu},
\end{align}
although the worldsheet instanton factors are non-diagonal
$e^{-T_1}\not=e^{-T_2}$.
Therefore, the quantum A-period is 
reduced to the classical period \eqref{clasicalP}.
In \cite{Hatsuda:2013gj}, it is conjectured for the ABJM case 
that the quantum period for 
the odd $k$ case, corresponding to
$q=-1$, is also expressed in terms of the hypergeometric function,
obtained by simply replacing $z\to z^2$ in \eqref{clasicalP}.
We expect that this is also the case for the ABJ theory.

\section{Exact results on ABJ partition function}
\label{app:values}
In this section,
we present exact results of the ABJ partition function for various $k$, $M$ and $N$.
Although we have indeed obtained the exact values 
for $(k,M, N_{\rm max} )=(2,1,65 )$, $(3,1,31 )$, $(4,1,62 )$, $(4,2, 29 )$, $(6,1, 23 )$, 
$(6,2, 21 )$ and $(6,3,22 )$,
we list these values up to $N=20$ at most.
This is because the expressions for larger $N$ are too long.
The other values are available upon request to the present authors.
Our values are consistent with the result obtained in the work \cite{Matsumoto:2013nya} by a different approach,
which has calculated the ABJ partition function exactly or numerically with high precision
for $(k,M, N_{\rm max} )=(2,1,4)$, $(3,1,3)$, $(4,1,2)$, $(4,2, 2)$, $(6,1, 3 )$,  $(6,2, 2 )$ and $(6,3,2 )$.

\begin{figure}[p]
 \rotatebox{90}{
 \begin{minipage}{1.36\linewidth}
\begin{\eqa}
&&\scriptscriptstyle \hat{Z}^{(1,2)}(2) = \frac{1}{4\pi} ,\quad
\hat{Z}^{(2,3)}(2) = \frac{1}{128}-\frac{1}{16 \pi ^2} ,\quad
\hat{Z}^{(3,4)}(2) = \frac{5 \pi ^2-48}{4608 \pi ^3} ,\quad
\hat{Z}^{(4,5)}(2) = \frac{9}{32768}+\frac{5}{3072 \pi ^4}-\frac{53}{18432 \pi ^2} ,\quad
 \hat{Z}^{(5,6)}(2) = \frac{6240-800 \pi ^2+17 \pi ^4}{29491200 \pi ^5},\NN\\
&&\scriptscriptstyle  \hat{Z}^{(6,7)}(2) = \frac{-218880+1413600 \pi ^2-1160264 \pi ^4+103275 \pi ^6}{8493465600 \pi ^6},\quad
  \hat{Z}^{(7,8)}(2) = \frac{-4677120-8631840 \pi ^2+14206864 \pi ^4-1345977 \pi ^6}{1664719257600 \pi ^7},\NN\\
&&\scriptscriptstyle \hat{Z}^{(8,9)}(2) = 
\frac{61608960-1051438080 \pi ^2+2363612608 \pi ^4-1477376224 \pi ^6+126511875 \pi ^8}{213084064972800 \pi ^8},\quad
 \hat{Z}^{(9,10)}(2) = 
\frac{633830400+6140897280 \pi ^2-22473501120 \pi ^4+16465544384 \pi ^6-1444050207 \pi ^8}{23013079017062400 \pi ^9},\NN\\
&&\scriptscriptstyle \hat{Z}^{(10,11)}(2) = 
\frac{-45945446400+1789603200000 \pi ^2-8040102013440 \pi ^4+12612559256000 \pi ^6
 -6759848237256 \pi ^8+563610403125 \pi ^{10}}{18410463213649920000 \pi ^{10}},\NN\\
&&\scriptscriptstyle \hat{Z}^{(11,12)}(2) = 
-\frac{1899826790400+58020665472000 \pi ^2-380149120834560 \pi ^4+713711875201600 \pi ^6
    -419001890609104 \pi ^8+35516095421175 \pi ^{10}}{8910664195406561280000 \pi ^{11}}
,\NN\\
&&\scriptscriptstyle \hat{Z}^{(12,13)}(2) = 
   \frac{1}{5132542576554179297280000 \pi ^{12}} 
\Bigl[
 89510709657600-7305762207744000 \pi ^2+55891102669839360 \pi ^4 
 -165671005272960000 \pi ^6  +212595597816249824 \pi ^8 \NN\\
&&\scriptscriptstyle -102685378935079440 \pi ^{10}+8388213629709375 \pi^{12}
\Bigr] ,\NN\\
&&\scriptscriptstyle \hat{Z}^{(13,14)}(2) = 
   \frac{1}{3469598781750625204961280000 \pi ^{13}}
\Bigl[
 4720114977177600+362547424176537600 \pi ^2  -3751225073125770240 \pi ^4  +13042072681841448960 \pi ^6 \NN\\
&&\scriptscriptstyle -19012094772801329056 \pi ^8  +9993495094908215904 \pi ^{10}-830549319123470625\pi ^{12}  \Bigr] 
,\NN\\
&& \scriptscriptstyle \hat{Z}^{(14,15)}(2) = 
   \frac{1}{5440330889784980321379287040000 \pi ^{14}}
\Bigl[
-555727897460736000 +89573494835323699200 \pi ^2     -1055636150467356057600 \pi ^4  +5131488836828022789120 \pi ^6 \NN\\
&&\scriptscriptstyle -12078328057432325328640 \pi ^8 +13537831707363614586208 \pi^{10} 
   -6051892803562043641080 \pi ^{12}+486239579473363340625 \pi ^{14} \Bigr] ,\NN\\
&&\scriptscriptstyle \hat{Z}^{(15,16)}(2) = 
  - \frac{1}{4896297800806482289241358336000000 \pi ^{15}}
\Bigl[
 36090194527715328000 +6104583949671567360000 \pi ^2  -92067509353118319820800 \pi ^4 \NN\\
&&\scriptscriptstyle  +507831737592928484736000 \pi ^6 -1344043476982266371351040 \pi ^8 +1708199914796799315018400\pi ^{10} 
  -841038818134977117865584 \pi ^{12}+69024176701151867566875 \pi ^{14} \Bigr] 
,\NN\\
&&\scriptscriptstyle \hat{Z}^{(16,17)}(2) = 
   \frac{1}{1253452237006459466045787734016000000 \pi ^{16}}
\Bigl[
 644515885825523712000-195903374317541130240000 \pi ^2  +3317788425511538166988800 \pi ^4 
 -24242575894767562235904000 \pi ^6 \NN\\
&&\scriptscriptstyle +91686579377609424295127040 \pi^8 
 -184621497276384941161625600 \pi ^{10}   +187135567910967396538249344 \pi ^{12} 
   -78705401521585216044052800 \pi ^{14}+6236022606745884843515625 \pi^{16} \Bigr] 
,\NN
\end{\eqa}
 \end{minipage}
 } 
 \end{figure}

\clearpage
\begin{figure}[p]
 \rotatebox{90}{
 \begin{minipage}{1.36\linewidth}
\begin{\eqa}
&&\scriptscriptstyle \hat{Z}^{(17,18)}(2) =
   \frac{1}{1448990785979467142748930620522496000000 \pi ^{17}}
\Bigl[
50222901705188179968000 +17114872531857226334208000 \pi ^2 -355825180591455246748876800 \pi ^4 \NN\\
&&\scriptscriptstyle +2838673897897708635616051200 \pi ^6
   -11715836542518334641324349440 \pi^8  +26594007390595358524134338560 \pi ^{10} \NN\\
&&\scriptscriptstyle -31088486157208526910587238784 \pi ^{12} 
  +14680941405810341458359816576 \pi ^{14}-1194793767361309903416444375 \pi^{16} \Bigr] ,\NN \\
&&\scriptscriptstyle \hat{Z}^{(18,19)}(2) = 
   \frac{1}{1251928039086259611335076056131436544000000 \pi ^{18}} 
 \Bigl[
-2837855912505174392832000  +1565466573304371781435392000 \pi ^2  
 -36201047925887447842868428800 \pi^4 \NN\\
&&\scriptscriptstyle +374743421357886210747698380800 \pi^6 
    -2099946681695866974987064688640 \pi^8 
 +6688454172020401470415744112640 \pi ^{10}-11982818897222541532284369726464 \pi ^{12} \NN\\
&&  \scriptscriptstyle +11220643955054903542467568447104 \pi ^{14}-4489098718626188671320477135000 \pi ^{16} 
 +351431054003164340356323046875 \pi ^{18}
\Bigr]  
,\NN\\
&&\scriptscriptstyle \hat{Z}^{(19,20)}(2) = 
  - \frac{1}{1807784088440558878767849825053794369536000000 \pi ^{19}}
 \Bigl[
260034050935690604052480000   +167378576740920004904091648000 \pi ^2 -4603213941146778919710228480000 \pi ^4 \NN\\
&&\scriptscriptstyle   +49864936569429230001889571635200 \pi^6  -292626274613554624545116349235200 \pi ^8 
  +1022337025900122231369611246684160 \pi ^{10} -2112649945836780855818878981703680 \pi ^{12} \NN\\
&&\scriptscriptstyle   +2341691134873926453650025102600576 \pi^{14} -1075830030189292612090801154991984 \pi ^{16} 
 +87057436298005995587368943405625 \pi ^{18}
 \Bigr] , \NN\\
&&\scriptscriptstyle \hat{Z}^{(20,21)}(2) = 
   \frac{1}{2892454541504894206028559720086070991257600000000 \pi ^{20}}
 \Bigl[
25690000707001171197296640000 
 -24885517234201682474749132800000 \pi ^2+755668170954645465216072941568000 \pi ^4 \NN\\
&&\scriptscriptstyle  -10621602174332426380613505515520000 \pi^6  +83475984203142035463930152647065600 \pi ^8 
 -388930780899716024500536537133056000 \pi ^{10}+1087554872133572209767569467463813120 \pi ^{12} \NN\\
&&\scriptscriptstyle -1775932910388692220449035532375705600\pi ^{14}+1558438899830276774076529628858407584 \pi ^{16} 
 -597787290215170549303861405923030000 \pi ^{18}+46306312830726949307050906271484375 \pi^{20}
\Bigr] . \NN
\end{\eqa}
 \end{minipage}
 } 
 \end{figure}

\clearpage
\begin{figure}[p]
 \rotatebox{90}{
 \begin{minipage}{1.36\linewidth}
\begin{\eqa *}
&&\scriptscriptstyle \hat{Z}^{(1,2)}(3) = \frac{1}{12} \left(2 \sqrt{3}-3\right) ,\quad
\hat{Z}^{(2,3)}(3) = \frac{1}{432} \left(-27+14 \sqrt{3}+\frac{9}{\pi }\right) ,\quad
\hat{Z}^{(3,4)}(3) =-\frac{45+18 \sqrt{3}-14 \sqrt{3} \pi }{1728 \pi } ,\quad
\hat{Z}^{(4,5)}(3) = \frac{702+84 \left(27+2 \sqrt{3}\right) \pi +\left(1152 \sqrt{3}-2881\right) \pi ^2}{248832 \pi ^2}, \\
&&\scriptscriptstyle  \hat{Z}^{(5,6)}(3)
=\frac{54 \left(14 \sqrt{3}-37\right)+840 \sqrt{3} \pi +\left(5797-3574 \sqrt{3}\right) \pi ^2}{995328 \pi ^2} ,\quad
\hat{Z}^{(6,7)}(3)
= \frac{17982+162 \left(182 \sqrt{3}-1647\right) \pi +27 \left(2304 \sqrt{3}-5905\right) \pi ^2-7 \left(42110 \sqrt{3}-78327\right) \pi ^3}{322486272 \pi ^3} ,\\
&&\scriptscriptstyle  \hat{Z}^{(7,8)}(3)
=\frac{-2430 \left(61+18 \sqrt{3}\right)+83916 \sqrt{3} \pi +27 \left(28553+16770 \sqrt{3}\right) \pi ^2+\left(78732-335594 \sqrt{3}\right) \pi ^3}{1289945088 \pi
   ^3} ,\\\
&&\scriptscriptstyle  \hat{Z}^{(8,9)}(3) 
=\frac{1472580+9072 \left(2295+74 \sqrt{3}\right) \pi +324 \left(91008 \sqrt{3}-105709\right) \pi ^2-168 \left(793233+60254 \sqrt{3}\right) \pi
   ^3+\left(178071703-76499136 \sqrt{3}\right) \pi ^4}{371504185344 \pi ^4} ,\\
&&\scriptscriptstyle  \hat{Z}^{(9,10)}(3) 
=\frac{2916 \left(774 \sqrt{3}-2857\right)+5533920 \sqrt{3} \pi -324 \left(152814 \sqrt{3}-521209\right) \pi ^2-48 \left(1483097 \sqrt{3}-511758\right) \pi
   ^3+\left(226863738 \sqrt{3}-370195279\right) \pi ^4}{1486016741376 \pi ^4} ,\\
&&\scriptscriptstyle  \hat{Z}^{(10,11)}(3)
=\frac{299312820+72900 \left(7070 \sqrt{3}-171747\right) \pi +24300 \left(382464 \sqrt{3}-411157\right) \pi ^2-2700 \left(6698762 \sqrt{3}-62488989\right) \pi ^3-9
   \left(5065099200 \sqrt{3}-9212744479\right) \pi ^4+25 \left(6475592722 \sqrt{3}-11826421389\right) \pi ^5}{4012245201715200 \pi ^5}  ,\\
&&\scriptscriptstyle  \hat{Z}^{(11,12)}(3)
=\frac{-131220 \left(28925+6642 \sqrt{3}\right)+2915854200 \sqrt{3} \pi +24300 \left(2404421+1236762 \sqrt{3}\right) \pi ^2-5400 \left(8730505
   \sqrt{3}-7125246\right) \pi ^3-9 \left(40242657395+27610808958 \sqrt{3}\right) \pi ^4+50 \left(3671699105 \sqrt{3}-1298211948\right) \pi ^5}{16048980806860800\pi ^5} ,\\
&&\scriptscriptstyle  \hat{Z}^{(12,13)}(3)
=\frac{1}{6933159708563865600 \pi ^6} \Bigl[
25651672920+3674160 \left(281907+4562 \sqrt{3}\right) \pi +218700 \left(11516544 \sqrt{3}-6754681\right) \pi ^2
  -226800 \left(106545861+3385430 \sqrt{3}\right)\pi ^3 -162 \left(232246756800 \sqrt{3}-175868541043\right) \pi ^4 \\
&&\scriptscriptstyle +36 \left(3302763448131+214002197506 \sqrt{3}\right) \pi ^5+25 \left(2873091390912
   \sqrt{3}-6479908382207\right) \pi ^6 
\Bigr] ,\\
&&\scriptscriptstyle  \hat{Z}^{(13,14)}(3)
=\frac{1}{27732638834255462400 \pi ^6} 
\Bigl[ 787320 \left(66366 \sqrt{3}-316045\right)+212550156000 \sqrt{3} \pi -218700 \left(12640806 \sqrt{3}-69160621\right) \pi ^2
 -64800 \left(147495509 \sqrt{3}-104910390\right) \pi ^3 \\
&&\scriptscriptstyle +162 \left(296550133938 \sqrt{3}-1076243018035\right) \pi ^4 
 +360 \left(205884495833 \sqrt{3}-99859338540\right) \pi ^5-175
   \left(1189802574054 \sqrt{3}-1946635606421\right) \pi ^6 
\Bigr] ,\\
&&\scriptscriptstyle  \hat{Z}^{(14,15)}(3)
= \frac{1}{146761124710879907020800 \pi^7}
 \Bigl[ 9762153103080+38578680 \left(456134 \sqrt{3}-24118263\right) \pi +19289340 \left(70712064 \sqrt{3}-37224121\right) \pi ^2
  -3572100 \left(371449442  \sqrt{3}-9412651737\right) \pi ^3 \\
&&\scriptscriptstyle -23814 \left(1164740241600 \sqrt{3}-760466097211\right) \pi ^4 
   +7938 \left(4920941754202 \sqrt{3}-40985286126129\right) \pi ^5
 +9 \left(10577414304217152 \sqrt{3}-16694829965655623\right) \pi ^6-1225 \left(241451318806186 \sqrt{3}-436239059157621\right) \pi ^7 
   \Bigr],\\
&&\scriptscriptstyle  \hat{Z}^{(15,16)}(3)
=\frac{1}{587044498843519628083200 \pi^7}
\Bigl[ -49601160 \left(3662825+677322 \sqrt{3}\right) +170696384888400 \sqrt{3} \pi +19289340 \left(210402593+126703602 \sqrt{3}\right) \pi^2
  -7144200 \left(729601789 \sqrt{3}-1620343926\right) \pi ^3  \\
&&\scriptscriptstyle -23814 \left(2803076569895+2326367522214 \sqrt{3}\right) \pi ^4
 +79380 \left(992950495129 \sqrt{3}-1601735224140\right) \pi ^5+9
   \left(65702735219040679+49510164883503726 \sqrt{3}\right) \pi ^6-2450 \left(134862880599065 \sqrt{3}-58222941612138\right) \pi ^7
   \Bigr] ,
\end{\eqa *}
 \end{minipage}
 } 
 \end{figure}
 
\clearpage
\begin{figure}[p]
 \rotatebox{90}{
 \begin{minipage}{1.36\linewidth}
\begin{\eqa *}
&&\scriptscriptstyle  \hat{Z}^{(16,17)}(3)
= \frac{1}{338137631333867305775923200 \pi ^8}
\Bigl[ 873064837174320+1851776640 \left(44415567+393626 \sqrt{3}\right) \pi 
 +154314720 \left(1916774784 \sqrt{3}-606477925\right) \pi ^2 -240045120\left(17798150169+279968270 \sqrt{3}\right) \pi ^3 \\
&&\scriptscriptstyle -285768 \left(39244249540800 \sqrt{3}-13493732619847\right) \pi ^4
 +127008 \left(536988724735743+6895335985354   \sqrt{3}\right) \pi ^5 
+216 \left(518491284724453824 \sqrt{3}-326463576135631595\right) \pi ^6 \\
&&\scriptscriptstyle -336 \left(893382789200775309+46813383823326550 \sqrt{3}\right) \pi^7
-1225 \left(152389841509146240 \sqrt{3}-337119938695893131\right) \pi ^8 
   \Bigr] ,\\
&&\scriptscriptstyle \hat{Z}^{(17,18)}(3) 
= \frac{1}{1352550525335469223103692800 \pi ^8}
\Bigl[ 1488034800 \left( 1521342 \sqrt{3}-8972381\right) +13565467542816000 \sqrt{3} \pi
   -154314720 \left(1439560350 \sqrt{3}-11885783617\right) \pi ^2 \\
&&\scriptscriptstyle -68584320 \left(20996707697 \sqrt{3}-26100169758\right) \pi ^3 
 +1428840 \left(6210164253042 \sqrt{3}-37135089788099\right) \pi ^4+1270080 \left(26266226178293 
   \sqrt{3}-26656967579580\right) \pi ^5 \\
&&\scriptscriptstyle -216 \left(598709018914049250 \sqrt{3}-2213891956469549207\right) \pi ^6 
 -672 \left(304318107429747331
   \sqrt{3}-172874144162115774\right) \pi ^7+1225 \left(435235968952169778 \sqrt{3}-713337768334931003\right) \pi ^8
   \Bigr] ,\\
&&\scriptscriptstyle \hat{Z}^{(18,19)}(3) 
= \frac{1}{1314679110626076084856789401600 \pi ^9}
\Bigl[ 58464585220935600+18749238480 \left(5867234 \sqrt{3}-600328125\right) \pi 
 +1785641760 \left(15138920448 \sqrt{3}-4236518269\right) \pi ^2\\
&&\scriptscriptstyle -462944160\left(31133608058 \sqrt{3}-1722444578037\right) \pi ^3 
 -115736040 \left(10989313693632 \sqrt{3}-3280921971535\right) \pi ^4
  +2571912 \left(455517813319858 \sqrt{3}-7391659237811325\right) \pi ^5 \\
&&\scriptscriptstyle +5832 \left(2873264941254076992 \sqrt{3}-1573633144207062371\right) \pi ^6 
 -13608 \left(1569723702002837758 \sqrt{3}-11686691616115566447\right) \pi ^7 \\
&&\scriptscriptstyle  -81 \left(605022932882421984384 \sqrt{3}-886168883247494751971\right) \pi ^8
+1225 \left(114549503631688611466  \sqrt{3}-205470512408562766641\right) \pi ^9
   \Bigr] ,\\
&&\scriptscriptstyle \hat{Z}^{(19,20)}(3)
= \frac{1}{5258716442504304339427157606400 \pi ^9} 
\Bigl[ -2678462640 \left(575940725+87843042 \sqrt{3}\right) +1682253111024208800 \sqrt{3} \pi
   +1785641760 \left(9018837149+16522525962 \sqrt{3}\right) \pi^2 \\
&&\scriptscriptstyle -925888320 \left(64084310353 \sqrt{3}-436522842414\right) \pi ^3 
 -949035528 \left(1252129132998 \sqrt{3}-302840637145\right) \pi ^4
 +25719120 \left(63997148204653 \sqrt{3}-458742838215420\right) \pi ^5 \\
&&\scriptscriptstyle  +5832 \left(2814188682934565611+4284440486643065238 \sqrt{3}\right) \pi ^6 
  -27216 \left(1228646175500522483 \sqrt{3}-2983446322924960494\right) \pi ^7 
 -81 \left(3174112050264905901151+2556850825191662907462 \sqrt{3}\right) \pi ^8 \\
&&\scriptscriptstyle +2450 \left(63005382101415209525 \sqrt{3}-30828591348338574072\right) \pi ^9
   \Bigr] ,\\
&&\scriptscriptstyle \hat{Z}^{(20,21)}(3)
=\frac{1}{18931379193015495621937767383040000 \pi ^{10}}
\Bigl[ 27212508208631239200 +624974616000 \left(8466191307+43655330 \sqrt{3}\right) \pi
  +66961566000 \left(390117511296 \sqrt{3}-69963418417\right) \pi^2 \\
&&\scriptscriptstyle -138883248000 \left(3538375943949+30304795526 \sqrt{3}\right) \pi ^3
 -231472080 \left(8146846411531200 \sqrt{3}-1434108747628219\right) \pi ^4 
-771573600 \left(283388114263510 \sqrt{3}-21213224962217631\right) \pi ^5 \\
&&\scriptscriptstyle +437400 \left(103210691255334516672 \sqrt{3}-29179474844566192319\right) \pi ^6+1360800
   \left(1852787727580925522 \sqrt{3}-161730054824069504817\right) \pi ^7 \\
&&\scriptscriptstyle -486 \left(770013459328381479964800 \sqrt{3}-439771266544674399965423\right) \pi ^8 
 +2100  \left(432570801010943237965497+17596866511600094228950 \sqrt{3}\right) \pi ^9 \\
&&\scriptscriptstyle +214375 \left(2686006964880752857344 \sqrt{3}-5861724703722040120369\right) \pi^{10} 
   \Bigr] .
\end{\eqa *}
 \end{minipage}
 } 
 \end{figure}

\clearpage
\begin{figure}[p]
 \rotatebox{90}{
 \begin{minipage}{1.36\linewidth}
\begin{\eqa *}
&&\scriptscriptstyle \hat{Z}^{(1,2)}(4) =\frac{\pi -2}{16 \pi } ,\quad
 \hat{Z}^{(2,3)}(4) = \frac{12+12 \pi -5 \pi ^2}{512 \pi ^2} ,\quad
 \hat{Z}^{(3,4)}(4) = \frac{-168+396 \pi +202 \pi ^2-99 \pi ^3}{73728 \pi ^3} ,\quad
  \hat{Z}^{(4,5)}(4) = \frac{1200+4320 \pi -3512 \pi ^2-4872 \pi ^3+1755 \pi ^4}{4718592 \pi ^4} ,\\
&&\scriptscriptstyle \hat{Z}^{(5,6)}(4) 
= \frac{-38880+241200 \pi +186000 \pi ^2-400200 \pi ^3-203494 \pi ^4+96975 \pi ^5}{1887436800 \pi ^5} ,\quad
\hat{Z}^{(6,7)}(4)
= \frac{953280+8320320 \pi -7378800 \pi ^2-36784800 \pi ^3+17373764 \pi ^4+27667476 \pi ^5-9333225 \pi ^6}{543581798400 \pi ^6} ,\\ 
&&\scriptscriptstyle \hat{Z}^{(7,8)}(4)
= \frac{-52536960+691346880 \pi +566479200 \pi ^2-2914304400 \pi ^3-2014346488 \pi ^4+3962357364 \pi ^5+2156964930 \pi ^6-995722875 \pi ^7}{426168129945600 \pi ^7} ,\\ 
&&\scriptscriptstyle \hat{Z}^{(8,9)}(4)
= \frac{478759680+8468167680 \pi -7157041920 \pi ^2-89293397760 \pi ^3+38961966624 \pi ^4+232256453184 \pi ^5-82822457776 \pi ^6-145218219408 \pi ^7+47021834475 \pi
   ^8}{54549520633036800 \pi ^8} ,\\
&&\scriptscriptstyle \hat{Z}^{(9,10)}(4)
=  \frac{1}{23565392913471897600 \pi ^9}
 \Bigl[ -12959654400+320811321600 \pi +249167439360 \pi ^2-2406136078080 \pi ^3-1813794333120 \pi ^4+7622732486880 \pi ^5 \\
&&\scriptscriptstyle +5866548067808 \pi ^6-10329554789424 \pi^7-6075970569810 \pi ^8+2721498152625 \pi ^9
\Bigr] ,\\
&&\scriptscriptstyle \hat{Z}^{(10,11)}(4)
= \frac{1}{18852314330777518080000 \pi ^{10}} 
\Bigl[ 646656998400+20855511936000 \pi -15908447520000 \pi ^2-423480742272000 \pi ^3+155455887162240 \pi ^4
 +2407085602588800 \pi ^5 \\
&&\scriptscriptstyle -690712514324000 \pi^6 
-4858102787889600 \pi ^7+1434686348402316 \pi ^8+2720310664056300 \pi ^9-855380089265625 \pi ^{10} 
\Bigr] ,\\
&&\scriptscriptstyle \hat{Z}^{(11,12)}(4)
= \frac{1}{36498080544385275002880000 \pi ^{11}}
  \Bigl[ -71248933324800+3066836963097600 \pi +2147272497216000 \pi ^2-32994976801248000 \pi ^3-26307684678401280 \pi ^4+169824342336485760 \pi ^5 \\
&&\scriptscriptstyle +173665340769940800\pi ^6-543644538181826400 \pi ^7-506552450721933352 \pi ^8+791312771801094444 \pi ^9+502106969790796050 \pi ^{10}-218816278991454375 \pi^{11} 
\Bigr] ,\\
&&\scriptscriptstyle \hat{Z}^{(12,13)}(4)
= \frac{1}{21022894393565918401658880000 \pi ^{12}}
 \Bigl[ 2305385523302400+126291787634073600 \pi -84666760738560000 \pi ^2-4394402461709568000 \pi ^3+1299328657279107840 \pi ^4 \\
&&\scriptscriptstyle +44622842590319938560 \pi^5 
 -9109124891322297600 \pi ^6-187333512163572614400 \pi ^7+38674141099980946736 \pi ^8+324147996295923358944 \pi ^9
-82963880513737784280 \pi^{10} \\
&&\scriptscriptstyle -168381450188362233000 \pi ^{11}+51759053721397378125 \pi ^{12} 
\Bigr] , \\
&&\scriptscriptstyle \hat{Z}^{(13,14)}(4)
= \frac{1}{56845906440202243358085611520000 \pi ^{13}}
 \Bigl[ -326696029973913600+23110746777403084800 \pi +14222217559476326400 \pi ^2-296594132415214233600 \pi ^3-262735740032464258560 \pi ^4 \\
&&\scriptscriptstyle +1724522555482742695680 \pi
   ^5+3140617642113715146240 \pi ^6-9346414694706594236160 \pi ^7-17675870289759454430944 \pi ^8+38402692345719161274672 \pi ^9+45493756685677679170896 \pi^{10} \\
&&\scriptscriptstyle -63043272699716161765224 \pi ^{11}-42976871049629192344650 \pi ^{12}+18272369792404283180625 \pi ^{13}
   \Bigr] , 
\end{\eqa *}
 \end{minipage}
 } 
 \end{figure}

\clearpage
\begin{figure}[p]
 \rotatebox{90}{
 \begin{minipage}{1.36\linewidth}
\begin{\eqa *}
&&\scriptscriptstyle \hat{Z}^{(14,15)}(4)
= \frac{1}{89134381298237117585478238863360000 \pi ^{14}}
\Bigl[ 26391913012168704000+2332482753805940736000 \pi -1355032723453445836800 \pi ^2-129495692619143333683200 \pi ^3 \\
&&\scriptscriptstyle +30046144200216759014400 \pi^4  +2094928031999642929689600 \pi ^5 -253496012528736284186880 \pi ^6
-15100547205665994294942720 \pi ^7+1646913997990346974848960 \pi ^8 \\
&&\scriptscriptstyle +52651301340949893975032640
   \pi ^9-8050948997442445017147952 \pi ^{10}  
 -81998763263694004366967328 \pi ^{11}+18677773839866223207584700 \pi ^{12}+40305044547604234787287500 \pi^{13} \\
&&\scriptscriptstyle -12163520270564951460046875 \pi ^{14}
   \Bigr] ,\\
&&\scriptscriptstyle \hat{Z}^{(15,16)}(4)
=\frac{1}{320883772673653623307721659908096000000 \pi ^{15}}
\Bigl[ -4633702702858285056000+516113389226566010880000 \pi +275371922971758412800000 \pi ^2-6222420431735196648960000 \pi ^3 \\
&&\scriptscriptstyle -7145638913552848995686400 \pi
   ^4-2693255806580867079552000 \pi ^5+154547350796588963653440000 \pi ^6+159652773393243655228704000 \pi ^7-1473487743866342908878894720 \pi^8 \\
&&\scriptscriptstyle +935904177913483525445899200 \pi ^9+6685851766942719820185100000 \pi ^{10}-9622431406752930664123352400 \pi ^{11}-15115257648023669929267894152 \pi^{12}+18658828832122713811445030700 \pi ^{13} \\
&&\scriptscriptstyle +13606289597410394667282123750 \pi ^{14}-5657632516340449201697765625 \pi^{15} 
   \Bigr] ,\\
&&\scriptscriptstyle \hat{Z}^{(16,17)}(4)
=  \frac{1}{82146245804455327566776744936472576000000 \pi ^{16}}
\Bigl[ 56772848827620237312000+7781492578357428682752000 \pi -3887052200944811458560000 \pi ^2-657782957741870623703040000 \pi ^3 \\
&&\scriptscriptstyle +117619027037528930526412800 \pi
   ^4+15734914034909020252626124800 \pi ^5-574014170297132056267776000 \pi ^6-174641421273933510335222784000 \pi ^7
+852999519098267742094333440 \pi^8 \\
&&\scriptscriptstyle +1019582557904026953326637987840 \pi ^9-41099081861260979377311526400 \pi ^{10}
-3129469328085580990310547801600 \pi ^{11}  +353854932895814323114927802304 \pi^{12} \\
&&\scriptscriptstyle  +4510179835184969512289869463424 \pi ^{13} 
 -930128008456563471286686727200 \pi ^{14}-2123885924262761275712164860000 \pi ^{15}
+631139928800895901116370265625 \pi ^{16}
   \Bigr] ,\\
&&\scriptscriptstyle \hat{Z}^{(17,18)}(4)
=  \frac{1}{379844240599801434668775668586249191424000000 \pi ^{17}}
 \Bigl[ -12013213941558716350464000+2036573414071306950647808000 \pi +935022496319363796664320000 \pi ^2 \\
&&\scriptscriptstyle  -12425536068551131275509760000 \pi
   ^3-32428640849237305307924889600 \pi ^4-785115838027162292557223116800 \pi ^5+1357434101342472557859366912000 \pi ^6+13228798815940219457808006144000 \pi^7 \\
&&\scriptscriptstyle -20719602189385098881987341716480 \pi ^8-73298297241245350647746631498240 \pi ^9
+144768362750377327584834458956800 \pi ^{10}+82890136643185050776939034201600\pi ^{11} \\
&&\scriptscriptstyle -533161924740328983354031383553408 \pi ^{12} 
 +486555711295564474193718217457856 \pi ^{13}+1066895651332706638406604528945600 \pi^{14}
-1180193000974188589401149640060000 \pi ^{15} \\
&&\scriptscriptstyle -916867447192175983960684691531250 \pi ^{16} +373638901549556570993440766015625 \pi^{17}
 \Bigr] ,
\end{\eqa *}
 \end{minipage}
 } 
 \end{figure}

\clearpage
\begin{figure}[p]
 \rotatebox{90}{
 \begin{minipage}{1.36\linewidth}
\begin{\eqa *}
 &&\scriptscriptstyle \hat{Z}^{(18,19)}(4)
= \frac{1}{328185423878228439553822177658519301390336000000 \pi ^{18}}
\Bigl[ 465855763117726264098816000+95961974798933629893771264000 \pi -41045994308278306379096064000 \pi ^2 \\
 &&\scriptscriptstyle -11958856225081234996476837888000 \pi^3 
 +1625708668057072168313718374400 \pi ^4+401149261361386900409804454297600 \pi ^5+22012349141916298410418523750400 \pi ^6 \\
&&\scriptscriptstyle -6386577865336077441393321780019200 \pi^7 
  -730454799806610208657032441415680 \pi ^8+56234718518007944271908329628129280 \pi ^9+4958475482710824132244381543595520 \pi^{10}\\
&&\scriptscriptstyle  -284006654332827248522353155536885760 \pi ^{11} 
  -4201646145009753606197144172314368 \pi ^{12}+793555280427244296686025422264848128 \pi^{13}
-64556095975146628958789565015676608 \pi ^{14} \\
&&\scriptscriptstyle  -1077034100575608827227760504479502976 \pi ^{15} 
+203513467805398089020909664008446500 \pi^{16} 
  +489980901156954879254298891944422500 \pi ^{17}-143684383828322958995542770961734375 \pi ^{18} 
   \Bigr] , \\
&&\scriptscriptstyle \hat{Z}^{(19,20)}(4)
=  \frac{1}{1895599008320647466862876898155607484830580736000000 \pi ^{19}}
 \Bigl[ -116306979030787541840363520000+29192296360318816884296417280000 \pi +11486184129763831327751847936000 \pi ^2 \\
&&\scriptscriptstyle +239320576444282253980789137408000 \pi
   ^3-513169127421214360070776209408000 \pi ^4-42743845381692489150560595959808000 \pi ^5
 +45251247570770573890585342105190400 \pi^6 \\
&&\scriptscriptstyle +934429033843195012274006453562163200 \pi ^7 -1052245455735711826873922214895718400 \pi ^8
  -9028950896774505741213994757643110400 \pi^9 +10746023163461949104861069204115563520 \pi ^{10} \\
&&\scriptscriptstyle   +41560493596576859624407451573236277760 \pi ^{11} 
 -57881393600330601611761255722545559040 \pi^{12}
-66186603676499391299245009164286083840 \pi ^{13} \\
&&\scriptscriptstyle +178545259495591193181144141473488440192 \pi ^{14}-96420843472907346704815702438798945344 \pi
   ^{15}-320195024501159969100699347694779917560 \pi ^{16} \\
&&\scriptscriptstyle +319436559534411766172420108596826374500 \pi ^{17}+263428226729315780237172594059811468750 \pi
   ^{18}-105400940044220014291579806083525390625 \pi ^{19}
   \Bigr] ,\\
&&\scriptscriptstyle \hat{Z}^{(20,21)}(4)
=  \frac{1}{3032958413313035946980603037048971975728929177600000000 \pi ^{20}}
\Bigl[ 7890027009727842671990538240000+2380484007269648549678894284800000 \pi \\
&&\scriptscriptstyle   -870094846069703809457430528000000 \pi ^2  -427316635043437533171964836249600000 \pi^3
+43696857006690484114204893093888000 \pi ^4 +19309012462801249322624401699307520000 \pi ^5 \\
&&\scriptscriptstyle  +3093442573893979347765702641664000000 \pi^6 
 -419353566853385599959711300749475840000 \pi ^7-108995682246502516003641497033589350400 \pi ^8
   +5204520388687342110106615498605459456000 \pi^9 \\
&&\scriptscriptstyle +1282260583281319661150390013932136960000 \pi ^{10} 
 -39083199328307624131617979444901171712000 \pi ^{11}-6587607360043389970335760445349628270080 \pi^{12} \\
&&\scriptscriptstyle  +177384958265607837901489684929544370841600 \pi ^{13} 
  +10909362164629355169418704132227582272000 \pi ^{14} 
 -460899618288436440498095906010085475827200 \pi^{15} \\
&&\scriptscriptstyle  +25328535748683448729484308375998661305744 \pi ^{16} 
   +596114539513624624789699821725588956317600 \pi ^{17} 
 -104110976642972847276582535965831687465000 \pi^{18} \\
&&\scriptscriptstyle  -263554335642165799982102557882064679375000 \pi ^{19}+76392973813974105821930412592541417578125 \pi^{20}
   \Bigr] .
\end{\eqa *}
 \end{minipage}
 } 
 \end{figure}

\clearpage
\begin{figure}[p]
 \rotatebox{90}{
 \begin{minipage}{1.36\linewidth}
\begin{\eqa *}
&&\scriptscriptstyle \hat{Z}^{(1,3)}(4) = -\frac{\pi -4}{16 \pi } ,\quad
\hat{Z}^{(2,4)}(4) = -\frac{24+8 \pi -5 \pi ^2}{512 \pi ^2} ,\quad
\hat{Z}^{(3,5)}(4) = \frac{-480+216 \pi +404 \pi ^2-135 \pi ^3}{73728 \pi ^3} ,\quad
\hat{Z}^{(4,6)}(4) = \frac{3648+1920 \pi -7280 \pi ^2-3056 \pi ^3+1611 \pi ^4}{4718592 \pi ^4} ,\\
&&\scriptscriptstyle \hat{Z}^{(5,7)}(4) =
\frac{149760-91200 \pi -542400 \pi ^2+290000 \pi ^3+421388 \pi ^4-145575 \pi ^5}{1887436800 \pi ^5} ,\quad
\scriptscriptstyle \hat{Z}^{(6,8)}(4) 
= -\frac{3985920+2695680 \pi -23150400 \pi ^2-14083200 \pi ^3+33643016 \pi ^4+17182584 \pi ^5-8199225 \pi ^6}{543581798400 \pi ^6} ,\\
&&\scriptscriptstyle \hat{Z}^{(7,9)}(4)
= \frac{-262563840+195310080 \pi +2234400000 \pi ^2-1536561600 \pi ^3-6423200672 \pi ^4+3327482984 \pi ^5+4550305860 \pi ^6-1570037175 \pi ^7}{426168129945600 \pi ^7} , \\
&&\scriptscriptstyle \hat{Z}^{(8,10)}(4)
= \frac{2612736000+2100510720 \pi -30784481280 \pi ^2-23158732800 \pi ^3+108930057600 \pi ^4+86101125376 \pi ^5-150569774944 \pi ^6-90240714720 \pi ^7+40191979275 \pi
   ^8}{54549520633036800 \pi ^8} ,\\
&&\scriptscriptstyle \hat{Z}^{(9,11)}(4)
= \frac{1}{23565392913471897600 \pi ^9} 
\Bigl[ 81765089280-70543872000 \pi -1275440947200 \pi ^2+1042115880960 \pi ^3+8366238687744 \pi ^4-4965466665600 \pi ^5
-19810036547200 \pi ^6 +9535303927008 \pi^7 \\
&&\scriptscriptstyle +12943323415332 \pi ^8-4424232492225 \pi ^9 \Bigr] ,\\
&&\scriptscriptstyle \hat{Z}^{(10,12)}(4)
 = -\frac{1}{18852314330777518080000 \pi ^{10}}
\Bigl[ 4457056665600+4088254464000 \pi -89106006528000 \pi ^2-77950494720000 \pi ^3+538063720104960 \pi ^4+604591011763200 \pi ^5-1701797414614400 \pi^6 \\
&&\scriptscriptstyle    -1807751653088000 \pi ^7+2473516719298584 \pi ^8+1696571352275400 \pi ^9-717382007645625 \pi ^{10}
   \Bigr] ,\\
&&\scriptscriptstyle \hat{Z}^{(11,13)}(4)
= \frac{1}{36498080544385275002880000 \pi ^{11}}
 \Bigl[ -555876886118400+539303856537600 \pi +13867726473216000 \pi ^2-12915778917888000 \pi ^3-173484665239818240 \pi ^4+101989570995164160 \pi ^5 \\
&&\scriptscriptstyle +874230130509964800
   \pi ^6-416580747274102400 \pi ^7-1759437074938728416 \pi ^8+781753129254675864 \pi ^9+1075073613621876900 \pi ^{10}-363298087269376875 \pi^{11} 
   \Bigr] ,\\
&&\scriptscriptstyle \hat{Z}^{(12,14)}(4)
= \frac{1}{21022894393565918401658880000 \pi ^{12}}
 \Bigl[ 19613815190323200+20011567900262400 \pi -597928491933696000 \pi ^2-588280335261696000 \pi ^3+4783833481684561920 \pi ^4 \\
&&\scriptscriptstyle +8202607253775728640 \pi^5 
 -23473859992513075200 \pi ^6-48581456996570572800 \pi ^7+83891453324831079488 \pi ^8+122557375894589640576 \pi ^9
-136781388348577452720 \pi^{10} \\
&&\scriptscriptstyle  -105487589166075975600 \pi ^{11}+42802678421835688125 \pi ^{12} 
   \Bigr] ,\\
&&\scriptscriptstyle \hat{Z}^{(13,15)}(4)
=  \frac{1}{56845906440202243358085611520000 \pi ^{13}}
 \Bigl[ 3101121451524096000-3314734767164620800 \pi -113492010554405683200 \pi ^2+117455538452668416000 \pi ^3+2510569904270741913600 \pi ^4 \\
&&\scriptscriptstyle -1255894305023202324480\pi ^5-22531100208573717995520 \pi ^6
+8145323538866671872000 \pi ^7+90366000850395251421440 \pi ^8-33958009378858482961472 \pi ^9-159580793080270585003968 \pi^{10} \\
&&\scriptscriptstyle +65614854027384539289360 \pi ^{11}+92249506159524359154900 \pi ^{12}-30813026723470622255625 \pi ^{13}
   \Bigr] ,
\end{\eqa *}
 \end{minipage}
} 
 \end{figure}

\clearpage
\begin{figure}[p]
 \rotatebox{90}{
 \begin{minipage}{1.36\linewidth}
\begin{\eqa *}
&&\scriptscriptstyle \hat{Z}^{(14,16)}(4)
= -\frac{1}{89134381298237117585478238863360000 \pi ^{14}}
 \Bigl[ 272456547603893452800+303909902249361408000 \pi -11796329959008731136000 \pi ^2-12779374972152486297600 \pi ^3 \\
&&\scriptscriptstyle +88402649575861390049280 \pi^4 
 +300945154994061638860800 \pi ^5-274652591332004876390400 \pi ^6-3040999319565357886095360 \pi ^7+2392964858229889270559232 \pi ^8
+14371451346116304035115520 \pi ^9 \\
&&\scriptscriptstyle -15226199420446315479330880 \pi ^{10} 
 -31624131908955527921476224 \pi ^{11}+29636047830935581156037880 \pi ^{12}+25367799398832012658019400 \pi
   ^{13}-9949768870178118843736875 \pi ^{14}
   \Bigr] ,\\
&&\scriptscriptstyle \hat{Z}^{(15,17)}(4)
= \frac{1}{320883772673653623307721659908096000000 \pi ^{15}}
 \Bigl[ -52816581126156976128000+61302723210876026880000 \pi +2671384160433137909760000 \pi ^2-3019623748856863948800000 \pi ^3 \\
&&\scriptscriptstyle -101801104140692502901555200 \pi^4
+34359394044258577502208000 \pi ^5+1479560245436840096065536000 \pi ^6-246607811634796018767360000 \pi ^7-9938438552495431469565634560 \pi^8 \\
&&\scriptscriptstyle +1932188409365037474069619200 \pi ^9+33320970708654049290601222400 \pi ^{10}-9815668497514759021394616000 \pi ^{11}-52977344146357554465169589856 \pi
   ^{12}+20252592342926015539537459800 \pi ^{13} \\
&&\scriptscriptstyle +29234697966297464473704847500 \pi ^{14}-9655969846089559832542828125 \pi^{15}
   \Bigr] ,\\
&&\scriptscriptstyle \hat{Z}^{(16,18)}(4)
= \frac{1}{82146245804455327566776744936472576000000 \pi ^{16}}
\Bigl[701688110191636119552000+845065298018511618048000 \pi -41015292300794234142720000 \pi ^2-48212524807418711900160000 \pi ^3 \\
&&\scriptscriptstyle +33377554746987386948812800 \pi
   ^4+1883902643769674227462963200 \pi ^5+7405301480379404495781888000 \pi ^6-30011307493082807320805376000 \pi ^7-58036636289136883355977359360 \pi^8 \\
&&\scriptscriptstyle +228863613933917454124313640960 \pi ^9+64591904895522703563532595200 \pi ^{10}-902096039731772098842612838400 \pi ^{11}+566188274817049226514236146944 \pi
   ^{12}+1775528104008903044065560976896 \pi ^{13} \\
&&\scriptscriptstyle -1427367442670304033284940916800 \pi ^{14}-1342811627750294054504431560000 \pi
   ^{15}+511788703740693849230401265625 \pi ^{16}
   \Bigr] ,\\
&&\scriptscriptstyle \hat{Z}^{(17,19)}(4)
= \frac{1}{379844240599801434668775668586249191424000000 \pi ^{17}}
\Bigl[ 162643672023548836184064000-202787863845382838550528000 \pi -10883450153360530278973440000 \pi ^2 \\
&&\scriptscriptstyle +13270738435564987408711680000 \pi
   ^3+713290097741706884519401881600 \pi ^4-86557136201968160430961459200 \pi ^5-15744942652447803166976507904000 \pi ^6-1036043702904303284818378752000 \pi^7 \\
&&\scriptscriptstyle +161322408118212869486991167815680 \pi ^8+6514261302033641531650754519040 \pi ^9-872204344115141992581158642073600 \pi ^{10}+64697127143895332194417934643200  \pi ^{11} \\
&&\scriptscriptstyle +2544365487742659413806808205145088 \pi ^{12}-583488151420607839842058460457216 \pi ^{13}-3713942049814044603212945067705600 \pi
   ^{14}+1326599155381451599614802920984000 \pi ^{15} \\
&&\scriptscriptstyle +1970134581088094635028490942262500 \pi ^{16}-643854435117209856303331458515625 \pi^{17} 
\Bigr] ,
\end{\eqa *}
 \end{minipage}
} 
 \end{figure}

\clearpage
\begin{figure}[p]
 \rotatebox{90}{
 \begin{minipage}{1.36\linewidth}
\begin{\eqa *}
&&\scriptscriptstyle \hat{Z}^{(18,20)}(4)
= \frac{1}{328185423878228439553822177658519301390336000000 \pi ^{18}}
 \Bigl[
 -6818632796571774159421440000-8782758289271637153939456000 \pi +518045091547114329555861504000 \pi ^2 \\
&&\scriptscriptstyle +653646753485853096620851200000 \pi
   ^3+9767061332872923054131380224000 \pi ^4-42638245624623521034090735206400 \pi ^5-514541701082903817796128630374400 \pi ^6 \\
&&\scriptscriptstyle +1015545049014275699845796462592000 \pi
   ^7+6505372217358445011376553120563200 \pi ^8-11533140594833244700822296274206720 \pi ^9
-34269764753420797828841473986846720 \pi^{10} \\
&&\scriptscriptstyle +70674019950235884345942638314291200 \pi ^{11}+60599270372093604085666262521169920 \pi ^{12}-241041349850696485773413550483658752 \pi^{13} \\
&&\scriptscriptstyle +80659913268744637454130425647020288 \pi ^{14}+432554302544884468404160110954324480 \pi ^{15}-303221748347485891159335256566333000 \pi^{16}\\
&&\scriptscriptstyle -311117607632817246963387522855375000 \pi ^{17}+115681163171007840645199063462734375 \pi ^{18}
\Bigr] , \\
&&\scriptscriptstyle \hat{Z}^{(19,21)}(4)
= -\frac{1}{1895599008320647466862876898155607484830580736000000 \pi ^{19}}
 \Bigl[ 1853195352508743683514826752000-2461526439562410471551139840000 \pi -158687022555520006087507968000000 \pi ^2 \\
&&\scriptscriptstyle +206780011137517738243185967104000 \pi^3
+18019000624301807180292379430092800 \pi ^4+2116367559414370251822553104384000 \pi ^5
 -573443968518124891921630490001408000 \pi^6 \\
&&\scriptscriptstyle -170450475627637867821056731407974400 \pi ^7+8427278175378501317604499242099671040 \pi ^8+2332094105405401372879212793090867200 \pi
   ^9-67491543737021248955405868724821196800 \pi ^{10} \\
&&\scriptscriptstyle -11719930691169437550192833629694238720 \pi ^{11}+310374638857838871590330438246560784384 \pi
   ^{12}+4551241481609414422878350063706449920 \pi ^{13}-811336473990649324356818711228496184320 \pi ^{14} \\
&&\scriptscriptstyle +142954478542259347741858835399237637888 \pi
   ^{15}+1103438222962196441806408655034632274336 \pi ^{16}-369962673389562437047259920111271157000 \pi ^{17} \\
&&\scriptscriptstyle -565761818418237777723415575356664697500 \pi
   ^{18}+183064696868576267532252866292344578125 \pi ^{19}
   \Bigr] ,\\
&&\scriptscriptstyle \hat{Z}^{(20,22)}(4)
= \frac{1}{3032958413313035946980603037048971975728929177600000000 \pi ^{20}}
\Bigl[ 135524964632671292532342128640000+185319535250874368351482675200000 \pi \\
&&\scriptscriptstyle  -12994366429554983728012748390400000 \pi ^2  -17453990126765531115036868608000000 \pi
   ^3-826703439458812008006079387533312000 \pi ^4 \\
&&\scriptscriptstyle +1929881191736265974030465583022080000 \pi ^5+44706162801630300252375901997629440000 \pi^6 
 -65718247693094483918733454344192000000 \pi ^7 \\
&&\scriptscriptstyle -799061755111411391982253822833288806400 \pi ^8
+1054034365475863702663283362230042624000 \pi^9 
  +6996975360556420530421061961032220672000 \pi ^{10} \\
&&\scriptscriptstyle  -9339768640879776470200524264075018240000 \pi ^{11}
 -31113410956395586861188424176997444116480 \pi^{12} 
  +48405914301987862664070787917746017894400 \pi ^{13} \\
&&\scriptscriptstyle   +57217667515278489114848713777537890611200 \pi ^{14} 
 -146938050001901934655037647024142366208000 \pi^{15} 
  +18745796253838016877363977524061318956224 \pi ^{16} \\
&&\scriptscriptstyle  +243984108761765592799935642993419521219200 \pi ^{17} 
 -151068247341247732708406772367272929970000 \pi^{18} \\
&&\scriptscriptstyle  -168021797387686596207728363557037009250000 \pi ^{19}+61136238207447676268730807457781098828125 \pi^{20}
   \Bigr] .
\end{\eqa *}
 \end{minipage}
} 
 \end{figure}

\clearpage
\begin{figure}[p]
 \rotatebox{90}{
 \begin{minipage}{1.36\linewidth}
\begin{\eqa *}
&&\scriptscriptstyle \hat{Z}^{(1,2)}(6) =\frac{9-\sqrt{3} \pi }{108 \pi } ,\quad
\hat{Z}^{(2,3)}(6) = -\frac{432+408 \sqrt{3} \pi -269 \pi ^2}{31104 \pi ^2} ,\quad
\hat{Z}^{(3,4)}(6) = \frac{-9720+12960 \sqrt{3} \pi +4347 \pi ^2-1931 \sqrt{3} \pi ^3}{10077696 \pi ^3} ,\quad
\hat{Z}^{(4,5)}(6)
= \frac{536544+1441152 \sqrt{3} \pi -557280 \pi ^2-3523152 \sqrt{3} \pi ^3+1912867 \pi ^4}{5804752896 \pi ^4} ,\\
&&\scriptscriptstyle \hat{Z}^{(5,6)}(6)
= \frac{87130080-307346400 \sqrt{3} \pi +310165200 \pi ^2+911260800 \sqrt{3} \pi ^3+70893099 \pi ^4-108143075 \sqrt{3} \pi ^5}{15672832819200 \pi ^5} ,\\
&&\scriptscriptstyle  \hat{Z}^{(6,7)}(6)
=-\frac{5391567360+29989543680 \sqrt{3} \pi +47373919200 \pi ^2-245744668800 \sqrt{3} \pi ^3-93947321232 \pi ^4+416879149464 \sqrt{3} \pi ^5-207252738175 \pi
   ^6}{13541327555788800 \pi ^6} ,\\\
&&\scriptscriptstyle  \hat{Z}^{(7,8)}(6)
= -\frac{1525334872320-10774871009280 \sqrt{3} \pi +44069352052320 \pi ^2+87765668272800 \sqrt{3} \pi ^3-162341159067864 \pi ^4-195569135070624 \sqrt{3} \pi
   ^5+2889114383061 \pi ^6+21333980075275 \sqrt{3} \pi ^7}{71660705425234329600 \pi ^7} ,\\
&&\scriptscriptstyle  \hat{Z}^{(8,9)}(6)
= \frac{1}{82553132649869947699200 \pi ^8}
  \Bigl[ 103900541875200+1038163124920320 \sqrt{3} \pi +5608295669975040 \pi ^2-18539127946644480 \sqrt{3} \pi ^3-50054331039733440 \pi ^4+93338872774430976 \sqrt{3}\pi ^5 \\
&&\scriptscriptstyle +62169932532485952 \pi ^6-135509099423062944 \sqrt{3} \pi ^7+63809267778892475 \pi ^8
   \Bigr] ,\\
&&\scriptscriptstyle  \hat{Z}^{(9,10)}(6)
= \frac{1}{80241644935673589163622400 \pi ^9}
 \Bigl[ 4888772865930240-60647422455567360 \sqrt{3} \pi +556321804406000640 \pi ^2+933542468377989120 \sqrt{3} \pi ^3-5164192759594454208 \pi ^4 \\
&&\scriptscriptstyle -5149624432989572928
   \sqrt{3} \pi ^5+12236223416455926432 \pi ^6+10793989672813518336 \sqrt{3} \pi ^7-712918751457709125 \pi ^8
-1128948143041047875 \sqrt{3} \pi^9 
\Bigr] ,\\
&&\scriptscriptstyle  \hat{Z}^{(10,11)}(6)
= -\frac{1}{577739843536849841978081280000 \pi ^{10}}
\Bigl[ 1808196272496230400+29820491666574336000 \sqrt{3} \pi +341976908820339072000 \pi ^2-956671865037923328000 \sqrt{3} \pi ^3
-6334631015171771612160 \pi^4 \\
&&\scriptscriptstyle +9678244617726722956800 \sqrt{3} \pi ^5+30714971087435892321600 \pi ^6-39110535608560387142400 \sqrt{3} \pi ^7-32542174192979041907184 \pi
   ^8+52142498074569979739400 \sqrt{3} \pi ^9-23628686735164823781875 \pi ^{10}
   \Bigr] ,\\
&&\scriptscriptstyle  \hat{Z}^{(11,12)}(6)
= -\frac{1}{7549904275339553734969566167040000 \pi ^{11}}
\Bigl[ 1050190755685895577600-21133567503116999884800 \sqrt{3} \pi +353851100633289764736000 \pi ^2
 +512058440947537604736000 \sqrt{3} \pi^3 \\
&&\scriptscriptstyle -5892361026837026926947840 \pi ^4-4862552393702426691502080 \sqrt{3} \pi ^5+34407529240248996249844800 \pi ^6
+24361850020186437764068800 \sqrt{3} \pi^7-70137894935039233623459336 \pi ^8 \\
&&\scriptscriptstyle -51946867156243127085448992 \sqrt{3} \pi ^9+5204432021907025587870225 \pi ^{10}
 +5303975247315120333409375 \sqrt{3} \pi^{11}
   \Bigr] , \\
&&\scriptscriptstyle  \hat{Z}^{(12,13)}(6)
=\frac{1}{13046234587786748854027410336645120000 \pi ^{12}}
\Bigl[ 83490815735172963532800+2145434180984613421056000 \sqrt{3} \pi +43668077436860469910732800 \pi ^2
-110745866224587673626624000 \sqrt{3} \pi^3 \\
&&\scriptscriptstyle -1387778216242127674344145920 \pi ^4+1882767167057484541106380800 \sqrt{3} \pi ^5
 +13805689691880169831901767680 \pi ^6
-14594418424900287992172211200 \sqrt{3} \pi ^7-51398948522742435781662738528 \pi ^8 \\
&&\scriptscriptstyle +52247271668612310462692995200 \sqrt{3} \pi ^9+49652161964137007705409085152 \pi ^{10}-66095382311707064295866007600
   \sqrt{3} \pi ^{11}+29085854454495182982256994375 \pi ^{12}
   \Bigr] ,\\
&&\scriptscriptstyle  \hat{Z}^{(13,14)}(6)
=\frac{1}{238119873696283740083708293464446730240000 \pi ^{13}}
 \Bigl[ 62849266996403142807552000-1943173953584639751809433600 \sqrt{3} \pi +53385958444572041103393177600 \pi ^2 \\
&&\scriptscriptstyle  +66023578152379508996682547200 \sqrt{3} \pi
   ^3-1363985460563660248877968204800 \pi ^4-877294872785590956560315612160 \sqrt{3} \pi ^5+14005002680919502346212592348160 \pi ^6 \\
&&\scriptscriptstyle +7269144146566425060461610577920 \sqrt{3} \pi ^7 
  -68507512250412289427276031105120 \pi ^8-38164057060083435366285546885024 \sqrt{3} \pi ^9
 +130699522405140370854851521228464 \pi^{10} \\
&&\scriptscriptstyle +86146219768443297716247806118528 \sqrt{3} \pi ^{11}   -10796547621935370680916662084625 \pi ^{12} 
 -8667048310788410692867517174375 \sqrt{3} \pi^{13}
 \Bigr] ,
\end{\eqa *}
 \end{minipage}
 } 
 \end{figure}

\clearpage
\begin{figure}[p]
 \rotatebox{90}{
 \begin{minipage}{1.36\linewidth}
\begin{\eqa *}
&&\scriptscriptstyle \hat{Z}^{(14,15)}(6) 
= -\frac{1}{3360347657601956140061291437370272257146880000 \pi^{14}}
\Bigl[ 37280700172463779534248345600+1430800961782505255711716147200 \sqrt{3} \pi +46959531974488320087821650329600 \pi ^2 \\
&&\scriptscriptstyle -110734509411935392506802126848000
   \sqrt{3} \pi ^3-2318704455841811380006441016279040 \pi ^4+2843694054617147244933909889720320 \sqrt{3} \pi ^5+38883041245158786599977296753269760 \pi^6 \\
&&\scriptscriptstyle -35592996246082299070711317247795200 \sqrt{3} \pi ^7-287479361396599057309776104235419136 \pi ^8+238328824190189888015057811740974848 \sqrt{3} \pi
   ^9+913452426869641020030660743649197664 \pi ^{10} \\
&&\scriptscriptstyle -793701024923422507490563913890381440 \sqrt{3} \pi ^{11}-828502804920131860258087704578943216 \pi
   ^{12}+970010173088674006823737901405370600 \sqrt{3} \pi ^{13}-416923797096980269596922609044983125 \pi ^{14}
   \Bigr] ,\\
&&\scriptscriptstyle \hat{Z}^{(15,16)}(6) 
= -\frac{1}{27218816026575844734496460642699205282889728000000 \pi ^{15}}
\Bigl[ 11655581136109833031168131072000-531945723270633974788317511680000 \sqrt{3} \pi \\
&&\scriptscriptstyle   +22459346747183334888044730224640000 \pi^2 
  +23259651401441142743355457597440000 \sqrt{3} \pi ^3-798649571110462414122365309935411200 \pi ^4
-354038668700109749709151039008768000 \sqrt{3} \pi^5 \\
&&\scriptscriptstyle +12285854289287121883964091952708224000 \pi ^6 
  +3605043505396224389483991099878784000 \sqrt{3} \pi ^7-103949252472270033877052684126355068160 \pi
   ^8-34348790079113820831695828753606707200 \sqrt{3} \pi ^9 \\
&&\scriptscriptstyle +471139137709414089529378610157414429600 \pi ^{10} 
  +215618250565476073304142901337348085600 \sqrt{3} \pi
   ^{11}-873785177563799625900238612448808749496 \pi ^{12} \\
&&\scriptscriptstyle -528804213795199675848467935208130602400 \sqrt{3} \pi ^{13} 
  +76699182755636328453427304836395778125 \pi^{14} 
 +52706415790368230420152468852553421875 \sqrt{3} \pi ^{15} 
   \Bigr] ,\\ 
&&\scriptscriptstyle \hat{Z}^{(16,17)}(6) 
= \frac{1}{62712152125230746268279845320778968971777933312000000 \pi ^{16}}
\Bigl[ 1045787987824187839056252370944000+58014129313312794056859372748800000 \sqrt{3} \pi \\
&&\scriptscriptstyle +2890261326910604797067083799592960000 \pi
   ^2-6413144343783076626904574440243200000 \sqrt{3} \pi ^3-208261527204980447864355597281044070400 \pi ^4+231220844332636664649027039842598912000 \sqrt{3} \pi^5 \\
&&\scriptscriptstyle +5310167958631405213406042497343963136000 \pi ^6-4199879795694261520913887713026088960000 \sqrt{3} \pi ^7
-64636033654509388747663468270676996229120 \pi^8 \\
&&\scriptscriptstyle +44476379938352175694352768854448245555200 \sqrt{3} \pi ^9 
 +397555135299362306766111560458758659174400 \pi ^{10}-274214511409705427065744106708215415040000
   \sqrt{3} \pi ^{11} \\
&&\scriptscriptstyle -1133224637471468481853392175295954525526912 \pi ^{12} 
 +873883980542535798386471638841639699950080 \sqrt{3} \pi
   ^{13}+980923655230612055725505504867743328784000 \pi ^{14} \\
&&\scriptscriptstyle -1043126836337282812319811534016246860360000 \sqrt{3} \pi
   ^{15}+439654999165985712784947408927095351921875 \pi ^{16}
   \Bigr] ,\\
&&\scriptscriptstyle \hat{Z}^{(17,18)}(6) 
= \frac{1}{1957371692132702052525550532152153179547132854534144000000 \pi ^{17}}
\Bigl[ 1187188616500188104746391667867648000-77536133658334911501554228771291136000 \sqrt{3} \pi \\ 
&&\scriptscriptstyle +4810739638569891348192726073877987328000 \pi
   ^2+4054475696913622311424902227070812160000 \sqrt{3} \pi ^3-221704036730716852743787889531211025612800 \pi ^4 \\
&&\scriptscriptstyle -48653620995618065625861298517125506662400 \sqrt{3}\pi ^5 
 +4593816991830289286895674382544697479987200 \pi ^6+40988555881701866013263774924560859136000 \sqrt{3} \pi ^7 \\
&&\scriptscriptstyle -57235359314747167831428016693861057649264640 \pi ^8 
 -1398797694358764418873290184223707594782720 \sqrt{3} \pi ^9+449486157560945178309022953237331490324981760 \pi
   ^{10} \\
&&\scriptscriptstyle +81736004904460580036374006298400554152550400 \sqrt{3} \pi ^{11} 
 -1991149843318171818517305815241134645209551744 \pi
   ^{12}-763094817486525779984776249604309279112991872 \sqrt{3} \pi ^{13} \\
&&\scriptscriptstyle +3665688600395452658283818679073093240377865536 \pi^{14} 
 +2076250465297094073190381210322975790916915200 \sqrt{3} \pi ^{15}-334639061519883196854100662637622379849583125 \pi^{16}\\
&&\scriptscriptstyle -205705317537621467241886165786736713854921875 \sqrt{3} \pi ^{17}
   \Bigr] ,
\end{\eqa *}
 \end{minipage}
 } 
 \end{figure}

\clearpage
\begin{figure}[p]
 \rotatebox{90}{
 \begin{minipage}{1.36\linewidth}
\begin{\eqa *}
&&\scriptscriptstyle \hat{Z}^{(18,19)}(6) 
= -\frac{1}{5073507426007963720146226979338381041386168358952501248000000 \pi ^{18}}
 \Bigl[ 112169667425907909754827947020124160000+8761009682611327305521536546857025536000 \sqrt{3} \pi \\
 &&\scriptscriptstyle +635332157212957254490147188814503346176000 \pi
   ^2-1335978507129723918851682512043214110720000 \sqrt{3} \pi ^3-63975469393358265106212510128857407553536000 \pi ^4 \\
&&\scriptscriptstyle +63980320173702877942252093404352691070566400
   \sqrt{3} \pi ^5+2323326104095816088430370498810608085116518400 \pi ^6
-1557134878497265699955959034939273491709952000 \sqrt{3} \pi^7 \\
   &&\scriptscriptstyle -42141943716694674901775928843034754518482124800 \pi ^8
+23415863479297662752162691548313403601531781120 \sqrt{3} \pi
   ^9+418918067236502871108979717071171320007370475520 \pi ^{10} \\
&&\scriptscriptstyle -227597935466261852240500715081248018735154380800 \sqrt{3} \pi
   ^{11}-2266737868395036103765629341742177918107413370880 \pi ^{12}+1345357449740561434087113171243789155651838200832 \sqrt{3} \pi
   ^{13} \\
   &&\scriptscriptstyle +5963113721796320541990915258671906871421660762752 \pi ^{14}-4177809446408033818696318710544584410314568640000 \sqrt{3} \pi
   ^{15}-4976937991230192360734943699613720820024729305200 \pi ^{16} \\
&&\scriptscriptstyle +4905747103798712053260254753001363348153771525000 \sqrt{3} \pi
   ^{17}-2033421644433369114039111419321255669378175484375 \pi ^{18}
   \Bigr] ,\\
&&\scriptscriptstyle \hat{Z}^{(19,20)}(6) 
= -\frac{1}{197805907525198489521061097470444800041563931978840118657024000000 \pi ^{19}}
 \Bigl[ 150731009346983096078813746873519570944000-13741022465134014916631041692255538642944000 \sqrt{3} \pi  \\
&&\scriptscriptstyle +1212371300578919477303064934414268977250304000 \pi
   ^2+799266787251875790995461175224188516827136000 \sqrt{3} \pi ^3-68282304589870523791644727874266879697643110400 \pi
   ^4 \\
&&\scriptscriptstyle +2675706080954580204096533668621981643597414400 \sqrt{3} \pi ^5+1739598622719853666266534255255813331924048281600 \pi
   ^6-667076322839036484948869902342131197617265049600 \sqrt{3} \pi ^7 \\
&&\scriptscriptstyle -28241976718243698252529841837165906982659188408320 \pi
   ^8+12809034118276764577594461636592975258176319979520 \sqrt{3} \pi ^9+329458061046271598906606273285659450534891584890880 \pi
   ^{10} \\
&&\scriptscriptstyle -85031268046571243105039459934792110968191825684480 \sqrt{3} \pi ^{11}-2602571628198068645085756800316002172912931851512832 \pi
   ^{12}-118519781889886160154402538830539131859841308540928 \sqrt{3} \pi ^{13} \\
&&\scriptscriptstyle +11679507931222724417959690597437582046450926254080128 \pi
   ^{14}+3770457806715136548854832612771345317163076080849792 \sqrt{3} \pi ^{15}-21623284602056709093448036211277283146884464073758728 \pi
   ^{16} \\
&&\scriptscriptstyle -11609532246621221063550632589619964332437599404471200 \sqrt{3} \pi ^{17}+2029168372762486258176852925611507246328865226208125 \pi
   ^{18}+1145894148639381402666892230564021688810808531921875 \sqrt{3} \pi ^{19}
   \Bigr] ,\\
&&\scriptscriptstyle \hat{Z}^{(20,21)}(6) 
= \frac{1}{2848405068362858249103279803574405120598520620495297708661145600000000 \pi ^{20}}
 \Bigl[ 74642925464838449547547304788720185507840000+8032644269746433088802891089045584294707200000 \sqrt{3} \pi  \\
&&\scriptscriptstyle +821914392151038888548315083775217513686630400000 \pi
   ^2-1646627810094118339754333000594576617871769600000 \sqrt{3} \pi ^3-112072122835845105598595776552412346796091768832000 \pi
   ^4 \\
&&\scriptscriptstyle +100078922572326373426385515564632193478405652480000 \sqrt{3} \pi ^5+5537575176556884981245417353643513884947169935360000 \pi
   ^6-3038395184049792755100522916786179694453964144640000 \sqrt{3} \pi ^7 \\
&&\scriptscriptstyle -140471558890294574386350627112552273024109506250342400 \pi
   ^8+59237557362219389544947632545655820631827990544384000 \sqrt{3} \pi ^9+2049262991400237693715056253843845599196260137771008000 \pi
   ^{10}\\
&&\scriptscriptstyle -819186787947057841398031363162773098425313241403392000 \sqrt{3} \pi ^{11}
-17666867538774157240250172783244591468417805177490094080 \pi^{12} \\
&&\scriptscriptstyle +7768230814206640784225774649515404371869882999737958400 \sqrt{3} \pi ^{13} 
 +86889595320616483406745543398276844937549903089128140800 \pi^{14} \\
&&\scriptscriptstyle -45333491603535871102721096808975506192752031259363379200 \sqrt{3} \pi ^{15} 
 -214781285954871457582914152451483767401235871899570969376 \pi^{16} \\
&&\scriptscriptstyle +138900150552269068901280646293106327944880481639935881600 \sqrt{3} \pi ^{17} 
 +174055184316886248981890256621406019075124996641115060000 \pi^{18} \\
&&\scriptscriptstyle -161241149803358809890390659226153140332556916922691250000 \sqrt{3} \pi ^{19} 
 +65870864016187141559943692468582292090670138827658984375 \pi^{20}
   \Bigr] ,
\end{\eqa *}
 \end{minipage}
 } 
 \end{figure}


\clearpage
\begin{figure}[p]
 \rotatebox{90}{
 \begin{minipage}{1.36\linewidth}
\begin{\eqa *}
&&\scriptscriptstyle \hat{Z}^{(1,3)}(6)  = \frac{7}{72}-\frac{1}{2 \sqrt{3} \pi } ,\quad
\hat{Z}^{(2,4)}(6) = \frac{54+108 \sqrt{3} \pi -65 \pi ^2}{5184 \pi ^2} ,\quad
\hat{Z}^{(3,5)}(6) = \frac{-4212 \sqrt{3}+3969 \pi +8010 \sqrt{3} \pi ^2-4583 \pi ^3}{1679616 \pi ^3} ,\quad
\hat{Z}^{(4,6)}(6) = \frac{1944+21600 \sqrt{3} \pi -360 \pi ^2-33024 \sqrt{3} \pi ^3+17017 \pi ^4}{35831808 \pi ^4} ,\\ 
&&\scriptscriptstyle \hat{Z}^{(5,7)}(6)
= \frac{-94828320 \sqrt{3}+117369000 \pi +661802400 \sqrt{3} \pi ^2-344725200 \pi ^3-1082915916 \sqrt{3} \pi ^4+594333475 \pi ^5}{5224277606400 \pi ^5} ,\\
&&\scriptscriptstyle \hat{Z}^{(6,8)}(6)
= \frac{7873200+304780320 \sqrt{3} \pi +154693800 \pi ^2-1695427200 \sqrt{3} \pi ^3-396367290 \pi ^4+1897354716 \sqrt{3} \pi ^5-914518625 \pi ^6}{41794220851200 \pi^6} ,\\
&&\scriptscriptstyle \hat{Z}^{(7,9)}(6)
= \frac{-508378822560 \sqrt{3}+758572584840 \pi +7829183038320 \sqrt{3} \pi ^2+1097913511800 \pi ^3-39580063351668 \sqrt{3} \pi ^4+7998307736517 \pi ^5+61102529757546
   \sqrt{3} \pi ^6-32343075907625 \pi ^7}{5971725452102860800 \pi ^7} ,\\
&&\scriptscriptstyle \hat{Z}^{(8,10)}(6)
=\frac{1}{764380857869166182400 \pi ^8}
\Bigl[ 374984769600+40399417336320 \sqrt{3} \pi +36931216049280 \pi ^2-500894881966080 \sqrt{3} \pi ^3-365736091175280 \pi ^4
+1829353921448256 \sqrt{3} \pi^5 \\
&&\scriptscriptstyle +641781150699024 \pi ^6-1813331818276224 \sqrt{3} \pi ^7+839054170880975 \pi ^8
   \Bigr] ,\\
&&\scriptscriptstyle \hat{Z}^{(9,11)}(6)
=\frac{1}{13373607489278931527270400 \pi ^9}
\Bigl[-3917340893088000 \sqrt{3}+6730001653919040 \pi +106813099769080320 \sqrt{3} \pi ^2+195151699634814720 \pi ^3
-1093067403611480640 \sqrt{3} \pi^4 \\
&&\scriptscriptstyle  -1393374237131561328 \pi ^5+4806535472237733696 \sqrt{3} \pi ^6
+225249339667869216 \pi ^7-7180800897172940892 \sqrt{3} \pi ^8
+3687887513094183875 \pi^9
\Bigr] ,\\
&&\scriptscriptstyle \hat{Z}^{(10,12)}(6)
= \frac{1}{534944299571157261090816000 \pi ^{10}}
\Bigl[546727794076800+142828556011142400 \sqrt{3} \pi +176372211457512000 \pi ^2-3171251958177945600 \sqrt{3} \pi ^3-3459219039539776800 \pi ^4+24059184583496569920
   \sqrt{3} \pi ^5 \\
&&\scriptscriptstyle  +21076122335055152400 \pi ^6-73691721088732385280 \sqrt{3} \pi ^7
-31900423861637229402 \pi ^8+68744639166219603180 \sqrt{3} \pi
   ^9-30899352797215317625 \pi ^{10}
   \Bigr] ,\\
&&\scriptscriptstyle \hat{Z}^{(11,13)}(6)
= \frac{1}{314579344805814738957065256960000 \pi^{11}}
 \Bigl[ -249184525751800531200 \sqrt{3}+479113179190351675200 \pi +10626560651978376336000 \sqrt{3} \pi ^2
+55814843148794591400000 \pi ^3 \\
&&\scriptscriptstyle -182621146752160036240320
   \sqrt{3} \pi ^4-863282999197051079003280 \pi ^5+1549450367229406203856800 \sqrt{3} \pi ^6+3440470376638837660734000 \pi ^7-6279315699721705626883788 \sqrt{3} \pi
   ^8-1560037420137424782621177 \pi ^9 \\
&&\scriptscriptstyle +9188182264304984094662550 \sqrt{3} \pi ^{10}-4599968240521538521223125 \pi ^{11}
   \Bigr] ,\\
&&\scriptscriptstyle \hat{Z}^{(12,14)}(6)
=\frac{1}{20133078067572143293252176445440000 \pi ^{12}}
 \Bigl[ 35723194064978112000+20574767841066073958400 \sqrt{3} \pi +30820827346747879795200 \pi ^2
-721932436402072068096000 \sqrt{3} \pi ^3 \\
&&\scriptscriptstyle -992248303085584848072000
   \pi ^4+9271329452609638568394240 \sqrt{3} \pi ^5+11872031790931249180763520 \pi ^6-56803831889139009422668800 \sqrt{3} \pi ^7
-57575369358848103759307080 \pi^8 \\
&&\scriptscriptstyle +159173921437790833538037216 \sqrt{3} \pi ^9+79490830119527280447250488 \pi ^{10}-143563966093917770471011200 \sqrt{3} \pi ^{11}
+63100250254956643590929375 \pi^{12}
   \Bigr] ,\\
&&\scriptscriptstyle \hat{Z}^{(13,15)}(6)
=\frac{1}{19843322808023645006975691122037227520000 \pi ^{13}}
\Bigl[-34907347797869988541900800 \sqrt{3}+73693183644929154699993600 \pi +2149737429734240208774912000 \sqrt{3} \pi ^2 
 +24838948775258544412425139200 \pi^3 \\
&&\scriptscriptstyle -55738213089290281965751015680 \sqrt{3} \pi ^4-645992614854581260207543842240 \pi ^5+768698397238330587899341785600 \sqrt{3} \pi
   ^6+5582474179276010303493566680320 \pi ^7  -5816270665660415851684740035232 \sqrt{3} \pi ^8 \\
&&\scriptscriptstyle -17637884672485021695464836647096 \pi
   ^9+22365964892909712636179174887200 \sqrt{3} \pi ^{10}+9328711450693810178999289937968 \pi ^{11} \\
&&\scriptscriptstyle -32281884420313655284512321105300 \sqrt{3} \pi^{12}
+15811161255105991372453098368125 \pi ^{13}
   \Bigr] ,
\end{\eqa *}
 \end{minipage}
 } 
 \end{figure}

\clearpage
\begin{figure}[p]
 \rotatebox{90}{
 \begin{minipage}{1.36\linewidth}
\begin{\eqa *}
&&\scriptscriptstyle \hat{Z}^{(14,16)}(6)  
=\frac{1}{1111226077249324120390638702834084741120000 \pi ^{14}}
 \Bigl[ 2934088821332912251008000+3471466366119761151158016000 \sqrt{3} \pi +5979100640202388166192524800 \pi ^2 \\
&&\scriptscriptstyle -177514052745738454080323788800 \sqrt{3} \pi^3 
 -285984225780228175715064950400 \pi ^4+3415226130798575502906706809600 \sqrt{3} \pi ^5+5622685581419961608175573631680 \pi ^6 \\
&&\scriptscriptstyle -34327073094708414904281870151680  \sqrt{3} \pi ^7  -52187593470056431682665384974960 \pi ^8 
 +189123976816027469499535245433440 \sqrt{3} \pi ^9+218154420144737971927296937945272 \pi^{10} \\
&&\scriptscriptstyle -505688124606911960251521392502912 \sqrt{3} \pi ^{11} 
 -281681657336883025847215476645786 \pi ^{12}  +447296525946748740805375820288700 \sqrt{3} \pi
   ^{13}-193055602699026128461096502430625 \pi ^{14}
   \Bigr] ,\\
&&\scriptscriptstyle \hat{Z}^{(15,17)}(6)  
= \frac{1}{567058667220330098635342930056233443393536000000 \pi ^{15}}
\Bigl[ -1875571281287764093758396672000 \sqrt{3}+4288363162915162174640596800000 \pi \\
&&\scriptscriptstyle +157771603497621602127078760320000 \sqrt{3} \pi^2 
 +3548618715386777350340651428800000 \pi ^3-5754835690501444799002473199276800 \sqrt{3} \pi ^4-138990686851706487306226329011400000 \pi^5 \\
&&\scriptscriptstyle +116918755931077495558712117320752000 \sqrt{3} \pi ^6 
 +2024984536602142743849987695409480000 \pi ^7-1398791982574491766134898885056836640 \sqrt{3} \pi
   ^8-13394306269662689467865491966274031000 \pi ^9 \\
&&\scriptscriptstyle +9828187125855416763760085386671946800 \sqrt{3} \pi ^{10} +36913920716620168252191261044923707000 \pi
   ^{11}-36529125136359005195440425021097931724 \sqrt{3} \pi ^{12} \\
&&\scriptscriptstyle -20531503844503774756475712806217635325 \pi ^{13}  +52250200075877305640284378871036133750 \sqrt{3}
   \pi ^{14}-25106273358080443247860845512855921875 \pi ^{15}
   \Bigr] ,\\
&&\scriptscriptstyle \hat{Z}^{(16,18)}(6)  
=\frac{1}{145167018808404505250647790094395761508745216000000 \pi ^{16}}
 \Bigl[ 499088508508728373896460800000+1148069990630487253328755531776000 \sqrt{3} \pi \\
&&\scriptscriptstyle +2204123177433820377635667486720000 \pi ^2  -80832166137304211747295259361280000
   \sqrt{3} \pi ^3-146870981709838226042443135614720000 \pi ^4+2137662704524708086942013328769638400 \sqrt{3} \pi ^5 \\
&&\scriptscriptstyle +4333507855909733053165002541062912000 \pi^6 
 -31163561487184652028114399401822208000 \sqrt{3} \pi ^7-65027653921400377692940744622821488000 \pi ^8
+278481843400147886441641453342855787520 \sqrt{3} \pi^9\\
&&\scriptscriptstyle +507010393853337987925351395288072940800 \pi ^{10} 
 -1461516543904680903342407393215495987200 \sqrt{3} \pi ^{11}-1898383703828940279197072967138796980000 \pi^{12} \\
&&\scriptscriptstyle +3819576362919149697738269777021225146752 \sqrt{3} \pi ^{13} 
 +2325473090095435961117528605868105925600 \pi ^{14}  \\
&&\scriptscriptstyle -3341177451063568937491260972373242720000
   \sqrt{3} \pi ^{15}+1420183890747869311574490615653158765625 \pi ^{16}
   \Bigr] ,\\
&&\scriptscriptstyle \hat{Z}^{(17,19)}(6)  
= \frac{1}{81557153838862585521897938839673049147797202272256000000 \pi ^{17}}
\Bigl[ -439112983195936701145758975879168000 \sqrt{3}+1076278556792698817609766869586432000 \pi  \\
&&\scriptscriptstyle +48425065777444518067378103193378816000 \sqrt{3} \pi^2 
 +1972020549826087010008799913627955200000 \pi ^3-2365131596008101666098080900747089715200 \sqrt{3} \pi ^4 \\
&&\scriptscriptstyle -108694741517612648496548591058231413299200 \pi
   ^5+66373891890813761809976250869194457702400 \sqrt{3} \pi ^6+2377998408363926888156593340758416422400000 \pi ^7 \\
&&\scriptscriptstyle -1135620915824050752955993703324745803573760
   \sqrt{3} \pi ^8-26101912531467412821019263038647372999820160 \pi ^9+12385805906155630070979951640823682035205120 \sqrt{3} \pi^{10} \\
   &&\scriptscriptstyle +146630395568684606876738582176899688773312000 \pi ^{11}
-83256519394360944715825475953694806545127296 \sqrt{3} \pi
   ^{12}-368288661331332026053921317506530276708167456 \pi ^{13} \\
&&\scriptscriptstyle +303047900159493793427505427439460474181844352 \sqrt{3} \pi
   ^{14}+208923915906694334043721043321380114609080000 \pi ^{15} \\
&&\scriptscriptstyle -430978533650748595293988346116063042963822500 \sqrt{3} \pi
   ^{16}+203609092101422112674799451473899775456453125 \pi ^{17}
   \Bigr] ,
\end{\eqa *}
 \end{minipage}
 } 
 \end{figure}

\clearpage
\begin{figure}[p]
 \rotatebox{90}{
 \begin{minipage}{1.36\linewidth}
\begin{\eqa *}
&&\scriptscriptstyle \hat{Z}^{(18,20)}(6)  
=\frac{1}{652457230710900684175183510717384393182377618178048000000 \pi ^{18}}
\Bigl[ 2596258421262405001009389081600000+11107318095054118626882364305598464000 \sqrt{3} \pi \\ 
&&\scriptscriptstyle  +23303802373258992725678492029211136000 \pi
   ^2-1032472987639798739422860608045285376000 \sqrt{3} \pi ^3-2066679171508559795859901678104015360000 \pi ^4 \\
&&\scriptscriptstyle  +35002927743249184121372554322430774681600 \sqrt{3}
   \pi ^5+86747537184964597899307875350457221990400 \pi ^6-671606849547530437668819265336557905510400 \sqrt{3} \pi ^7 \\
&&\scriptscriptstyle  -1930754758262512055140873634541876915552000
   \pi ^8+8607725165427461940693864532763132320135680 \sqrt{3} \pi ^9+23839410014500396295200031362549568318267520 \pi
   ^{10}\\
&&\scriptscriptstyle  -74059628907079159742149540945021561913118720 \sqrt{3} \pi ^{11}-162616346042989972966619824807672192126200000 \pi
   ^{12}+384490645501206647632780617040348429495701888 \sqrt{3} \pi ^{13}\\
&&\scriptscriptstyle  +558049059278343856685878677173614804972605792 \pi
   ^{14}-997256308062756874082224554265091923222520832 \sqrt{3} \pi ^{15}-654657249271827076492025979509844811361784750 \pi
   ^{16}\\
&&\scriptscriptstyle  +867271418676735667061337537804433499754022500 \sqrt{3} \pi ^{17}
-363809598566867914834683537318687584410296875 \pi^{18}
   \Bigr] ,\\
&&\scriptscriptstyle \hat{Z}^{(19,21)}(6)  
=\frac{1}{2060478203387484265844386431983800000432957624779584569344000000 \pi ^{19}}
\Bigl[ -15904898042990123714979080360811939840000 \sqrt{3}+41458207856000494519885189421996775936000 \pi  \\
&&\scriptscriptstyle +2228077128735294863352982005596622824448000 \sqrt{3} \pi
   ^2+156389029666944843181362566870072954534400000 \pi ^3-140445105490805128670677239365856032074752000 \sqrt{3} \pi
   ^4 \\
&&\scriptscriptstyle -11568091795226820855199468075702123190060697600 \pi ^5
+5201496744352232030759451599541534650404147200 \sqrt{3} \pi
   ^6+354300941005918706752011427143461053447183104000 \pi ^7 \\
&&\scriptscriptstyle -117886750475579382879817606973641348752374131200 \sqrt{3} \pi
   ^8-5773112523840387992310360484317254287758208913280 \pi ^9+1797986964256111297801334005303799888186871087360 \sqrt{3} \pi
   ^{10} \\
&&\scriptscriptstyle +53078810858202617452122520867312702876287973430400 \pi ^{11}-18732679458183045964589094996361098775876099382400 \sqrt{3} \pi
   ^{12}-266391165039384616977490145534832173622056515400928 \pi ^{13}\\
&&\scriptscriptstyle +123549800133198867990270624714551291727463854565056 \sqrt{3} \pi
   ^{14}+625034140809457227883058274369425662988379365494560 \pi ^{15}
-444848450819261816977787586565824092069446662873012 \sqrt{3} \pi^{16} \\
&&\scriptscriptstyle -357338154197504213041080622569260681514491459495775 \pi ^{17}+630506310122982604989122190086441370480914314406250 \sqrt{3} \pi
   ^{18}-293394288653142371890408570666467619983024237109375 \pi ^{19}
   \Bigr] ,\\
&&\scriptscriptstyle \hat{Z}^{(20,22)}(6)  
= \frac{1}{131870605016798993014040731646963200027709287985893412438016000000 \pi ^{20}}
 \Bigl[546603785972164689368511932172472320000+4192168056529723783790216058523083423744000 \sqrt{3} \pi \\
&&\scriptscriptstyle +9482299572825339475444117827201467043840000 \pi
   ^2-498108691352894211083449648358215562428416000 \sqrt{3} \pi ^3-1081335323665801739438126236956879021076992000 \pi
   ^4 \\
&&\scriptscriptstyle +20276162513157527117918998975210781124786585600 \sqrt{3} \pi ^5+62497517617213410112927515282728748416219136000 \pi
   ^6-462067434529796376095127437790328449331799654400 \sqrt{3} \pi ^7 \\
&&\scriptscriptstyle -1959897567297279770591596595532500391588192998400 \pi
   ^8+7575401392049154930324319106891476707211700244480 \sqrt{3} \pi ^9+35201870642033135407771614064221957353311528780800 \pi
   ^{10}\\
&&\scriptscriptstyle -96286971563338005690087005676610320839863334584320 \sqrt{3} \pi ^{11}-372617594227615952559501314845689891287793942368640 \pi
   ^{12}+855033703121758304385175728327046651001417496265728 \sqrt{3} \pi ^{13}\\
&&\scriptscriptstyle +2281813231663965280857323099973323319891591826279680 \pi
   ^{14}-4521100166132713416237220277120541818980188569413632 \sqrt{3} \pi ^{15}-7288708796939658770952298573616982695487567305021112 \pi
   ^{16} \\
&&\scriptscriptstyle +11761074955706242770234659317182460615849362412096800 \sqrt{3} \pi ^{17}+8243510903418084641252252704534410307081605078045000 \pi
   ^{18}\\
&&\scriptscriptstyle -10204293076060846374166557075101543051605762249920000 \sqrt{3} \pi ^{19}
+4231285988915442409270700563223067256625379469859375 \pi^{20}
   \Bigr] ,
\end{\eqa *}
 \end{minipage}
 } 
 \end{figure}

\clearpage
\begin{figure}[p]
 \rotatebox{90}{
 \begin{minipage}{1.36\linewidth}
\begin{\eqa *}
&&\scriptscriptstyle \hat{Z}^{(1,4)}(6)   =\frac{5}{12 \pi }-\frac{2}{9 \sqrt{3}} ,\quad
 \hat{Z}^{(2,5)}(6) = -\frac{1728+192 \sqrt{3} \pi -281 \pi ^2}{31104 \pi ^2} ,\quad
 \hat{Z}^{(3,6)}(6) = \frac{-83592+10368 \sqrt{3} \pi +127035 \pi ^2-22840 \sqrt{3} \pi ^3}{10077696 \pi ^3} ,\\
&&\scriptscriptstyle  \hat{Z}^{(4,7)}(6)
 =\frac{4455648+414720 \sqrt{3} \pi -11109312 \pi ^2-971904 \sqrt{3} \pi ^3+1592539 \pi ^4}{5804752896 \pi ^4} ,\quad
 \hat{Z}^{(5,8)}(6)
= \frac{238295520-21461760 \sqrt{3} \pi -1411557840 \pi ^2+93070080 \sqrt{3} \pi ^3+1737995139 \pi ^4-302777720 \sqrt{3} \pi ^5}
{3134566563840 \pi ^5} ,\\
&&\scriptscriptstyle \hat{Z}^{(6,9)}(6)
= -\frac{76892820480+5576325120 \sqrt{3} \pi -573013245600 \pi ^2-1383091200 \sqrt{3} \pi ^3+1090674751104 \pi ^4+79621753536 \sqrt{3} \pi ^5-148557967675 \pi
   ^6}{13541327555788800 \pi ^6} ,\\
  &&\scriptscriptstyle \hat{Z}^{(7,10)}(6)
  = \frac{-31425707738880+2113494405120 \sqrt{3} \pi +429913865757600 \pi ^2+7801923628800 \sqrt{3} \pi ^3-1742440664795544 \pi ^4+32206476790656 \sqrt{3} \pi
   ^5+1964049821256195 \pi ^6-332651386503800 \sqrt{3} \pi ^7}{71660705425234329600 \pi ^7} ,\\ 
&&\scriptscriptstyle \hat{Z}^{(8,11)}(6)
  = \frac{1}{82553132649869947699200 \pi ^8}
\Bigl[ 2315467859604480+130161909104640 \sqrt{3} \pi -34843187276513280 \pi ^2+3014457537208320 \sqrt{3} \pi ^3+177112509501098304 \pi ^4
-11881252279830528 \sqrt{3}\pi ^5 \\
 &&\scriptscriptstyle  -310252204178676864 \pi ^6-19421735763949824 \sqrt{3} \pi ^7+41007215197066475 \pi ^8
   \Bigr] ,\\
&&\scriptscriptstyle \hat{Z}^{(9,12)}(6)
= -\frac{1}{80241644935673589163622400 \pi ^9}
\Bigl[ -145843276422251520+7480839015014400 \sqrt{3} \pi +3671605788162923520 \pi ^2+351856049737973760 \sqrt{3} \pi ^3
-31386652836855414336 \pi^4 \\
&&\scriptscriptstyle -3629211613868090880 \sqrt{3} \pi ^5 
 +106948701526929355296 \pi ^6+2180932927606245888 \sqrt{3} \pi ^7-115224671246959029591 \pi ^8+19058564900547151000
   \sqrt{3} \pi ^9
   \Bigr] ,\\
&&\scriptscriptstyle \hat{Z}^{(10,13)}(6)
= -\frac{1}{577739843536849841978081280000 \pi ^{10}}
\Bigl[59424658405126963200+2607345528496128000 \sqrt{3} \pi -1494646843349681280000 \pi ^2+276712093111394304000 \sqrt{3} \pi ^3
+14241443740755440670720 \pi^4 \\
&&\scriptscriptstyle  -3145180931422213017600 \sqrt{3} \pi ^5-63325975881575145816000 \pi ^6+7942059992375119411200 \sqrt{3} \pi ^7+107725328454640124507328 \pi
   ^8+5879363304117906600000 \sqrt{3} \pi ^9-13946884924872838124375 \pi ^{10}
   \Bigr] ,\\
&&\scriptscriptstyle \hat{Z}^{(11,14)}(6)
= \frac{1}{7549904275339553734969566167040000 \pi ^{11}}
\Bigl[-43996418266722182553600+1750333991776351027200 \sqrt{3} \pi +1794947928415916351616000 \pi ^2
+315585216449482484736000 \sqrt{3} \pi^3 \\
&&\scriptscriptstyle -26519121593636888111132160 \pi ^4 
 -6791071204650309034106880 \sqrt{3} \pi ^5+180384961146149907828964800 \pi ^6+36985994723795947364620800 \sqrt{3} \pi
   ^7-559735818006615235937698824 \pi ^8 \\
&&\scriptscriptstyle -29954344057179479732310912 \sqrt{3} \pi ^9+588264631762732876297770975 \pi ^{10}
-95359518807271769754935000 \sqrt{3} \pi^{11}
   \Bigr] ,\\
&&\scriptscriptstyle \hat{Z}^{(12,15)}(6)
= \frac{1}{13046234587786748854027410336645120000 \pi ^{12}}
\Bigl[ 3888350787580134543360000+134424416727794633932800 \sqrt{3} \pi -145906896136766039964057600 \pi ^2
+44000698719609952911360000 \sqrt{3} \pi^3 \\
&&\scriptscriptstyle +2186295711339895922635545600 \pi ^4-885133454804481258074603520 \sqrt{3} \pi ^5-17675240499320755967078031360 \pi ^6+6056394726785187025916928000 \sqrt{3} \pi
   ^7+74493064178295232470870149280 \pi ^8\\
&&\scriptscriptstyle -13122219995749934095847851008 \sqrt{3} \pi ^9-125673879667279669106172105024 \pi ^{10}-6046875111706034787686217600
   \sqrt{3} \pi ^{11}+16019905458889520591171759375 \pi ^{12}
 \Bigr] , \\
&&\scriptscriptstyle \hat{Z}^{(13,16)}(6) 
= -\frac{1}{238119873696283740083708293464446730240000 \pi ^{13}}
\Bigl[-3608102942195863693175193600+112879582873953846696345600 \sqrt{3} \pi +220052533396372639101890150400 \pi ^2 \\
&&\scriptscriptstyle +57827617497511962505858252800 \sqrt{3} \pi
   ^3-5005777725306638239230403292160 \pi ^4-2088006427359976403276063907840 \sqrt{3} \pi ^5+56245971637991179160763170135040 \pi
   ^6 \\
&&\scriptscriptstyle +23616031401582363829127165460480 \sqrt{3} \pi ^7-337585840274226906674842179016224 \pi ^8-97993822616708448714474411277056 \sqrt{3} \pi
   ^9+994123552057806633404778410526576 \pi ^{10} \\
&&\scriptscriptstyle +82135186115141484988489908686592 \sqrt{3} \pi ^{11}-1030590646305537582693111694626675 \pi
   ^{12}+164172996736677957360032776075000 \sqrt{3} \pi ^{13}
   \Bigr] ,
\end{\eqa *}
 \end{minipage}
 } 
 \end{figure}

\clearpage
\begin{figure}[p]
 \rotatebox{90}{
 \begin{minipage}{1.36\linewidth}
\begin{\eqa *}
&&\scriptscriptstyle \hat{Z}^{(14,17)}(6)   
= -\frac{1}{3360347657601956140061291437370272257146880000 \pi^{14}}
 \Bigl[ 2389057014651126852082021171200+65698433766906751948161024000 \sqrt{3} \pi -123781064682186877429231217049600 \pi ^2 \\
&&\scriptscriptstyle +55046198799925296179864430182400
   \sqrt{3} \pi ^3+2601568429000132673475755992350720 \pi ^4-1669968254800625566356738898329600 \sqrt{3} \pi ^5
-32540450556450080556412464001397760 \pi^6 \\
&&\scriptscriptstyle +20081095016873762428751582007459840 \sqrt{3} \pi ^7+247041181687428123019584262350455808 \pi ^8-112038933045827512726470954756311040 \sqrt{3} \pi
   ^9-1018899684818888046104189378086085664 \pi ^{10} \\
&&\scriptscriptstyle +225297105676708301290659045256897536 \sqrt{3} \pi ^{11}+1720832966823395628384922630399311360 \pi
   ^{12}+73569901730947262474592714028180800 \sqrt{3} \pi ^{13}-216674275307532485144334955022325625 \pi ^{14}
   \Bigr] ,\\
&&\scriptscriptstyle \hat{Z}^{(15,18)}(6)   
= \frac{1}{27218816026575844734496460642699205282889728000000 \pi ^{15}}
\Bigl[ -898700383096439185117648945152000+22368420103478267720549007360000 \sqrt{3} \pi \\
&&\scriptscriptstyle +77531309401121148711700384542720000 \pi^2 
 +27892526887178816642101198356480000 \sqrt{3} \pi ^3-2514585516783591144734581795723468800 \pi ^4-1534388901202423550357496451252224000 \sqrt{3} \pi^5 \\
&&\scriptscriptstyle  +41581557778769108504531140550689152000 \pi ^6 
 +28888856344525009970811969977730048000 \sqrt{3} \pi ^7-401636993683654539142726841461969194240 \pi^8 \\
&&\scriptscriptstyle  -239983696226888526294118453394719641600 \sqrt{3} \pi ^9 
  +2256740846813144661718725723167844040800 \pi ^{10} 
 +844010037209044375981979849638890451200 \sqrt{3}\pi ^{11} \\
&&\scriptscriptstyle  -6463981988123546979501753446214630152184 \pi ^{12} 
  -703163144799081000418299245890630185600 \sqrt{3} \pi ^{13} \\
&&\scriptscriptstyle  +6654334406359127352880965494490407266875
   \pi ^{14}-1043901176569132351962808253588525875000 \sqrt{3} \pi ^{15}
   \Bigr] ,\\
&&\scriptscriptstyle \hat{Z}^{(16,19)}(6)   
= \frac{1}{62712152125230746268279845320778968971777933312000000 \pi ^{16}}
\Bigl[ 90127565300227557177080441339904000+1989219180562744837319361036288000 \sqrt{3} \pi \\
&&\scriptscriptstyle -6086879461747599524298861104332800000 \pi^2 
 +3806151766381697481961667202908160000 \sqrt{3} \pi ^3+163759850350427760782660390415325593600 \pi ^4 \\
&&\scriptscriptstyle -158654026975586065480121866421325004800 \sqrt{3} \pi^5 
 -2865413537203518504566095432347648000000 \pi ^6 
 +2842199991676492247320779897442861056000 \sqrt{3} \pi ^7\\
&&\scriptscriptstyle +33964951775142977349118882712190993070080 \pi^8  -27502499397548699839371434246254034288640 \sqrt{3} \pi ^9 
-251246273219944513504881609440166948864000 \pi ^{10} \\
&&\scriptscriptstyle +138781542559564325556139299013421350502400\sqrt{3} \pi ^{11} 
 +1026881021321101118405595247489748594683008 \pi ^{12}
 -268452394166167533096672887170738280976384 \sqrt{3} \pi^{13} \\
&&\scriptscriptstyle  -1743942369638653966665930861839919484396800 \pi ^{14} 
 -66621085645103713260100911501297756480000 \sqrt{3} \pi
   ^{15}+217370492060820059041340470348100509421875 \pi ^{16}
   \Bigr] ,\\
&&\scriptscriptstyle \hat{Z}^{(17,20)}(6)   
= -\frac{1}{1957371692132702052525550532152153179547132854534144000000 \pi ^{17}}
\Bigl[ -120889230958601642729079302038880256000+2417861827849522283898055481622528000 \sqrt{3} \pi  \\
&&\scriptscriptstyle +14177644320711654416353827568445816832000 \pi
   ^2+6646979543701355919426306325767782400000 \sqrt{3} \pi ^3-619756855153875209831179686368838362726400 \pi ^4 \\
&&\scriptscriptstyle -525243030733861472854812318548395740364800 \sqrt{3}\pi ^5 
 +13844946268419810027656083744161646898380800 \pi ^6+14838581561193547881840059784280432312320000 \sqrt{3} \pi^7 \\
&&\scriptscriptstyle -191215581312309114558205663398303363105822720 \pi ^8 
 -200941199475875574644835787362865219121725440 \sqrt{3} \pi
   ^9+1719852396699294761246503154056922291135907840 \pi ^{10} \\
&&\scriptscriptstyle +1375660362580908279421469688224562968506368000 \sqrt{3} \pi^{11} 
 -9416863288002561362472259570534833666206708352 \pi ^{12}-4313105597616062970317147796981468155934956544 \sqrt{3} \pi^{13}\\
&&\scriptscriptstyle  +26662903847476393636552341568283318328033812544 \pi ^{14} 
 +3538067167007835312886011607310986492528358400 \sqrt{3} \pi^{15} \\
&&\scriptscriptstyle  -27373374398914281806076421955274149565839184375 \pi ^{16}
+4235931041565321332788848816966816968971375000 \sqrt{3} \pi^{17}
   \Bigr] ,
\end{\eqa *}
 \end{minipage}
 } 
 \end{figure}
 
\clearpage
\begin{figure}[p]
 \rotatebox{90}{
 \begin{minipage}{1.36\linewidth}
\begin{\eqa *}
&&\scriptscriptstyle \hat{Z}^{(18,21)}(6)   
= -\frac{1}{5073507426007963720146226979338381041386168358952501248000000 \pi ^{18}}
\Bigl[12762704492165762167566045678674116608000+227940214368036116111307200230588416000 \sqrt{3} \pi \\
&&\scriptscriptstyle -1073019104874844780079224731004898770944000 \pi
   ^2+920129921200905040115941751618764013568000 \sqrt{3} \pi ^3+33733217566169149972247827330549202760499200 \pi ^4\\
&&\scriptscriptstyle -49295388650125071812886895378229386385817600
   \sqrt{3} \pi ^5-768061093550480111115970561472885905873305600 \pi ^6+1188404345252945117937749465012554208850739200 \sqrt{3} \pi
   ^7\\
&&\scriptscriptstyle +13312804959484884601524704234134056446640783360 \pi ^8-16953234735035511015931927281485577730643066880 \sqrt{3} \pi
   ^9-154752665920866216680530674067573428782116116480 \pi ^{10} \\
&&\scriptscriptstyle +148356030943006803493298579170267543543041884160 \sqrt{3} \pi
   ^{11}+1120045834395682981662352660022192105893173866496 \pi ^{12}-714013156682187907777183011448198829100614197248 \sqrt{3} \pi
   ^{13}\\
&&\scriptscriptstyle -4551258049671777965384602224773009232539020844928 \pi ^{14}+1355530290169841776773920299880833498153682989056 \sqrt{3} \pi
   ^{15}+7788219021062834643491455989237338074218002184000 \pi ^{16}\\
&&\scriptscriptstyle +266923009461650175343998565765421267155971240000 \sqrt{3} \pi
   ^{17}-962464176228450056617521658966970102792458046875 \pi ^{18}
   \Bigr] ,\\
&&\scriptscriptstyle \hat{Z}^{(19,22)}(6)   
= \frac{1}{197805907525198489521061097470444800041563931978840118657024000000 \pi ^{19}}
\Bigl[-19981123334801427324621351891364947689472000+323945999526022043371943110994118574080000 \sqrt{3} \pi  \\
&&\scriptscriptstyle +3092789475541536089368801097017661745659904000 \pi
   ^2+1829476491182639556134215736606934602612736000 \sqrt{3} \pi ^3-174308434231860166101259653296438625327041740800 \pi
   ^4\\
&&\scriptscriptstyle -199454803294979645458629275540414993780441088000 \sqrt{3} \pi ^5+4870014401509569712787679918374600266658404761600 \pi
   ^6+7935266692884137206096483026916400718816765542400 \sqrt{3} \pi ^7 \\
&&\scriptscriptstyle -86749663392106079559474492695577178206656175882240 \pi
   ^8-158433738645555744645008864656886251928909768294400 \sqrt{3} \pi ^9+1114316067886446441725313566133112578033589992314880 \pi
   ^{10}\\
&&\scriptscriptstyle +1738334342409463058996459850762830181123435687198720 \sqrt{3} \pi ^{11}-9985684385001231296853671212270544596443376549653504 \pi
   ^{12}-10387869124326065679191717460222243658668751539978240 \sqrt{3} \pi ^{13}\\
&&\scriptscriptstyle +54972723791046308381814328130595618554038458142739328 \pi
   ^{14}+29890762330352061454223597628253936614966067204764672 \sqrt{3} \pi ^{15}-155630454510719902872838531346205403899500458775297928 \pi
   ^{16}\\
&&\scriptscriptstyle -24098057729919583688588685887945453548598105132822400 \sqrt{3} \pi ^{17}+159800550025924262455294478582679302562583478394931875 \pi
   ^{18}-24426064582146572034331657158127984482336945966875000 \sqrt{3} \pi ^{19}
   \Bigr] ,\\
&&\scriptscriptstyle \hat{Z}^{(20,23)}(6)   
=\frac{1}{2848405068362858249103279803574405120598520620495297708661145600000000 \pi ^{20}}
\Bigl[11042399320250999977738360524623065110282240000+160779743303448635817401329998420875673600000 \sqrt{3} \pi \\
&&\scriptscriptstyle  -1111240919339964695807122138390885519746662400000
   \pi ^2+1290387344237375919029063587879685567859916800000 \sqrt{3} \pi ^3+36227247672942025578005936404767379915852873728000 \pi
   ^4 \\
&&\scriptscriptstyle  -84145049993118302211708791398228708747389173760000 \sqrt{3} \pi ^5-1016872663423927222797985605774115756901148917760000 \pi
   ^6+2514841616520920355571454957180936883845380177920000 \sqrt{3} \pi ^7 \\
&&\scriptscriptstyle  +26230586759198915239786619932679505826191812213145600 \pi
   ^8-47382596771754379856743909009782451680354056798208000 \sqrt{3} \pi ^9
-456589355576035555873016927702192668810195028164608000 \pi^{10} \\
&&\scriptscriptstyle  +616213562433320523399391392342128599848688909910016000 \sqrt{3} \pi ^{11}+5033980445032586108846443861682103764245441310034462720 \pi
   ^{12}-5234927919419561245190940585371633939033356902993100800 \sqrt{3} \pi ^{13} \\
&&\scriptscriptstyle  -35129675227123350335845131292774314691392777548145868800 \pi
   ^{14}+24850467532183735592099483897155331294239998303605145600 \sqrt{3} \pi ^{15}+141704781206899451941039918909424668261482696598254880224 \pi
   ^{16}\\
&&\scriptscriptstyle  -46891482293862283956508791767367882552728431680105139200 \sqrt{3} \pi ^{17}-244534768579791078543821544134733211770818618315903640000 \pi
   ^{18} \\
&&\scriptscriptstyle  -7540069315966367353932952193293700525820039335970000000 \sqrt{3} \pi ^{19}
+29996619925240338939317620477337462628372360432030859375 \pi^{20}
   \Bigr] .
\end{\eqa *}
 \end{minipage}
 } 
 \end{figure}

\clearpage
\providecommand{\href}[2]{#2}\begingroup\raggedright\endgroup

\end{document}